\documentclass[fleqn,usenatbib]{mnras}

\usepackage{newtxtext,newtxmath}

\usepackage{url}            
\usepackage{booktabs}       
\usepackage{amsfonts}       
\usepackage{nicefrac}       
\usepackage{microtype}      
\usepackage{lipsum}
\usepackage[position =top]{subfig}
\usepackage{graphicx}
\usepackage{amsmath}
\usepackage{xspace}
\usepackage{multirow}
\usepackage{tablefootnote}
\usepackage{latexsym}
\usepackage{stackengine}

\usepackage[T1]{fontenc}

\DeclareRobustCommand{\VAN}[3]{#2}
\let\VANthebibliography\thebibliography
\def\thebibliography{\DeclareRobustCommand{\VAN}[3]{##3}\VANthebibliography}

\newcommand{\dm}{\ensuremath{\mathbf{M_{DM_{\boldsymbol{\alpha}}}}}\xspace}
\newcommand{\dmInit}{\ensuremath{\mathbf{M_{DM_{\boldsymbol{\alpha_0}}}}}\xspace}

\newcommand{\wfs}{\ensuremath{\mathbf{M_{WFS}}}\xspace}
\newcommand{\Da}{\ensuremath{ \mathbf{D_{\boldsymbol{\alpha}}}}\xspace}
\newcommand{\DaInit}{\ensuremath{ \mathbf{D_{\boldsymbol{\alpha_0}}}}\xspace}
\newcommand{\Ra}{\ensuremath{ \mathbf{R_{\boldsymbol{\alpha}}}}\xspace}
\newcommand{\gA}{\ensuremath{\boldsymbol{\alpha}}\xspace}

\newcommand{\dD}{\ensuremath{\mathbf{\boldsymbol{\delta} D_{\boldsymbol{{\alpha_0}}}}(\varepsilon_{i})}\xspace}

\newcommand{\phiRes}[1]{\ensuremath{\boldsymbol{\phi^{res}_{#1}}}\xspace}

\newcommand{\gDelta}{\ensuremath{\boldsymbol{\delta}}}
\newcommand{\norm}[1]{\left\lVert#1\right\rVert}

\newcommand{\ggamma}{\ensuremath{\textbf{G}}\xspace}

\title[SPRINT: A fast online AO calibration strategy]{SPRINT: System Parameters Recurrent INvasive Tracking, a fast and least-cost online calibration strategy for adaptive optics.}
\author[C. T. Heritier et al.]{
C. T. Heritier$^{1}$\thanks{E-mail: cheritie@eso.org},
T. Fusco$^{2,3}$,
S. Oberti$^{1}$,
B. Neichel$^{2}$,
S. Esposito$^{4}$,
and P.-Y. Madec$^{1}$ 
\\
\\
$^{1}$ European Southern Observatory, Karl-Schwarzschild-str-2, 85748 Garching, Germany\\
$^{2}$ Aix Marseille Univ, CNRS, CNES, LAM, Marseille, France\\
$^{3}$ ONERA, DOTA, Unit\'{e} HRA, 29 avenue de la division Leclerc, 92322 Chatillon, France\\
$^{4}$ INAF - Osservatorio Astrofisico di Arcetri Largo E. Fermi 5, 50125 Firenze Italy\\
}

\date{Accepted 2021 April 21. Received 2021 April 16; in original form 2021 February 15}

\pubyear{2021}

\begin{document}
\label{firstpage}
\pagerange{\pageref{firstpage}--\pageref{lastpage}}
\maketitle


\begin{abstract}
The future large adaptive telescopes will trigger new constraints for the calibration of Adaptive Optics (AO) systems equipped with pre-focal Deformable Mirrors (DM). The image of the DM actuators grid as seen by the Wave-Front Sensor (WFS) may evolve during the operations due to the flexures of the opto-mechanical components present in the optical path. The latter will result in degraded AO performance that will impact the scientific operation. To overcome this challenge, it will be necessary to regularly monitor and compensate for these DM/WFS mis-registrations either by physically re-aligning some optical components or by updating the control matrix of the system. In this paper, we present a new strategy to track mis-registrations using a pseudo-synthetic model of the AO system. The method is based on an invasive approach where signals are acquired on-sky, before or during the scientific operations, and fed to the model to extract the mis-registration parameters. We introduce a method to compute the most sensitive modes to these mis-registrations that allows to reduce the number of degrees of freedom required by the algorithm and minimize the impact on the scientific performance. We demonstrate that, using only a few of these well selected signals, the method provides a very good accuracy on the parameters estimation, well under the targeted accuracy, and has a negligible impact on the scientific path. In addition, the method appears to be very robust to varying operating conditions of noise and atmospheric turbulence and performs equally for both Pyramid and Shack-Hartmann WFS.    

\end{abstract}




\section{Introduction}

A large fraction of ground-based astronomical observations relies on the performance of Adaptive Optics (AO) systems that allow to compensate in real time for the wave-front aberrations induced by the atmospheric turbulence. The principle of an AO system consists in measuring signals related to the phase using a Wave-Front-Sensor (WFS) that are converted by a Real-Time Computer (RTC) into commands to apply on a Deformable Mirror (DM) that cancels out the optical aberrations. This loop is usually operated in a feedback-loop running at least ten times faster than the typical evolution rate of the atmospheric turbulence. 

Within a few years, the new generation of telescopes with diameters up to 39 meters, the Extremely Large Telescopes (ELT), will start their scientific operations. These giants will address fundamental astrophysical science cases such as direct imaging and characterization of rocky exoplanets located close to their orbiting star or the study of bulk and evolution of the first galaxies (\citealt{cirasuolo2018elt}). The scientific potential of these new telescopes relies on challenging new AO systems' features such as DM integrated inside the telescope itself, turning the telescopes into adaptive telescopes (\citealt{arsenault2006AOF}, \citealt{riccardi2010adaptive}, \citealt{vernet2012specifications}). The colossal size of these large adaptive telescopes and the complexity of their scientific instruments compel us to a complete rethinking, in order to improve the overall performance, but more specifically the sensitivity and the robustness of the AO systems to maximize the astrophysical return of AO-assisted instruments.
\begin{figure*}
    \begin{center}
    \subfloat[No Mis-registration]{\includegraphics[width=0.25\textwidth]{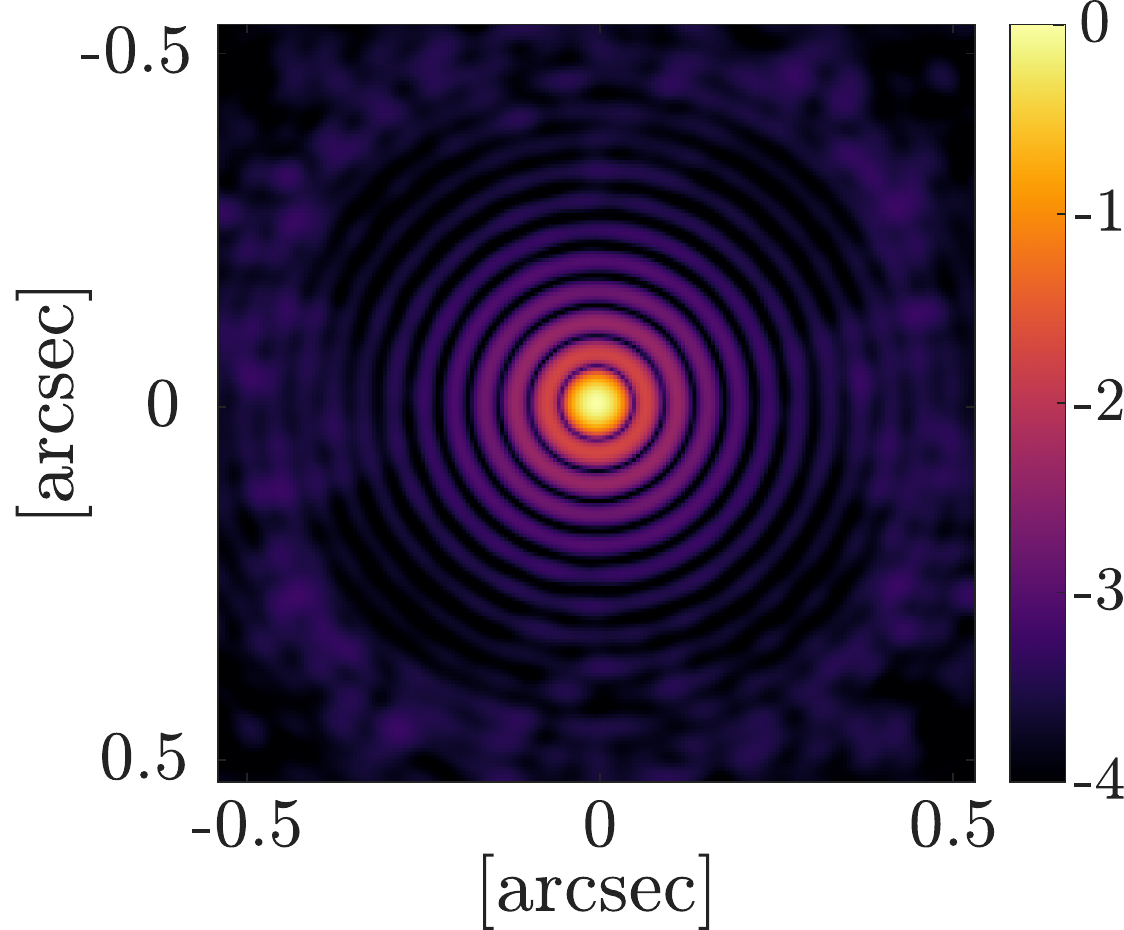}}
    \subfloat[Shift X: 15\% of a subap.]{\includegraphics[width=0.25\textwidth]{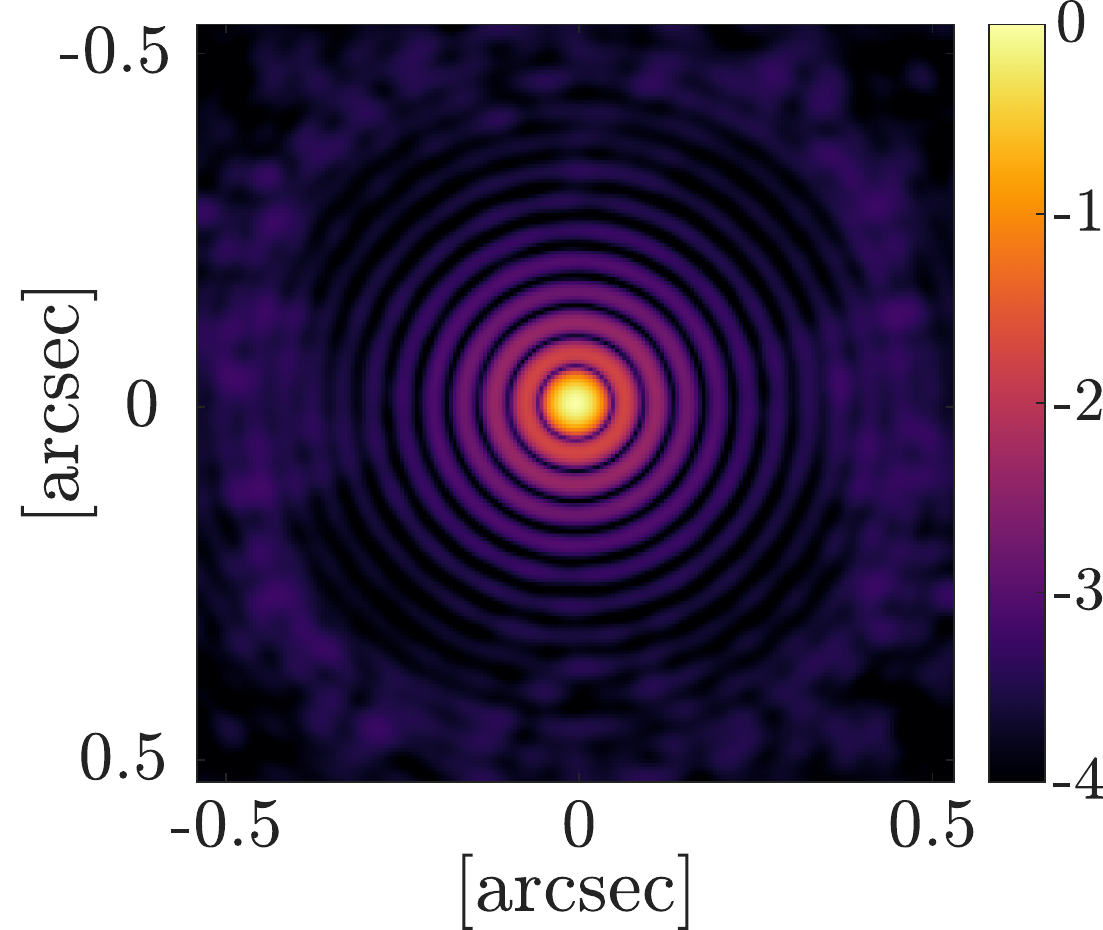}}
    \subfloat[Shift X: 30\% of a subap.]{\includegraphics[width=0.25\textwidth]{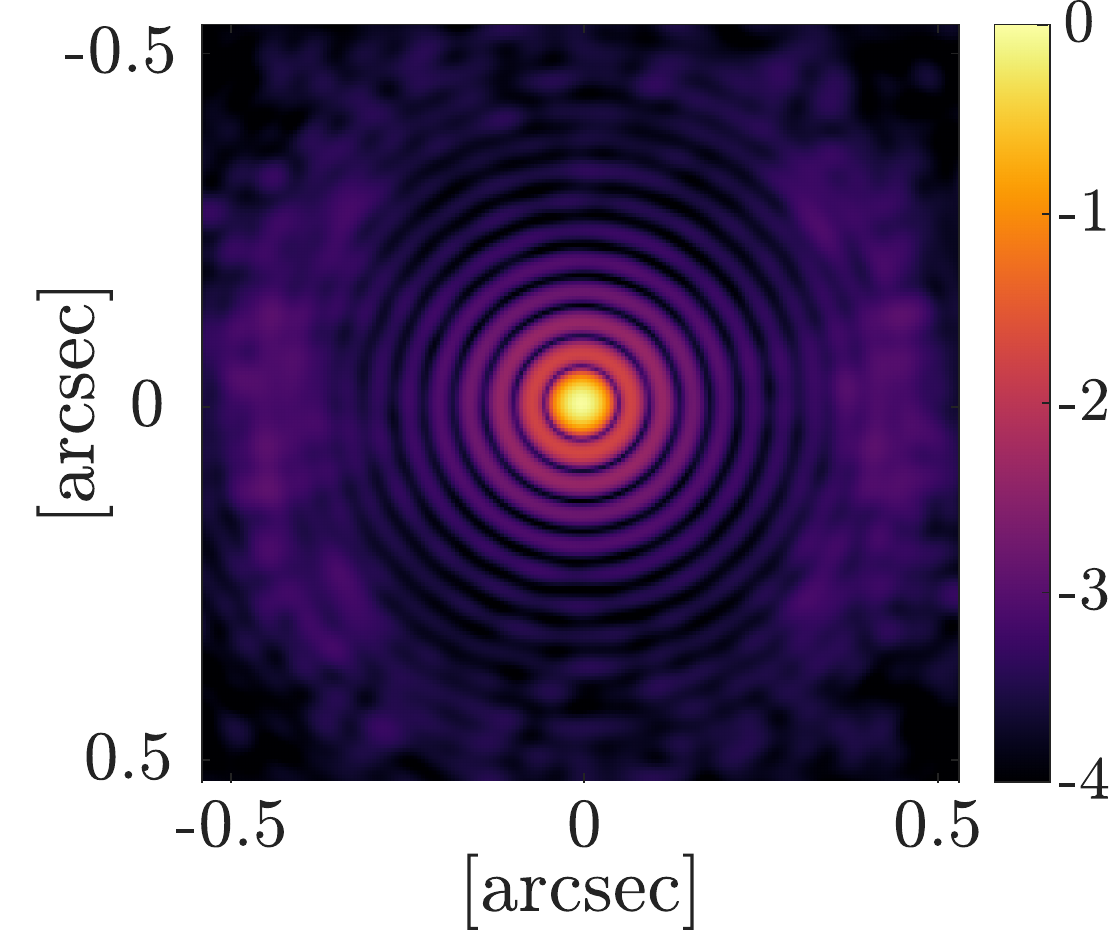}\label{psf_shift}}
    \subfloat[Shift X: 45\% of a subap.]{\includegraphics[width=0.25\textwidth]{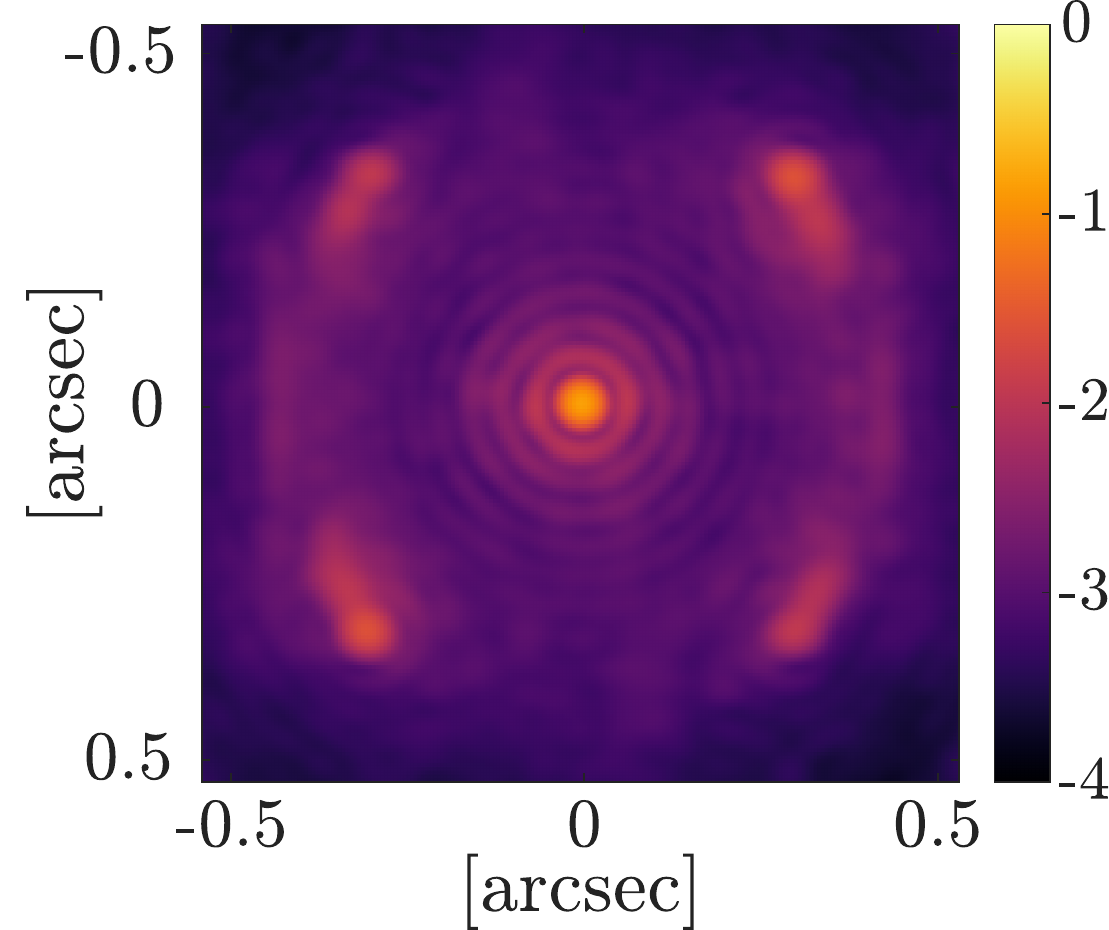}\label{psf_shift_div}}\\
    
    \subfloat[Strehl Ratio]{\includegraphics[height=4.8cm]{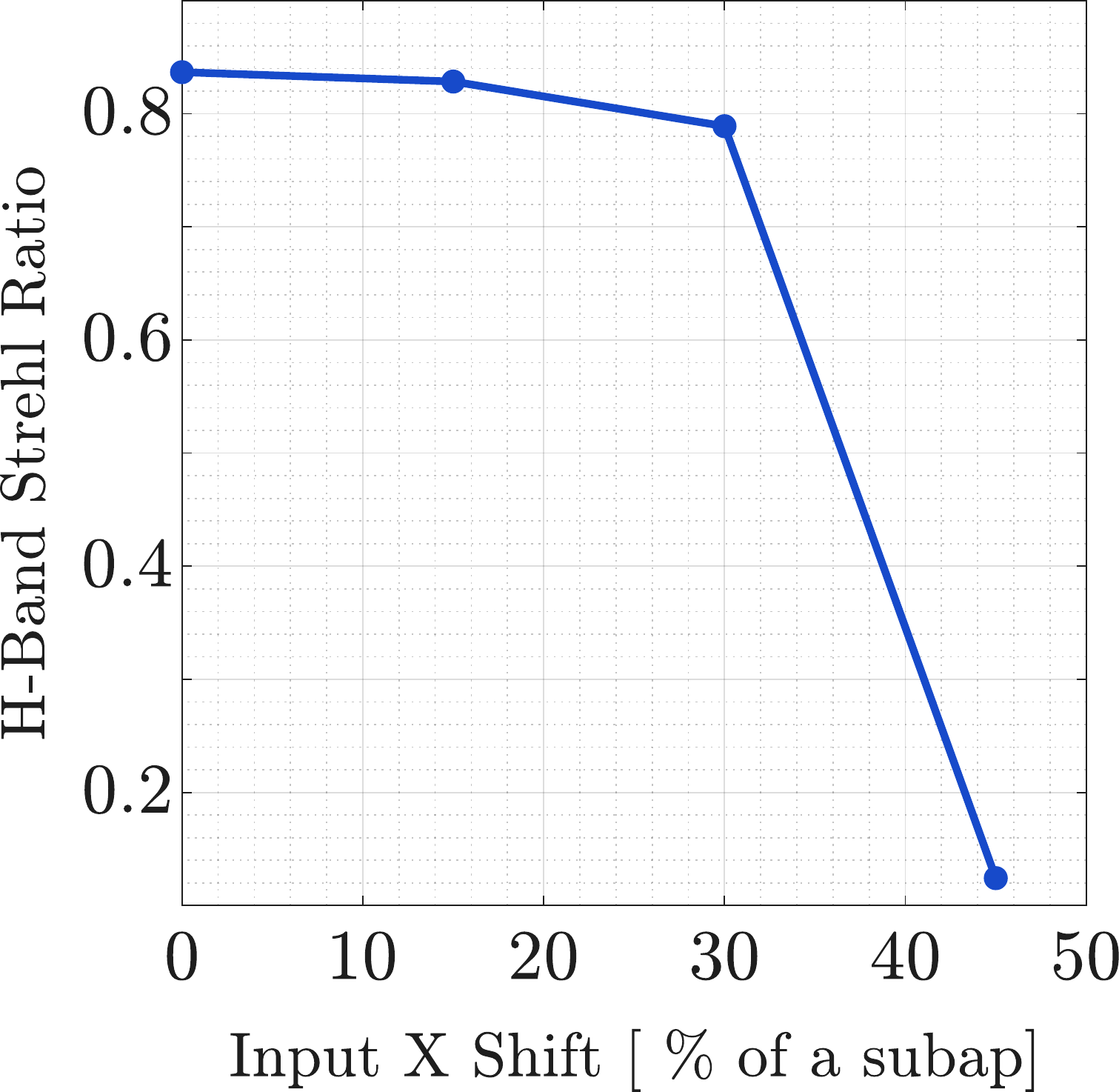}\label{wfe_shift}}
    \subfloat[Modal PSD]{\includegraphics[height=4.8cm]{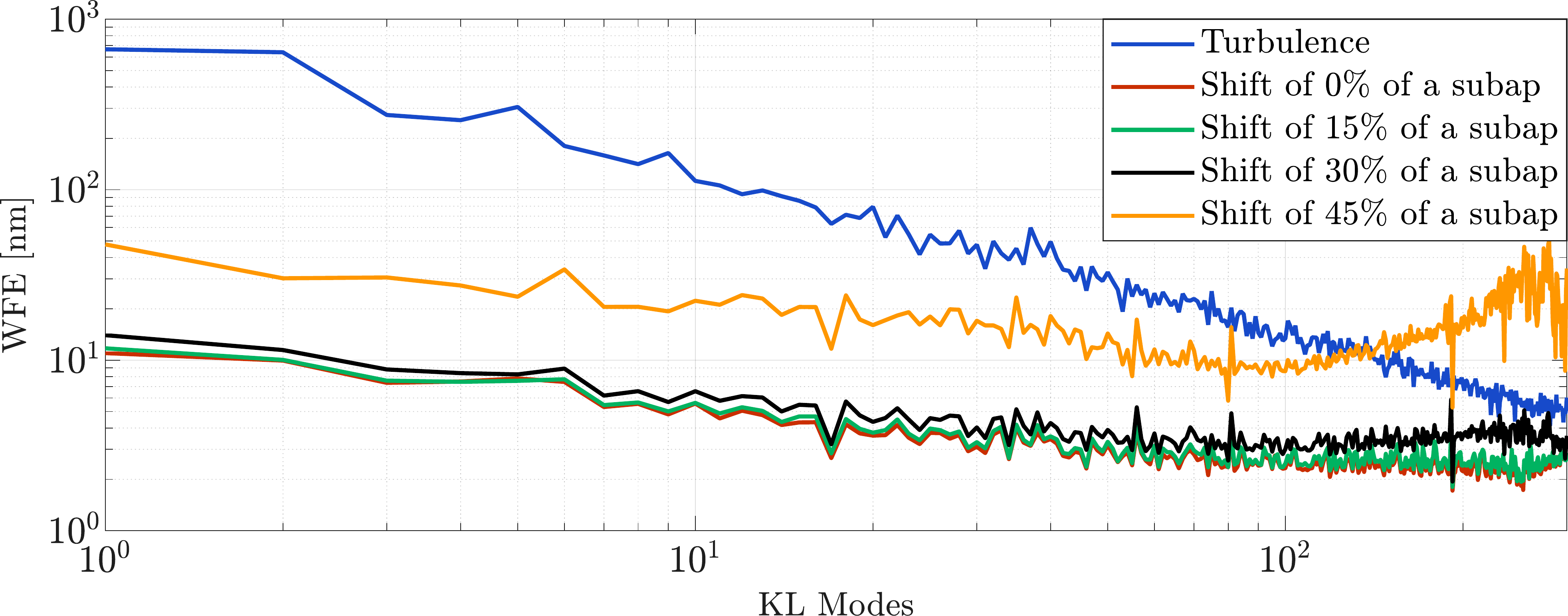}\label{psd_shift}}
    \caption{Impact of a mis-registration on the performance of an 8-meter class telescope equipped with a 20 by 20 sub-apertures AO system. Top: Impact of a shift on H-Band PSF. Bottom: Impact of a shift on the AO performance (Strehl Ratio (SR) and Modal Power Spectral Density (PSD)).}
    \label{fig:misRegImpactScience}
    \end{center}{}
\end{figure*}

In particular, the ESO-ELT (\citealt{gilmozzi2007european}), will provide a challenging environment for the AO systems. First of all, the calibration of a large number of degrees of freedom (around 5000 actuators (\citealt{vernet2012specifications})) with no external calibration source will be required. Furthermore, the use of a pre-focal DM far from the AO instruments with moving elements in the optical path may lead to regular evolution of the DM/WFS registration during the observations (rotation, shifts or higher order of pupil distortion of the DM actuators grid with respect to the WFS subapertures). These so-called mis-registrations have to be monitored and compensated as they will highly affect the AO performance or could create loop instabilities that will jeopardise the scientific observations.

An illustration of the effect of mis-registrations on the performance of a scientific instrument is given in Figure \ref{fig:misRegImpactScience} that provides simulated H-Band Point-Spread Functions (PSF) and closed-loop performance for an 8-meter telescope equipped with a 21 by 21 actuators DM and a 20 by 20 subapertures Pyramid WFS. This Figure shows that high-order modes are the first impacted by the mis-registrations. This is visible in the PSF (Figures \ref{psf_shift} and \ref{psf_shift_div}), that exhibit speckles on the external rings of the Airy pattern, and in the modal PSD of the AO residuals (Figure \ref{psd_shift}) that exhibits peaks going above the turbulence level for the high-order modes. For large values of shift (above 30\% of a subaperture), the loop becomes even unstable (Figure \ref{psf_shift_div} and \ref{wfe_shift}) and would require to interrupt the scientific operation to re-calibrate the system. 

Hence, to provide a nominal correction for the scientific instruments, the AO loop has to be properly calibrated before and during the operations. The accuracy required for the WFS/DM registration to prevent any impact on the scientific operation is system dependent and depends on the number of modes controlled by the AO loop. The value of 10 \% of a subaperture shift (and equivalent shift on the border of the pupil for the rotation) is usually taken as a reference (\citealt{bechet2012optimization}, \citealt{heritier2018new}) but should be carefully evaluated (see Section \ref{subsection_impact_science}). 

In this context, a scheme based on the use of pseudo-synthetic models to calibrate the AO systems has been proposed (\citealt{oberti2006large}, \citealt{bechet2012optimization}, \citealt{kolb2012calibration}, \citealt{heritier2017overview}, \citealt{heritier2018new}). It offers a fast way to numerically update the calibration of the system and relies only on the identification of a few model mis-registration parameters. By contrast, a scheme where the calibration is achieved using only on-sky measurements (\citealt{pinna2012first}) would require long telescope overheads (see Section  \ref{calibration_post_focal}).
\begin{figure*}
    \centering
    \includegraphics[width=\textwidth]{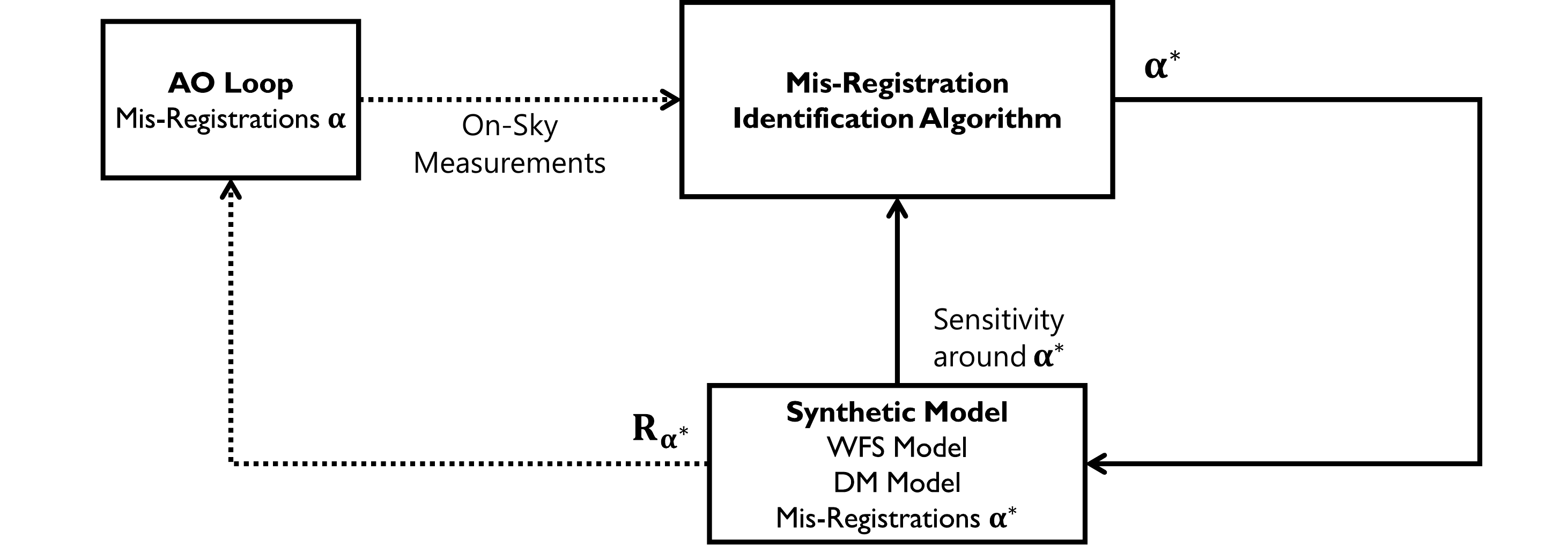}
    \caption{Principle of the Pseudo-Synthetic calibration. Experimental inputs representative of the real mis-registration state \gA are injected into a mis-registration identification algorithm that provides an estimation \gA$^*$. The identification algorithm is presented in section \ref{algo_principle} }
    \label{fig:PSIM}
\end{figure*}
A pseudo-synthetic calibration requires to develop a synthetic model that must be representative of the real AO system to allow computing numerically its associated interaction matrix (equation \ref{iMat_def}). A block diagram summarizing the principle of a pseudo-synthetic calibration is provided in Figure \ref{fig:PSIM}. It shows the key-elements of the model: 

\begin{itemize}
    \item The synthetic model, usually developed using an end-to-end simulator. It includes the DM and  WFS and the corresponding sensitivity to the different mis-registrations. In this paper we use the OOMAO simulator (\citealt{conan2014object}) that has been validated at the Large Binocular Telescope (\citealt{heritier2018new}).
    \item Experimental inputs to tune the model such as WFS valid pixels, DM influence functions, Modal Basis, WFS signals...
    \item A mis-registration identification algorithm fed on experimental inputs (DM commands and WFS signals that are representative of the real mis-registration \gA). In this paper, the strategy consists in projecting a given experimental signal that is composed of on-sky measurements of a few well selected modes onto sensitivity interaction matrices that correspond to these modes. More details can be found in section \ref{algo_principle}.
\end{itemize}

This paper focuses on the identification of the mis-registration parameters but more information about synthetic modelling of real AO systems can be found in \cite{oberti2006large}, \cite{pinna2012first},  \cite{kolb2012calibration} and \cite{heritier2018new}. 

Previous research works have already investigated the tracking of mis-registration parameters. For instance, at the Adaptive Optics Facility (\citealt{arsenault2006AOF}), the calibration baseline is to use a pseudo-synthetic model in which mis-registration parameters are identified from closed-loop data (\citealt{bechet1a2011identification}, \citealt{kolb2012calibration}, \citealt{oberti2018AOinTheAOF}). This identification strategy consists in correlating the incremental closed-loop DM commands to the incremental WFS signals to capture the interaction between WFS and DM. In \cite{neichel2012identification}, a strategy based on a Levenberg-Marquardt (\citealt{marquardt1963algorithm}) type algorithm is presented in the frame of tomographic AO systems but requires an experimental interaction matrix as a reference, that could for instance be measured on-sky. In the frame of Multi-Conjugate AO systems, the identification of the interaction matrix using on-sky signals has been investigated in \cite{chiuso2010dynamic} and more recently for Ground Layer AO systems in \cite{lai2020crime} that consists in a model-free method to reconstruct fully the interaction matrix injecting known random signals on the DM during the operation.

In this paper, we present a new strategy to monitor the mis-registrations during the operations applying known (and invisible to the science in most of the cases) perturbations on the DM and update the calibration using synthetic models of the AO systems. In particular, we investigate if the perturbation can be reduced to only a few well selected modes to provide a good estimation of the mis-registration parameters and minimize the impact on the science path to make it compatible with Single Conjugate AO applications.

For the sake of simplicity, we consider in this paper a simple AO system with 20 by 20 sub-apertures and 8 meter class telescope to properly validate and characterize the method. In addition, the experience acquired at the AOF showed that magnification X and Y were initially measured but eventually removed from the tracked parameters as they appeared to remain static over time. For this reason, we reduce the study to the tracking of only 3 mis-registration parameters (shift X and Y and rotation) as being the most occurring mis-registration errors (due to mechanical flexures and de-rotation errors). If required, the structure of the mis-registration identification algorithm (section \ref{algo_principle}) is fully compatible with any other type of mis-registration. It is then straightforward to include other types of mis-registrations depending on the system considered. 

The performance of the method for an ELT-like system will be investigated in a second step to include more complex challenges that will affect the closed-loop operations (\citealt{bonnet2018adaptive}, \citealt{le2016simulations}), in particular in presence of large optical gains variations when operating with PWFS (\citealt{deo2018modal},\citealt{deo2019telescope}, \citealt{chambouleyron2020pyramid}), pupil fragmentation effects due to the large thickness of the spiders holding the secondary mirror (\citealt{bonnefond2016wavefront}, \citealt{schwartz2018analysis}, \citealt{bertrou2020petalometry}), impact of phasing errors for segmented primary mirrors (\citealt{meimon2008phasing}, \citealt{briguglio2018possible}, \citealt{cheffot2020measuring}) and complex and large adaptive secondary mirrors (\citealt{biasi2010contactless}, \citealt{riccardi2010adaptive}, \citealt{madec2012overview}, \citealt{vernet2014way}, \citealt{briguglio2018optical}). 

We first recall the AO calibration procedures for post-focal AO systems and large adaptive telescopes (\ref{section_AO_calibration_procedure}). The strategy to monitor the mis-registrations using on-sky measurements is presented in \ref{section_Invasive} and an analysis of the accuracy and robustness of the method is provided in \ref{section_numerical_simulations} as well as an analysis of the impact on the scientific path. 

\section{Calibration of an AO system}
\label{section_AO_calibration_procedure}
The interaction matrix \Da of an AO system is defined as the image of the DM influence functions \dm as seen by the WFS:
\begin{equation}
    \Da=\wfs.\dm
    \label{iMat_def}
\end{equation}
where \wfs is the WFS measurement model and the notation $\boldsymbol{\alpha}$ corresponds to the DM/WFS registration state. The notations chosen here correspond to apply the mis-registrations in the DM space, considering that the WFS is fixed. In addition, we assume that the image of the pupil remains fixed on the WFS and that there are no errors of pupil stabilization. To operate the AO system in closed-loop, it is required to invert the interaction matrix to provide the command matrix \Ra. This is often achieved using a Truncated Singular Value Decomposition (TSVD), removing the modes associated to the lowest singular values (\citealt{boyer1990adaptive}). More advanced inversion methods such as adding priors on the noise and turbulence allows to improve the level of AO correction (\citealt{wallner1983optimal}, \citealt{fusco2001optimal}).

The interaction matrix of the system is usually projected onto a reduced modal basis that differs from the zonal actuation and that contains only the modes well seen by the WFS. This allows to preliminary filter out the modes badly seen by the AO system and provides a stable inversion of the interaction matrix. 
Typically, a common method consists in using Karhunen-Lo\`eves (KL) modes that are computed diagonalizing both the statistical covariance matrix of the atmosphere and the geometrical covariance matrix of the DM influence functions (\citealt{gendron1995optimisation}). In this paper, the closed loop control is always done using a number of KL modes slightly inferior to the number of actuators located in the pupil to ensure a small conditioning number for the interaction matrix (inferior to 10) and thus a stable inversion.

\subsection{Interaction matrix estimation of a post-focal AO system}

For a post-focal AO system, the interaction matrix is usually estimated prior to the operations using a set of calibration signals \textbf{U} that defines the control space of the AO loop, such as the KL Modes introduced in the previous section. In the following development, it is however assumed that the matrix \textbf{U} is full rank, which corresponds to calibrating all the degrees of freedom of the DM. The WFS measurements \textbf{Y} corresponding to the actuation signals \textbf{U} are given by: 
\begin{equation}
    \textbf{Y}=\wfs.[\dm.\textbf{U} + \Phi] + \eta
\end{equation}
where $\Phi$  is the local turbulence and $\eta$ the WFS noise (Photon and Readout noise).

Assuming that both the amplitude of the calibration signals and the local turbulence are small enough to remain in the linear regime of the WFS, the previous equation can be distributed such that the interaction matrix \Da of the system appears (equation \ref{iMat_def}) 
\begin{equation}
    \textbf{Y}=\Da.\textbf{U} +  \wfs.\Phi + \eta
\end{equation}
The interaction matrix of the system can be estimated by computing: 
\begin{equation}
    \Da = \textbf{Y}.\textbf{U}^\dag -  [\wfs.\Phi + \eta].\textbf{U}^\dag
\end{equation}
where $[\wfs.\Phi + \eta].\textbf{U}^\dag$ is the calibration error on the interaction matrix estimation and the symbol $\dag$ represents the pseudo-inverse.

For a post-focal AO system, where both DM and WFS are located on the same bench, the local turbulence can be considered as negligible and the use of a bright external calibration source allows to minimize the contribution of the measurement noise $\eta$ so that:
\begin{equation}
    \Da = \textbf{Y}.\textbf{U}^\dag
\end{equation}

In this situation, the evolution of the mis-registrations is slow since both DM and WFS are located on the same optical bench. The validity of \Da is then granted for long period of time and can easily be re-calibrated using day-time calibration, plugging an external calibration source.

\subsection{Calibration of an AO system for large adaptive telescopes}
\label{calibration_post_focal}
In the case of adaptive telescopes, frequent evolution of the DM/WFS registration are to be expected due to flexures and to the large distance separating the DM from the AO instrument. In addition, the access to an external calibration source is not often granted which leads to consider different approaches to calibrate the AO system. 

Regarding the first point, to provide nominal AO performance, it becomes necessary to regularly monitor and compensate for the mis-registrations during the operations. Considering the situation in which an AO system has been calibrated in a registration state $\alpha_0$ and evolved to a registration state $\alpha$, the compensation of the mis-registrations can be achieved following two strategies:
\begin{itemize}
    \item The compensation can be achieved updating the interaction matrix corresponding to the new DM/WFS registration. Using the notations introduced in this paper, it corresponds to the following transformation:
    \begin{equation}
        \DaInit \xrightarrow[]{} \Da
        \label{numerical_compensation}
    \end{equation}
    
    \item The compensation can be done optically, by physically re-aligning some elements in the optical path. This corresponds to the following transformation:
\begin{equation}
    \dm \xrightarrow[]{} \dmInit
    \label{optical_compensation}
\end{equation}
\end{itemize}

The update of the interaction matrix can be obtained experimentally by measuring a full interaction matrix during the operation (\citealt{oberti2004calibration}, \citealt{oberti2006large}, \citealt{pinna2012first}) or by computing the new interaction matrix numerically from a pseudo-synthetic model, fed with a few experimental parameters (\citealt{oberti2004calibration}, \citealt{kolb2012calibration}, \citealt{bechet2012optimization}, \citealt{oberti2018AOinTheAOF}, \citealt{heritier2018new}). 

On that aspect, in the context of the future large adaptive telescopes, fast evolution of potentially large mis-registrations are expected. Achieving a full measurement of the interaction matrix at each update comes then with a cost in terms of telescope operation, in particular for such a large number of degrees of freedom to calibrate (typically around 5000 modes (\citealt{vernet2012specifications}) in the case of the ESO-ELT). The corresponding time required to obtain high-SNR measurements of the interaction matrix becomes problematic especially because its validity is only ensured at the time of the measurement. Typically, for the de-commissioned FLAO system at LBT, a full measurement of the on-sky interaction matrix, multiplexing 2 modes at the same time, required 53 min of telescope time (including overheads) (\citealt{pinna2012first}). Extrapolating these numbers to an ELT-SCAO system (that contains about 10 times more degrees of freedom) leads to potentially several hours of calibration time, during which the measurements could be impacted by mis-registrations and eventual optical gains variations in the case of a Pyramid WFS (\citealt{korkiakoski2008improving}, \citealt{deo2018modal}, \citealt{deo2019close}, \citealt{chambouleyron2020pyramid}). 

As a consequence, in this context, this strategy does not seem suited to provide a regular tracking of the calibration during the operations. The method could however be useful at the beginning of the observations to provide a first interaction matrix that would be used as a reference for the next steps of calibration or to measure only a few reference signals and retrieve mis-registration parameters (see section \ref{section_Invasive}).

To comply with these requirements of fast and regular updates, a calibration based on the on-sky identification of few mis-registration parameters, making use of a synthetic model of the interaction matrix, appears to be a good candidate as it allows either to re-compute a whole interaction matrix(that will also have the advantage of being noise-free) or to re-align physically the optical system.

Updating the full interaction matrix \Da numerically (equation \ref{numerical_compensation}) provides the easiest solution in terms of software architecture since the only output is a reconstructor \Ra that is injected into the RTC. It requires a tuning of the model that can be achieved during the commissioning phase of the instrument using experimental measurements of the real system. In addition, it also allows to include high-order mis-registrations such as distortion that might not be correctable with an optical device. However, if too large mis-registrations occurred, the AO performance could be impacted due to an unusual DM/WFS registration. For instance, in a Fried geometry, locating the actuators of a system on the center of the WFS subapertures could result in a loss of sensitivity (\citealt{southwell1980wave}) depending on the AO system properties (actuator mechanical coupling and WFS sub-aperture size and number). In addition, depending on the optical design of the system, large drifts of the system could lead to pupil truncation effects, DM actuators not seen by the WFS sub-apertures and/or WFS sub-apertures corresponding to non-controlled actuators. All of these effects will impact the performance of the AO system. 

Alternatively, the second compensation strategy allows to optically re-align the system (equation \ref{optical_compensation}) and ensures to always operate around the nominal working point, \textit{i.e.} keeping the DM/WFS registration constant around the optimal configuration. In that case, the computation of only one numerical interaction matrix is required which reduces the computational load of the approach. It requires however a more complex software architecture to control opto-mechanical elements in the optical path that are more limited in terms of degrees of freedom than a numerical model.

The choice of the compensation strategy is however system-dependent and might be driven by other constraints in terms of software or hardware configuration. This paper focuses then only on the identification of the mis-registration parameters and will not investigate the AO performance when a given type of compensation is applied. In addition, it is assumed that the model of the system has been perfectly tuned to be representative of the real system considered. Therefore the results presented do not include other model errors than mis-registration estimation errors.


\section{Mis-registration identification strategy}
\label{section_Invasive}
In this section, we introduce a strategy that consists in acquiring on-sky signals to identify the mis-registration parameters. The particularity of this approach is that it can be applied either offline or during operation. In the latter case, it shall be verified that the actual impact on the science observations can be neglected as the signals could act as a noise on the scientific measurements depending on the signal amplitude. 

The strategy to acquire the signals required by the mis-registration identification algorithm is based on the same methods used for on-sky calibrations: dithering specific signals over the closed-loop DM commands using either fast push-pull measurements or sinusoidal modulation. More details about on-sky calibration can be found in \cite{oberti2004calibration}, \cite{oberti2006large}, \cite{pieralli2008sinusoidal} or \cite{pinna2012first}. 

If the acquisition has to be achieved during the operation, a scheme based on a sinusoidal modulation of the signals seems better suited since it allows to reduce the amplitude of the dithering signals to reach the same SNR obtained using a push pull approach, hence reducing the impact on the science. It requires however a good knowledge of the AO loop properties, in particular the loop delay (\citealt{pinna2012first}).  
For both methods, the SNR of the on-sky acquisition depends on the observing conditions (turbulence, level of noise) and on the AO system properties (number of actuators, bandwidth). The parameters for the signal measurements such as duration, frequency of modulation and amplitude of the signals have then to be tailored accordingly. Since both methods allow to retrieve signals with equivalent SNR, we propose to consider only the push-pull measurements to narrow down the analysis to the spatial properties of the signals. Some examples of the impact of the temporal modulation of signals on the science path are available in \cite{deo2019telescope} and in \cite{esposito2020sky}. In this section we propose to detail the 3 key steps of the identification strategy: mis-registration identification algorithm ( section \ref{algo_principle}), closed-loop signal acquisition strategy (section \ref{subsection_push_pull}) and choice of the signal (section \ref{subsection_PCA_modes}).
\subsection{Mis-Registration Identification Algorithm}
\label{algo_principle}
\subsubsection{Principle}

This section describes the principle of the mis-registration identification algorithm to extract the mis-registration parameters from a given interaction matrix that can be projected on a given modal basis. This algorithm is based on the work presented in \cite{kolb2012calibration} but we adapted it to take into consideration the PWFS optical gains and only a small number of signals from the interaction matrix. The general idea is to project an estimation of the interaction matrix on a set of sensitivity matrices that describe the sensitivity of the system to a given type of mis-registration around a given working point (in a small perturbation regime). This is based on the hypothesis that a given interaction matrix \Da can be decomposed as a linear combination of sensitivity matrices around the working point $\boldsymbol{\alpha_0}$: 
\begin{equation}
   \Da= \ggamma \left(\mathbf{D_{\alpha_0}} + \sum_{i}\alpha_i.\dD\right)
\label{model}
\end{equation}
where $\ggamma$ is a diagonal matrix accounting for the gain variations between \Da and $\mathbf{D_{\alpha_0}}$. The origin of these gain variations come from the mis-registrations and from the fact that \Da can be acquired on-sky and exhibit optical gains variations as it is the case for PWFS\footnote{This is based on the hypothesis that the optical gains matrix can be considered as diagonal (\citealt{deo2018modal},\citealt{chambouleyron2020pyramid})}. The sensitivity matrices \dD are defined as the partial directional derivative of the interaction matrix corresponding to a mis-registration of type $i$ (typically rotation and shifts):
\begin{equation}
\dD=\left(\frac{\mathbf{D}(\boldsymbol{{\alpha_0}} + \varepsilon_i)-\mathbf{D}(\boldsymbol{{\alpha_0}} - \varepsilon_i)}{2\varepsilon_i}\right)_{i=rot,X,Y,...}
\label{sensitivity_imat_def}
\end{equation}
where $\boldsymbol{\alpha_0}$ is the working point of the system, or in other words the vector of mis-registration amplitudes corresponding to the alignment of the system at this operating point, and $\varepsilon_{i}$ a vector of mis-registration amplitudes to compute the sensitivity matrices, chosen to be small enough to remain in the domain of validity of the hypothesis of linearity (typically one percent of subaperture). The notation $\varepsilon_i$ corresponds to the small mis-registration applied to compute the interaction matrix and is different from the parameter $\boldsymbol{\alpha_i}$ that corresponds to the mis-registration to identify.

For each type of mis-registration, the corresponding sensitivity matrices are concatenated in a meta sensitivity matrix $\boldsymbol{\Lambda_{\alpha_0}}$\footnote{In practice the matrices \dD are reshaped as vectors $\mathbf{\boldsymbol{\delta} d}_{\alpha_0}(\varepsilon_i)$, of length $N_S$=$N_{WFS}$ WFS signals $\times$ $N_{DM}$ sets of DM commands but for the sake of clarity, we prefer the matrix notation.}:
\begin{equation}
    \boldsymbol{\Lambda_{\alpha_0}} =
     {
     \begin{bmatrix} 
    {\mathbf{\boldsymbol{\delta} D}_{\alpha_0}(\varepsilon_{{rot}})} & {\mathbf{\boldsymbol{\delta} D}_{\alpha_0}(\varepsilon_{X})} &  {\mathbf{\boldsymbol{\delta} D}_{\alpha_0}(\varepsilon_{Y})} &\hdots\\
     \end{bmatrix}
     }
\end{equation}
Such that equation \ref{model} can be written:
\begin{equation}
   \Da= \ggamma \left(\mathbf{D_{\boldsymbol{\alpha_0}}} +\boldsymbol{\alpha}.\boldsymbol{\Lambda_{\alpha_0}}\right)
\label{model_algo}
\end{equation}
where $\boldsymbol{\alpha}$ is the vector of mis-registration parameters defined as 
\begin{equation}
    \boldsymbol{\alpha}=\left\{
    \begin{array}{l}
    \alpha_{rot}\\    
    \alpha_{X}\\    
    \alpha_{Y}\\    
    \hdots\\
    \end{array}
    \right\}
\end{equation}
The goal is to identify $\boldsymbol{\alpha*}$ that minimizes the criterion:
\begin{equation}
    \boldsymbol{\alpha^*}=\text{arg}\,\min \limits_{\ggamma, \boldsymbol{\alpha}}\  \norm{ \Da- \ggamma \left(\mathbf{D_{\alpha_0}}+\boldsymbol{\alpha}.\boldsymbol{\Lambda_{\alpha_0}}\right)}^2
        \label{convergence_criteria}
\end{equation}
The convergence conditions of the criterion are discussed in Section \ref{algorithm_large_mis_reg}. The solution of the least-square criterion is given by:
\begin{equation}
    \ggamma^*=\text{diag}\left((\mathbf{D_{\boldsymbol{\alpha_0}}}+ \boldsymbol{\alpha^*}\boldsymbol{\Lambda_{\alpha_0}})^\dag.\Da\right)
    \label{estim_ggamma}
\end{equation}
and
\begin{equation}
    \boldsymbol{\alpha^*}=(\boldsymbol{\Lambda_{\alpha_0}})^\dag.\left({\Da}{\ggamma^*}^{-1}-\mathbf{D_{\boldsymbol{\alpha_0}}} \right) 
        \label{estim_alpha}
\end{equation}

\subsubsection{Implementation}

Since the identification of both $\ggamma^*$ and $\boldsymbol{\alpha^*}$ are required, the algorithm has to be iterative. One iteration of the algorithm has two steps. The first step consists in estimating the optical gains taking $\boldsymbol{\alpha^*}=\mathbf{0}$ as starting point. By replacing in equation \ref{estim_ggamma}:
\begin{equation}
    \ggamma^*_1=\text{diag}\left(\mathbf{D_{\boldsymbol{\alpha_0}}}^\dag.\Da \right)
\end{equation}
The value of $\ggamma^*_1$ is injected in equation \ref{estim_alpha} to get the first estimation of $\boldsymbol{\alpha^*}$:
\begin{equation}
    \boldsymbol{\alpha^*_1}=(\boldsymbol{\Lambda_{\alpha_0}})^\dag.({\Da}{\ggamma^*_1}^{-1} -\mathbf{D_{\boldsymbol{\alpha_0}}}))
\end{equation}
From the estimation $\boldsymbol{\alpha^*_1}$, the matrix $\mathbf{D_{\boldsymbol{\alpha_0}}}$ can be updated by computing the new interaction matrix $\mathbf{D_{\boldsymbol{\alpha^*_1}}}$ corresponding to the mis-registration $\boldsymbol{\alpha^*_1}$ and get the next estimation of $\ggamma^*_2$ using equation \ref{estim_ggamma}. 
In practice, only a few iterations are required to converge. In our case, we consider that the algorithm has converged once the relative error between two successive estimation of the parameters is under 1\%.

\subsubsection{Convergence conditions}
\label{algorithm_large_mis_reg}
\begin{figure}
\centering
\subfloat[Rotation]{\includegraphics[width=0.24\textwidth]{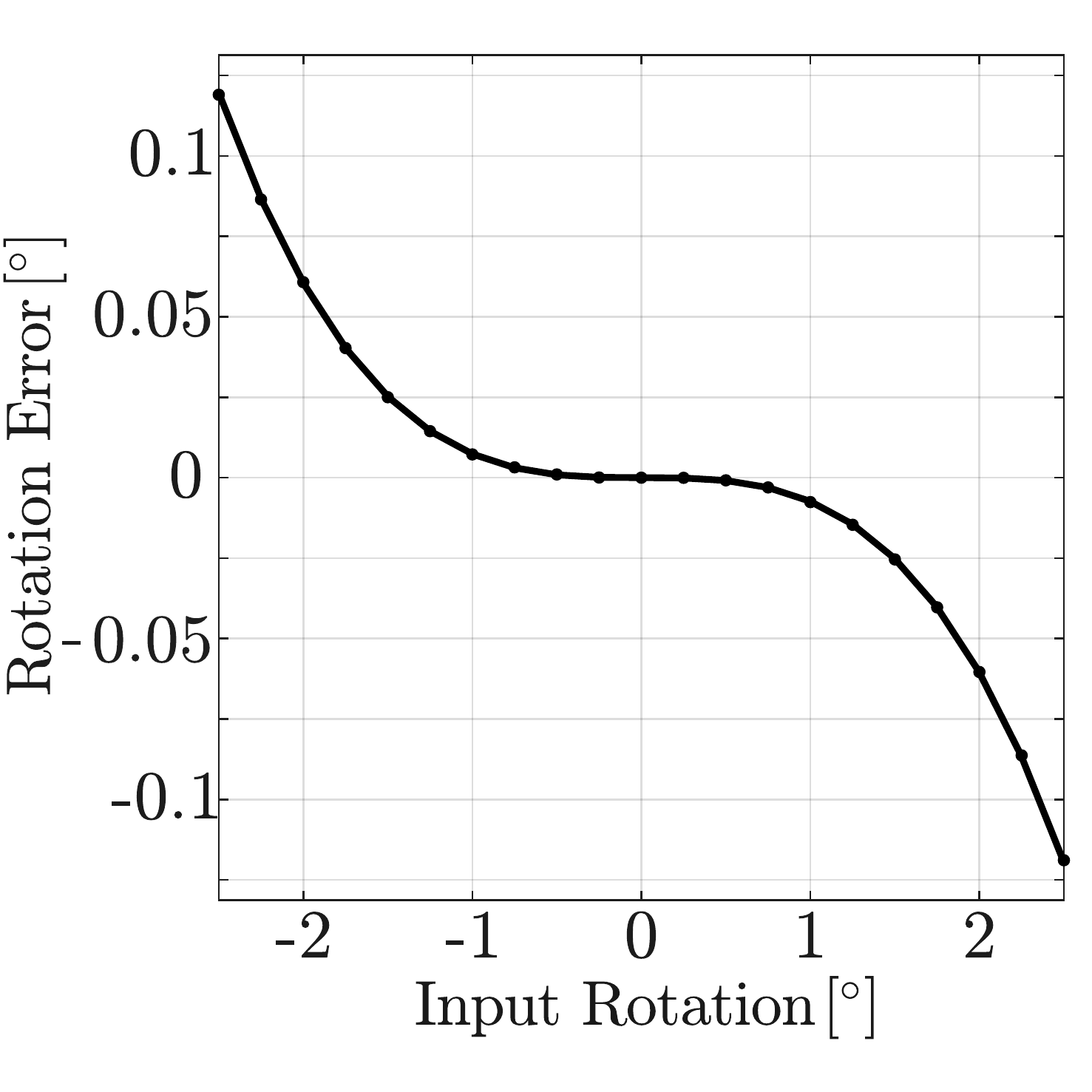}}
\subfloat[Shift X]{\includegraphics[width=0.24\textwidth]{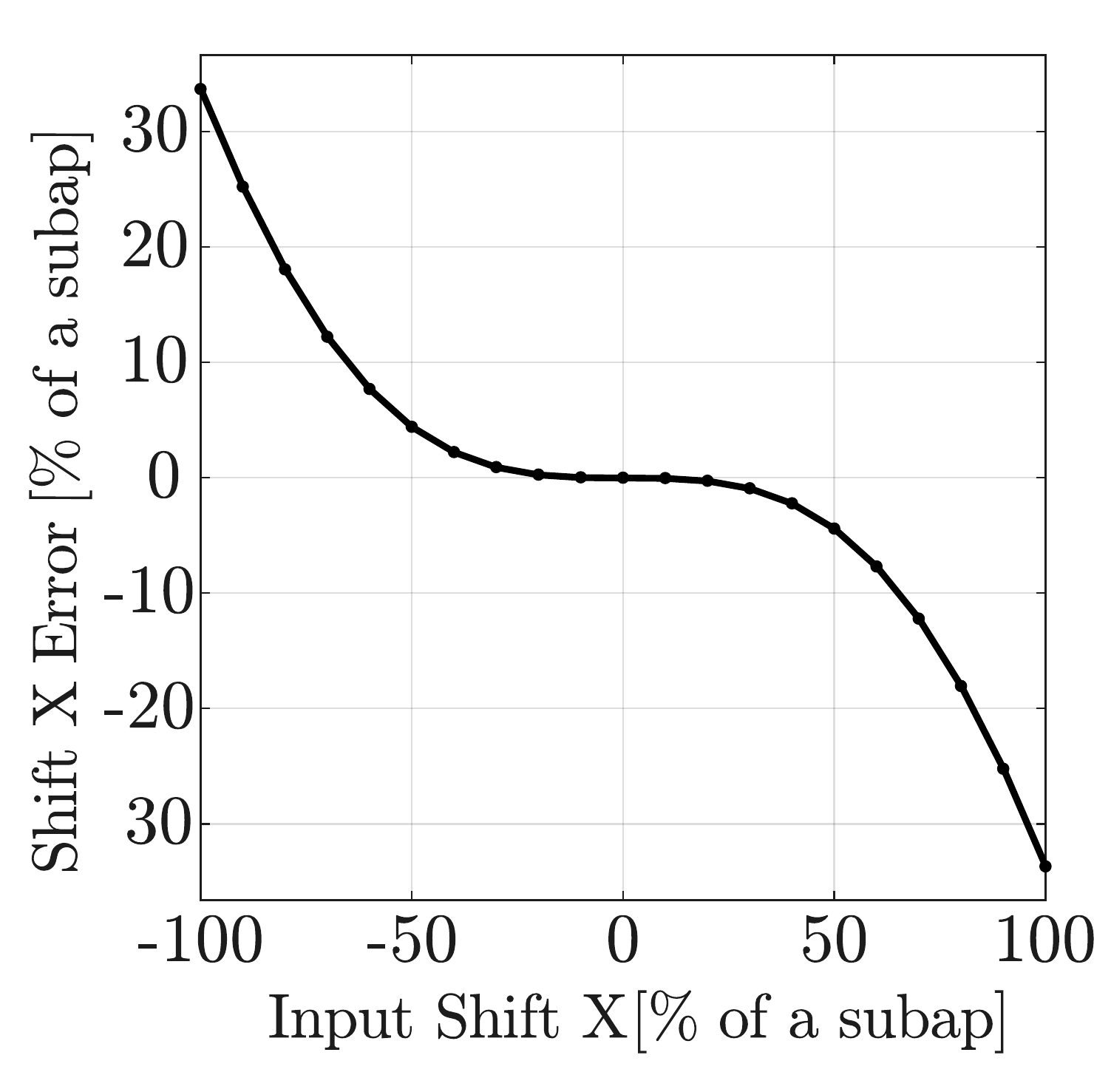}}\\
\caption{Linearity curves of the identification algorithm for the rotation(left) and shift X (right.) }
    \label{fig:linearity_algorithm}
\end{figure}
The criterion defined in equation \ref{convergence_criteria} is derived from the first order Taylor’s development of the interaction matrix (equation \ref{model}). By construction, its convexity and the uniqueness of the solution are only ensured within a limited linear range. We did not formally investigate the convergence limitations of the procedure but it has been verified that the algorithm converges for a large range of values around the operating point  (up to 70 \% of a sub-aperture shift and more than 2 degree of rotation). 

As a complement, we provide the linearity curves for both rotation and shift parameters in Figure \ref{fig:linearity_algorithm}. These plots show that within 60 \% of a sub-aperture shift and within 1.5 degree of rotation, the algorithm remains very linear. For larger mis-registrations, the working point $\boldsymbol{\alpha_0}$ around which the sensitivity matrices are computed must be updated. This is achieved by computing $\boldsymbol{\Lambda_{\alpha^*}}$ after each estimation\footnote{the estimation of $\boldsymbol{\alpha^*}$ is the convergence value of the iterative procedure to identify both scaling factors $\ggamma^*$ and mis-registration parameters $\boldsymbol{\alpha^*}$.} or by physically re-aligning the system after each estimation. This will allow the algorithm to converge to the right parameters using only a few iterations. Using this iterative approach, it has been empirically verified that the algorithm converges for very large mis-registrations (5 degrees of rotation and more than 100 \% of a sub-aperture shift).
 
If the mis-registration are too far out of the linear regime, typically 300 \% of a sub-aperture shift or 20 degrees of rotation, the convergence of the algorithm is not granted anymore. In addition, for such large mis-alignments, depending on the optical design, pupil truncation effects could alter the convergence of the algorithm but this topic is out of the scope of this paper. These very large mis-registrations should however only occur during commissioning phase where other methods less accurate could be used, for instance based on flux considerations, and provide a rough estimate of the mis-registration parameters. In closed-loop conditions, the system is not expected to drift too far from its initial working point, at least not on a fast timescale, and no problem of convergence is expected.
\subsection{On-sky Push-Pull measurement}
\label{subsection_push_pull}
In the previous section, the procedure to identify mis-registration parameters from a given interaction matrix has been presented. This section will now recall how to calibrate on-sky an interaction matrix, or a subset of it.
The measurement $\mathbf{y_k}$ of the WFS at the loop sample $k$ is given by: 
\begin{equation}
       \mathbf{y_k}=\wfs.\phiRes{k} + \boldsymbol{\eta_k} 
\end{equation}
Where \wfs defines the WFS measurement model (see equation \ref{iMat_def}). The push-pull measurement of a mode, represented by a phase vector ${\mathbf{b}}$ requires:
\begin{equation}
\mathbf{y_k^{+b}}=\wfs.\phiRes{k} + \boldsymbol{\eta_k} + a.\wfs.\mathbf{b}
\end{equation}
and 
\begin{equation}
\mathbf{y_k^{-b}}=\wfs.\phiRes{k+1} + \boldsymbol{\eta_{k+1}} - a\wfs.\mathbf{b}
\end{equation}
where $a$ is the amplitude of the mode considered. 
The sensitivity measurement $\mathbf{y_k^b}$ of the mode ${\mathbf{b}}$ is then given by: 
\begin{equation}
\begin{split}
  \mathbf{y_k}^\mathbf{b} &= \frac{\mathbf{y_k^{+b}}-\mathbf{y_k^{-b}}}{2a} \\ 
  &= \wfs.{\mathbf{b}}+\frac{\wfs.(\phiRes{k}-\phiRes{k+1})+\boldsymbol{\eta_{k}}-\boldsymbol{\eta_{k+1}} }{2a}   
  \label{pushPull}
\end{split}
\end{equation}
We can define the disturbance terms as : 
\begin{equation}
\boldsymbol{\xi_k}=\frac{-\wfs.\gDelta \phiRes{k}+\boldsymbol{\eta_{k}}-\boldsymbol{\eta_{k+1}}}{2a}
\label{push_pull_noise}
\end{equation}
where we define the incremental residual turbulence $\gDelta \phiRes{k}$ as:
\begin{equation}
\gDelta \phiRes{k}=\phiRes{k+1}-\phiRes{k}
\end{equation}
The push-pull measurement $\mathbf{y_k^b}$ is then given by:
\begin{equation}
\mathbf{y_k^b}=\wfs.\mathbf{b}+\boldsymbol{\xi_k}
     \label{pushPull_compact}
\end{equation}
In practice, to improve the SNR of the measurement we average $N$ push-pull measurements to estimate $\overline{\mathbf{y^b}}$: 
\begin{equation}
\overline{\mathbf{y^b}}=\frac{1}{N} \sum_k^N \mathbf{y_k^b}=\wfs.\mathbf{b}+ \frac{1}{N}\sum_k^N\boldsymbol{\xi_k}
     \label{pushPull_compact_average}
\end{equation}

The composition of $\boldsymbol{\xi_k}$ shows that the SNR of $\overline{\mathbf{y^b}}\xspace$ will depend on the level of noise $\boldsymbol{\eta_k}$, the amplitude of the signal $a$, the difference between two successive residual phases $\gDelta \phiRes{k}$ and the number of measurement $N$ averaged. The accuracy of the estimation of the mis-registration parameters will then depend on the brightness of the source (noise level), the amplitude of the modulation and the quality of the AO correction.

\subsection{An Optimal Modal Basis for the Mis-Registration Identification?}
\label{subsection_PCA_modes}
One important goal of this research is to make the calibration invisible to the science path while being as fast as possible. For this purpose, we propose to identify the most sensitive modes to a given mis-registration and thus to minimize the number of signals required to extract the mis-registration parameters. This will particularly be relevant for ELT AO instruments as the acquisition of an on-sky interaction matrix corresponding to several thousands of modes will require long measurement time, leading to large overheads and thus reducing significantly the time available for the scientific operation. 

To do so, we propose to apply a Principal Component Analysis (PCA) (\citealt{pearson1901principal}) on a sensitivity matrix \dD that is defined as the derivative of the interaction matrix with respect to a mis-registration of type $i$ (see equation \ref{sensitivity_imat_def}). A similar approach was proposed in \cite{oberti2018AOinTheAOF} but is based on flux considerations.

We proceed to the Singular Value Decomposition (SVD) of \dD\footnote{To apply the PCA, the mean value of each mode measurement (\textit{ie} a row of the sensitivity matrix) has to be subtracted to provide mean centered sensitivity matrices.} that contains the measurement for the $N_{DM}$ degrees of freedom of the DM (a zonal interaction matrix for instance):

\begin{equation}
     \dD =\mathbf{U}.\mathbf{S}.\mathbf{V}^T
\end{equation}
By definition, the variance contained in the signals of \dD and due to the input mis-registration $\varepsilon_{i}$ is given by $\lambda_i$:
\begin{equation}
    \lambda_i=\frac{1}{N_{DM}}s_i ^2
\end{equation}
where $s_i$ is the $i^{th}$ singular value of \dD and thus the $i^{th}$ element of the diagonal matrix $\mathbf{S}$. 

\begin{figure}
\centering
\subfloat[$\varepsilon_{i}$=1\% of a subap.]{\includegraphics[width=0.23\textwidth]{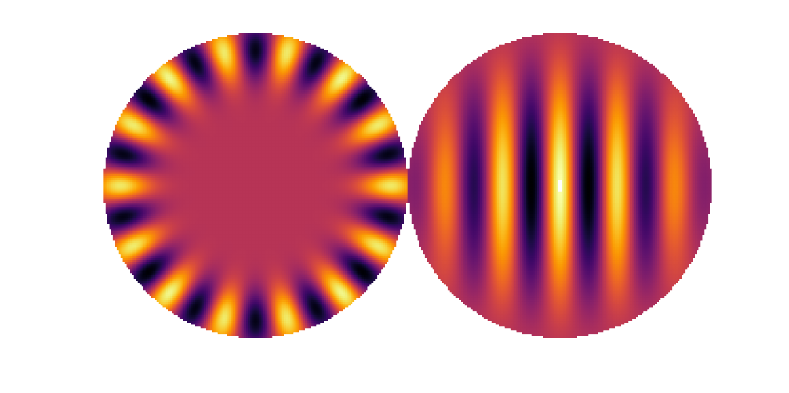}\label{PCA_1_per}}
\subfloat[$\varepsilon_{i}$=10\% of a subap.]{\includegraphics[width=0.23\textwidth]{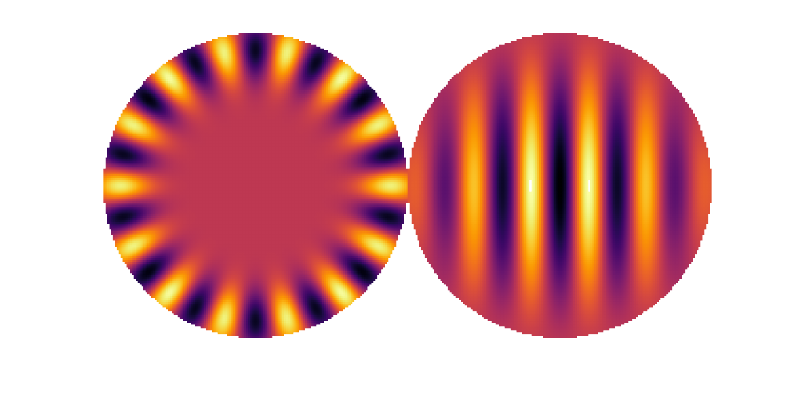}\label{PCA_10_per}}\\
\subfloat[$\varepsilon_{i}$=50\% of a subap.]{\includegraphics[width=0.23\textwidth]{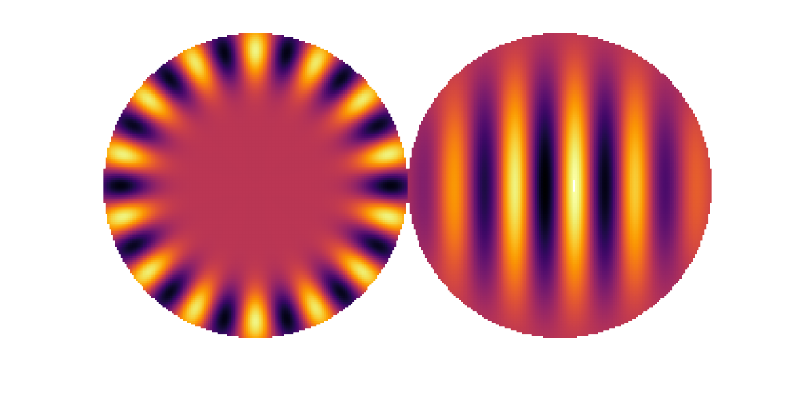}\label{PCA_50_per}}
\subfloat[$\varepsilon_{i}$=100\% of a subap.]{\includegraphics[width=0.23\textwidth]{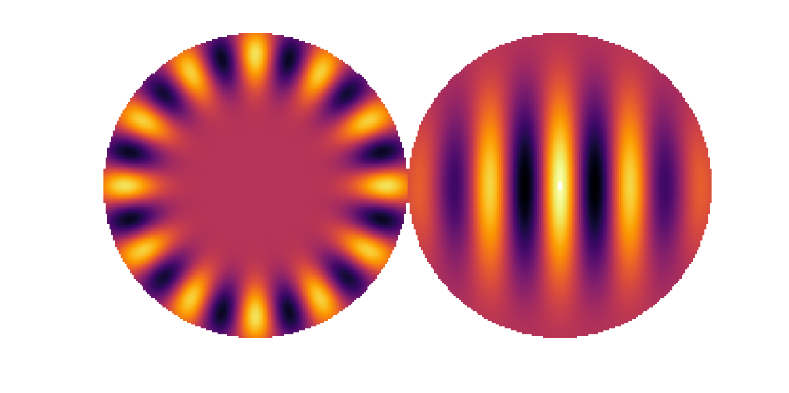}\label{PCA_100_per}}

\caption{Most sensitive mode corresponding to the rotation (left) and the shift X (right) derived from the PCA of \dD for different values of $\varepsilon_{i}$=1\%.}
    \label{fig:PCA_good_DM}
\end{figure}

\begin{figure}
\centering
\subfloat[PCA Mode: Rotation]{\includegraphics[width=0.23\textwidth]{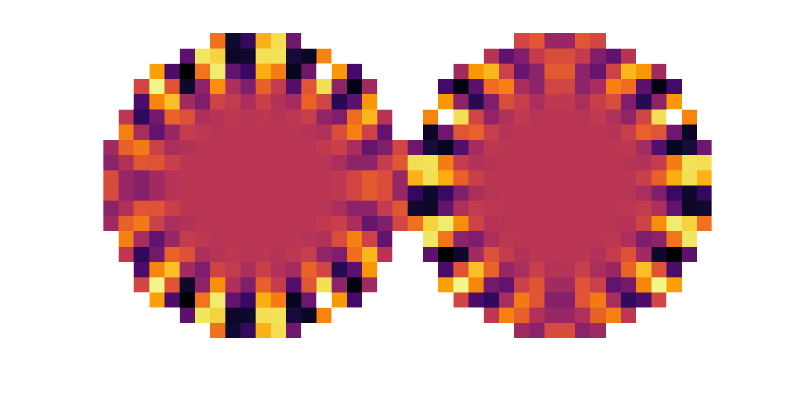}}
\subfloat[PCA Mode: Shift X]{\includegraphics[width=0.23\textwidth]{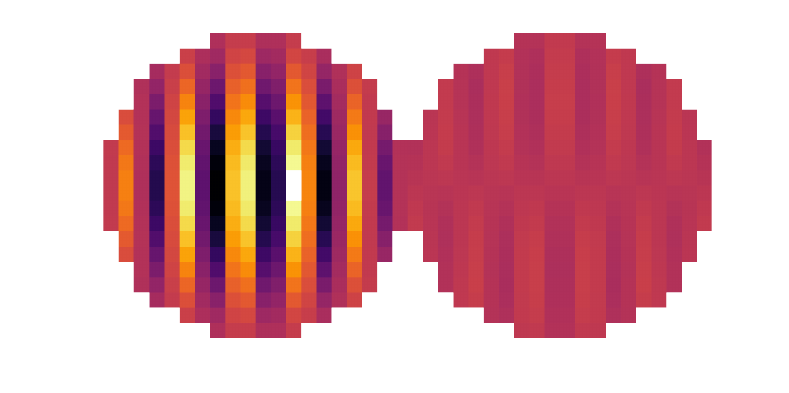}}\\

\caption{WFS measurement [slope X, slope Y] corresponding to the most sensitive modes for the rotation (a) and the shift X (b). In that example $\varepsilon_{i}$=1\% of a subaperture (Figure \ref{PCA_1_per}). }
    \label{fig:PCA_good_WFS}
\end{figure}

By construction, the first singular-modes $\mathbf{v}$ of $\mathbf{V}$ contain most of the variance due to the perturbation injected (the mis-registration $\varepsilon_{i}$). An illustration of these PCA Modes is given in the Figure \ref{fig:PCA_good_DM}, displaying the most sensitive modes to the rotation and shift X for different values of $\varepsilon_{i}$ (for the rotation the value represents the equivalent shift for the actuators located at the edge of the pupil). The corresponding WFS measurement is given in Figure \ref{fig:PCA_good_WFS}.

The modes of Figure \ref{fig:PCA_good_DM} are consistent with what one would expect: a radial non-symmetry with the signal localized on the edge of the pupil for the rotation and a Fourier-like Mode for the shift. We notice that the spatial frequency of the modes is actually quite low, we could have expected higher spatial frequencies considering that there are 20 actuators along one diameter. But this depends on the WFS geometry as well. It appears that using significantly larger values of mis-registrations has a small effect on the spatial frequency of the PCA modes and changes start to be visible when $\varepsilon_{i}$ is above 50 \% of a subaperture shift with PCA modes exhibiting a slightly lower spatial frequency (Figure \ref{PCA_50_per}). This result suggests that one single set of modes could be pre-computed and kept for both types of application: closed-loop operations that require a fine tuning of the parameters (since the mis-registrations should remain small enough to allow a stable closed-loop) and commissioning of an instrument where the mis-registrations to identify might be much larger. 
\begin{figure}
\centering
\subfloat[PCA Mode: Rotation]{\includegraphics[width=0.23\textwidth]{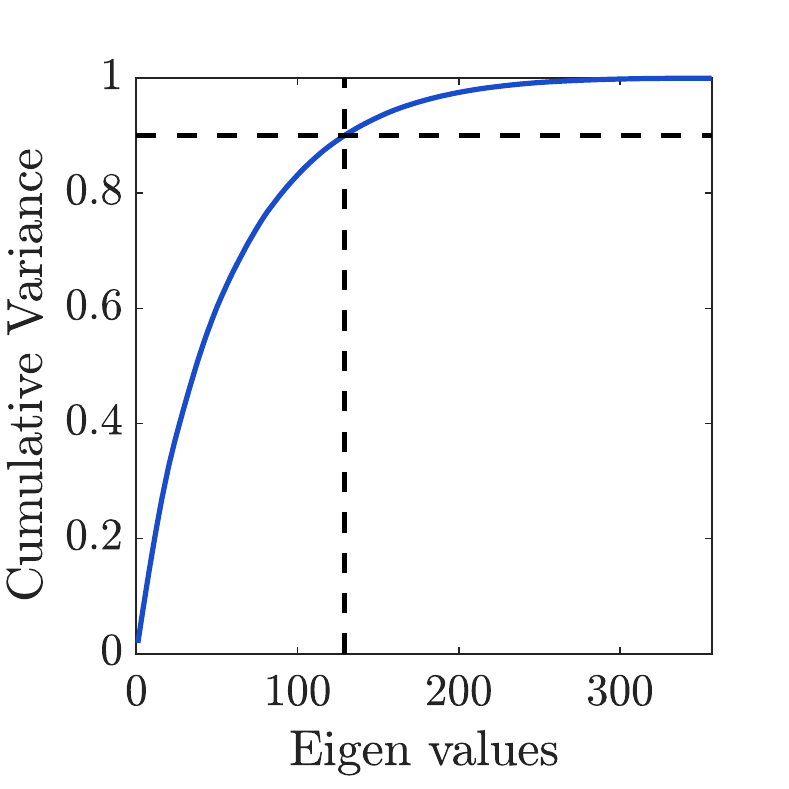}}
\subfloat[PCA Mode: Shift X]{\includegraphics[width=0.23\textwidth]{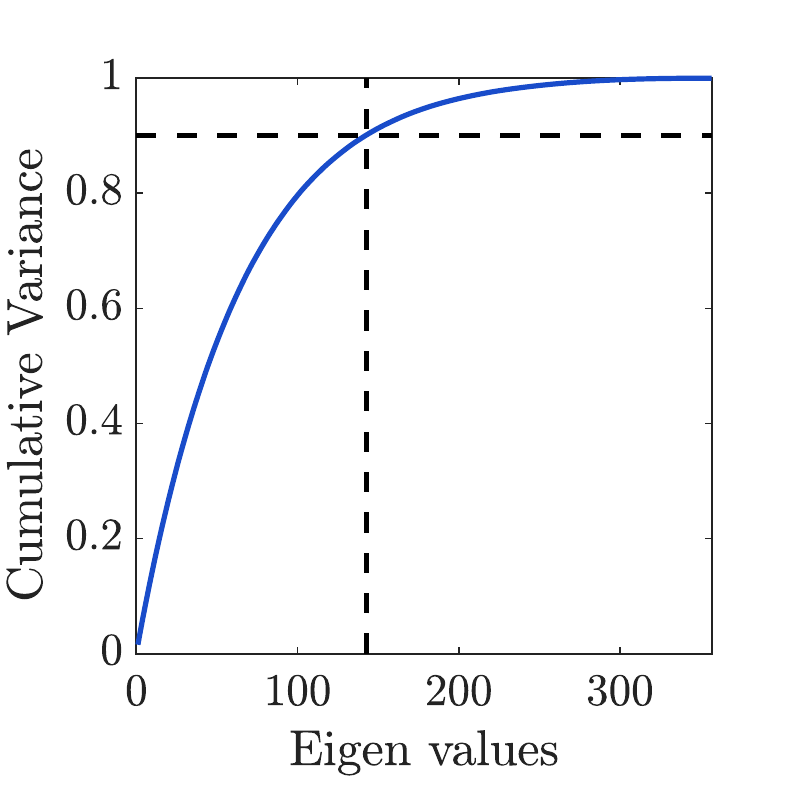}}
\caption{Normalized cumulative variance corresponding to the rotation (a) and shift X (b). In that example $\varepsilon_{i}$=1\% of a subaperture (Figure \ref{PCA_1_per}). }
    \label{fig:PCA_cumVar}
\end{figure}

At last, Figure \ref{fig:PCA_cumVar} gives the cumulative variance associated to the PCA modes. This figure shows an important information; more than 100 PCA modes are required to explain 90 \% of the variance due to a mis-registration in all cases. This shows that the reduction of dimensionality is not so efficient. In other words, this means that the sensitivity of the first PCA mode and any other of the 100 first PCA modes will be quite similar. This gives some flexibility in the choice of the mode to use and other criteria that are system dependent might also be considered. 

The advantage of this procedure remains that it clearly identifies a single mode per mis-registration that takes into consideration the AO system sensitivity according to the DM and WFS geometry, DM mechanical coupling and maximum stroke and WFS sampling and sensitivity. Moreover, the procedure is very general and can be used to define PCA modes for other types of mis-registrations that were not considered in this study (the anamorphosis or distortion for instance). It will be relevant to investigate how the method performs for these higher order mis-registrations and if coupling between the mis-registration parameters appear, typically between a magnification in X and a shift X.

\begin{table}
\small{
\begin{centering}
    \begin{tabular}{|c|c|c|}	
	\cline{1-3}
	\multirow{4}{*}{\textbf{Turbulence}}&Fried Parameter $r_0$ &8-15 cm @550 nm \\
	&Outer Scale $L_0$ &30 m\\
	&C$n^2$ profile&1 layer\\
	&Wind-speed& 10-30 m/s\\
	\cline{1-3}	
	\multirow{4}{*}{\textbf{Control}}&Frequency&1 kHz\\
	&Integrator& $g$=0.3-0.6\\
	& Rejection Bandwidth & 50-80 Hz \\
	& Int. Matrix & 300 KL modes \\
	\cline{1-3}	
    \multirow{3}{*}{\textbf{NGS}}&Wavelength&850 nm\\
    &Magnitude&7.5-13.5\\
    &Photons/subap.&10-500\\
    \cline{1-3}	
	\multirow{3}{*}{\textbf{Telescope}}&Diameter&8 m\\
    &Obstruction/Spider&None\\
    &Resolution&160 px\\
    \cline{1-3}	
    \multirow{4}{*}{\textbf{DM}}&Actuator&357\\
	&Geometry&Cartesian\\
	&Inf. Functions&Gaussian\\
	& Coupling&35\%\\
	\cline{1-3}	
    \multirow{5}{*}{\textbf{PWFS}}&Sub-apertures&20$\times$20 \\
	&Modulation&3 $\lambda$/D\\
	&RON&none\\
	&Photon Noise&Yes\\
	&Signal Processing& Slopes-Maps\\
	& Optical Gains control & Yes\tablefootnote{\citealt{chambouleyron2020pyramid}}\\
	\cline{1-3}	
    \multirow{4}{*}{\textbf{SH WFS}}&Sub-apertures&20$\times$20 \\
	&Pixel Scale&0.22" \\
	&RON&none\\
	&Photon Noise&Yes\\
	&Signal Processing&CoG\\
	\cline{1-3}	
	\end{tabular}
    \end{centering}}
    \caption{Numerical Simulations parameters}
    \label{tab:def_model}
\end{table}

\begin{figure}
    \begin{center}
    \subfloat[Amp 10 nm - 10 phot/subap]{\includegraphics[width=0.235\textwidth]{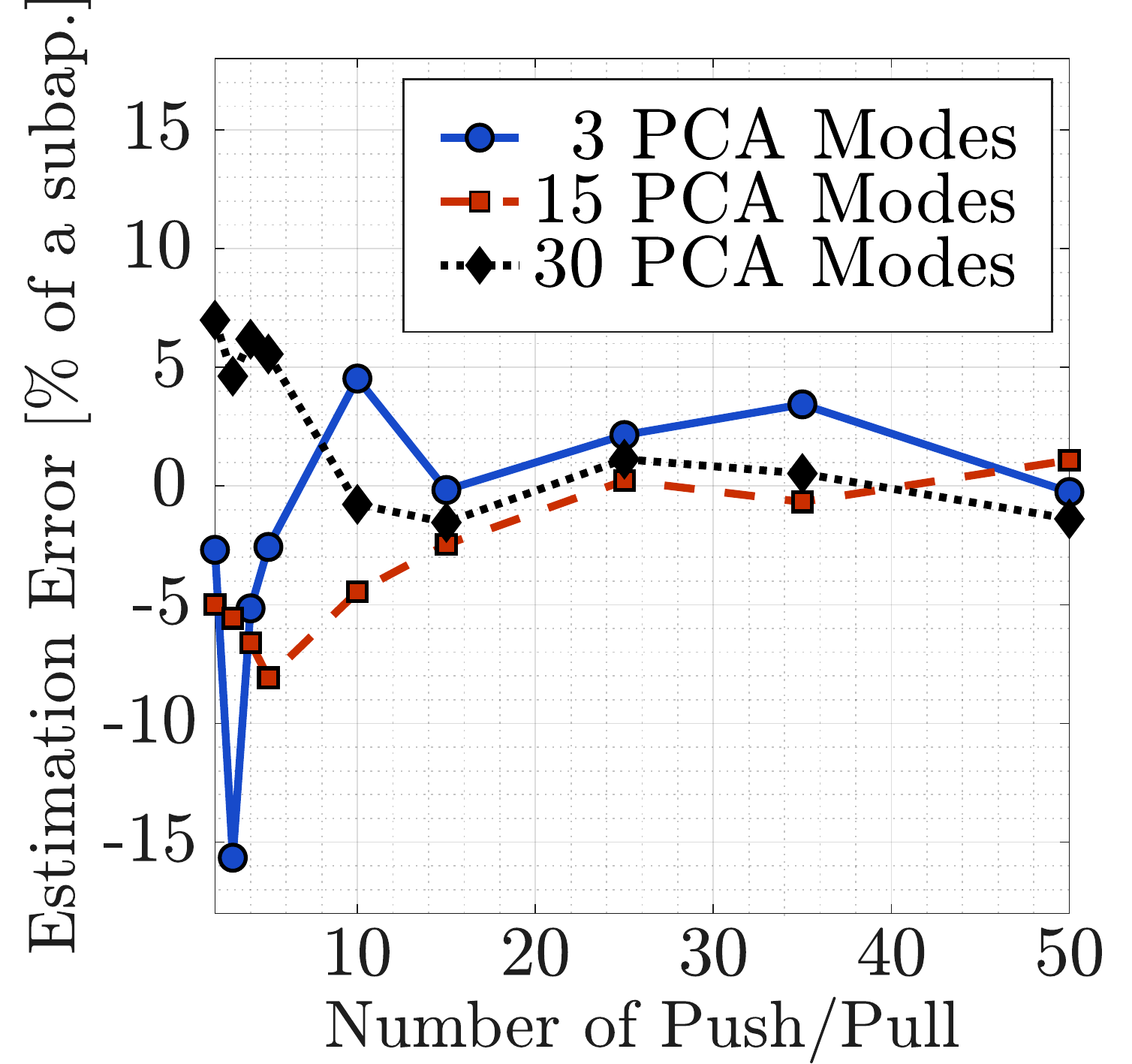}}
    \subfloat[Amp 10 nm - 500 phot/subap]{\includegraphics[width=0.235\textwidth]{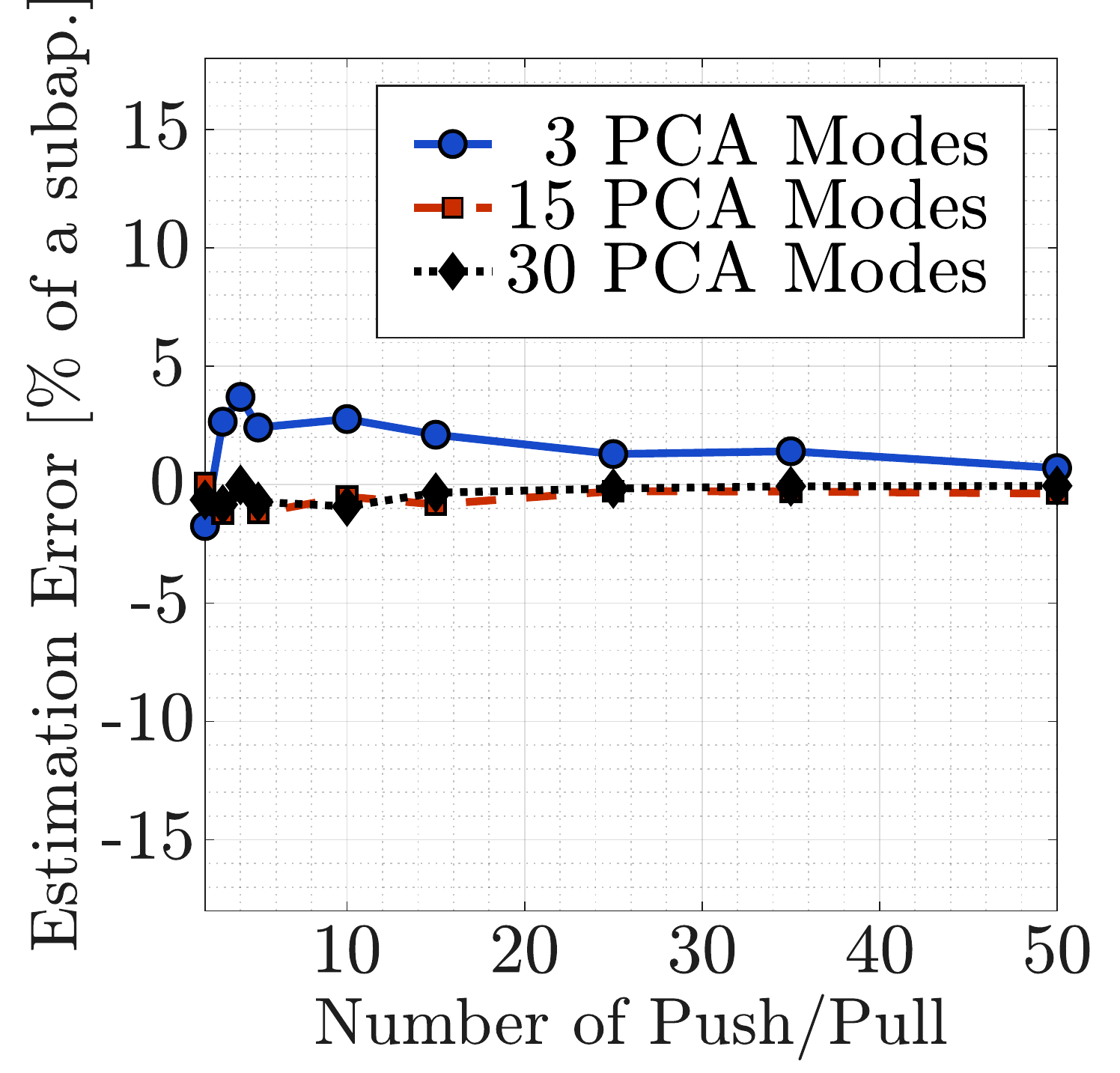}}\\
    \subfloat[Amp 50 nm - 10 phot/subap]{\includegraphics[width=0.235\textwidth]{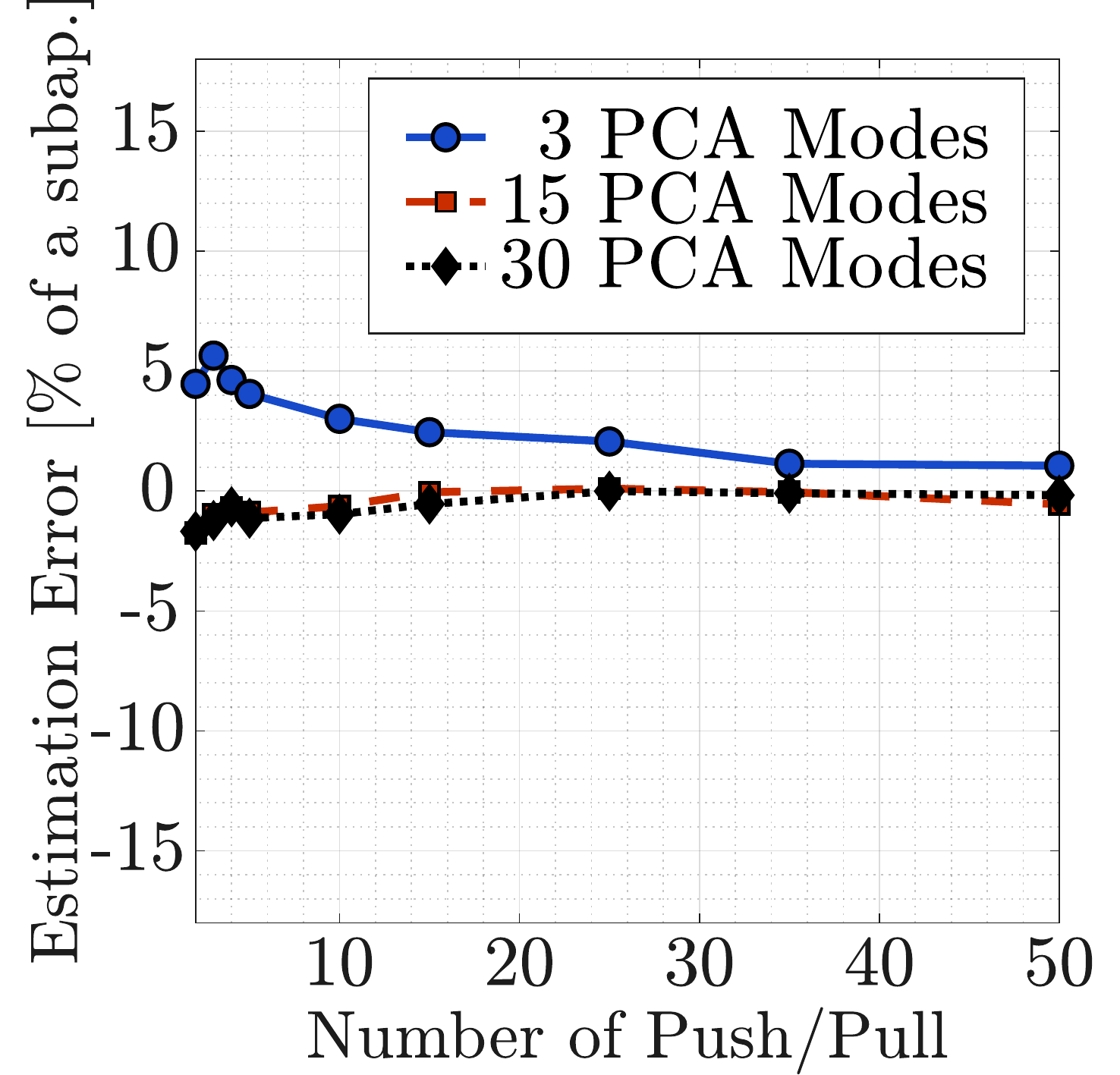}}
    \subfloat[Amp 50 nm - 500 phot/subap]{\includegraphics[width=0.235\textwidth]{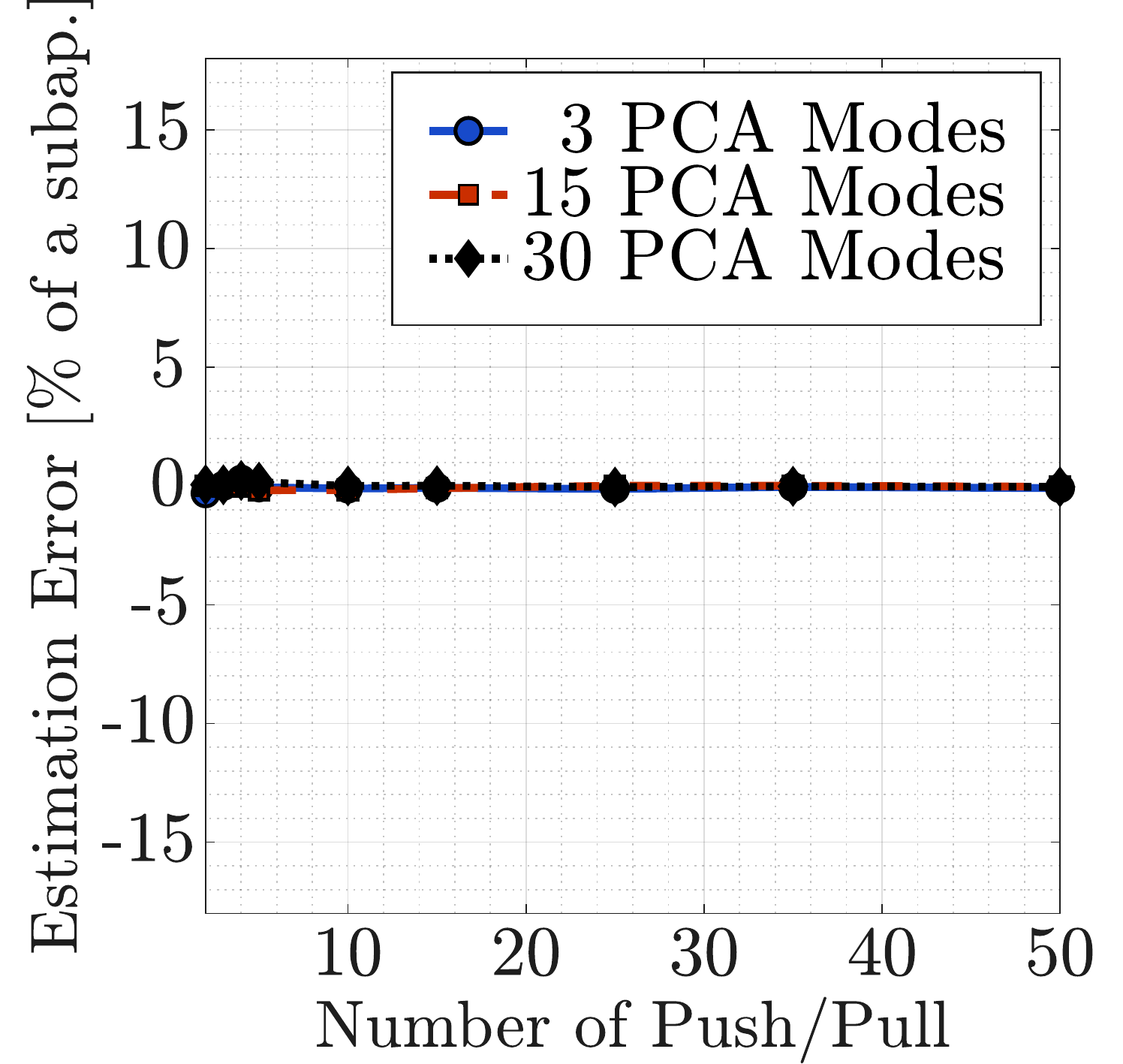}}
    \caption{Accuracy of the algorithm on the estimation of a static shift X as a function of the number of push/pull measurements averaged. The results are shown for different SNR conditions (a,b,c,d) and for different number of modes acquired on-sky.}
    \label{fig:invasive_impact_SNR}
    \end{center}{}
\end{figure}

\section{Application: Numerical simulations}
\label{section_numerical_simulations}
In this section, we investigate the feasibility of tracking mis-registration parameters using the strategy presented in Section \ref{section_Invasive}, using end-to-end simulations in the OOMAO simulator (\citealt{conan2014object}). We propose to explore different closed-loop conditions of noise and turbulence as discussed in section \ref{subsection_push_pull}, simulating a simple AO System which properties are summarized in Table \ref{tab:def_model}. In the analysis, we considered both Pyramid WFS (\citealt{ragazzoni1996pupil}) and Shack-Hartmann WFS (\citealt{hartmann1900bermerkungen}, \citealt{shack1971production}) to investigate if the non-linearities of the PWFS can impact the tracking of the mis-registrations (\citealt{korkiakoski2008improving}, \citealt{esposito2015NCPA}, \citealt{deo2018modal}, \citealt{fauvarque2019kernel}, \citealt{chambouleyron2020pyramid}) since the sensitivity matrices required by the model (equation \ref{sensitivity_imat_def}) are computed in a diffraction-limited regime that differs from the real operating point of the PWFS in which on-sky measurements are acquired. 

In this analysis, we consider one push-pull measurement acquired every 0.05 s (every 50 frames, one push-pull measurement at 1 kHz is acquired), releasing the controller to apply the push-pull commands above the static DM correction.

\begin{table}
    \centering
    \begin{tabular}{c|c|c|c}
         Amplitude $a$  & 30 Modes  & 15 Modes  & 3 Modes\\
         \hline
         50 $nm$ RMS    &300            &150             &100\\
         \hline
         10 $nm$ RMS    &>1500          &>750          &>150\\
         \hline
    \end{tabular}
    \caption{Total number of averaged push/pull measurements required to reach convergence (< 1\% of a subaperture) for the mis-registration estimation of a shift X in all the conditions of noise investigated (10 and 500 photons per subaperture per frame).}
    \label{tab:convergence_PCA_nAveraged}
\end{table}

\begin{figure*}
    \begin{center}
    \stackunder[10pt]{\subfloat[Push-Pull Amplitude 10 nm]{\includegraphics[width=0.31\textwidth]{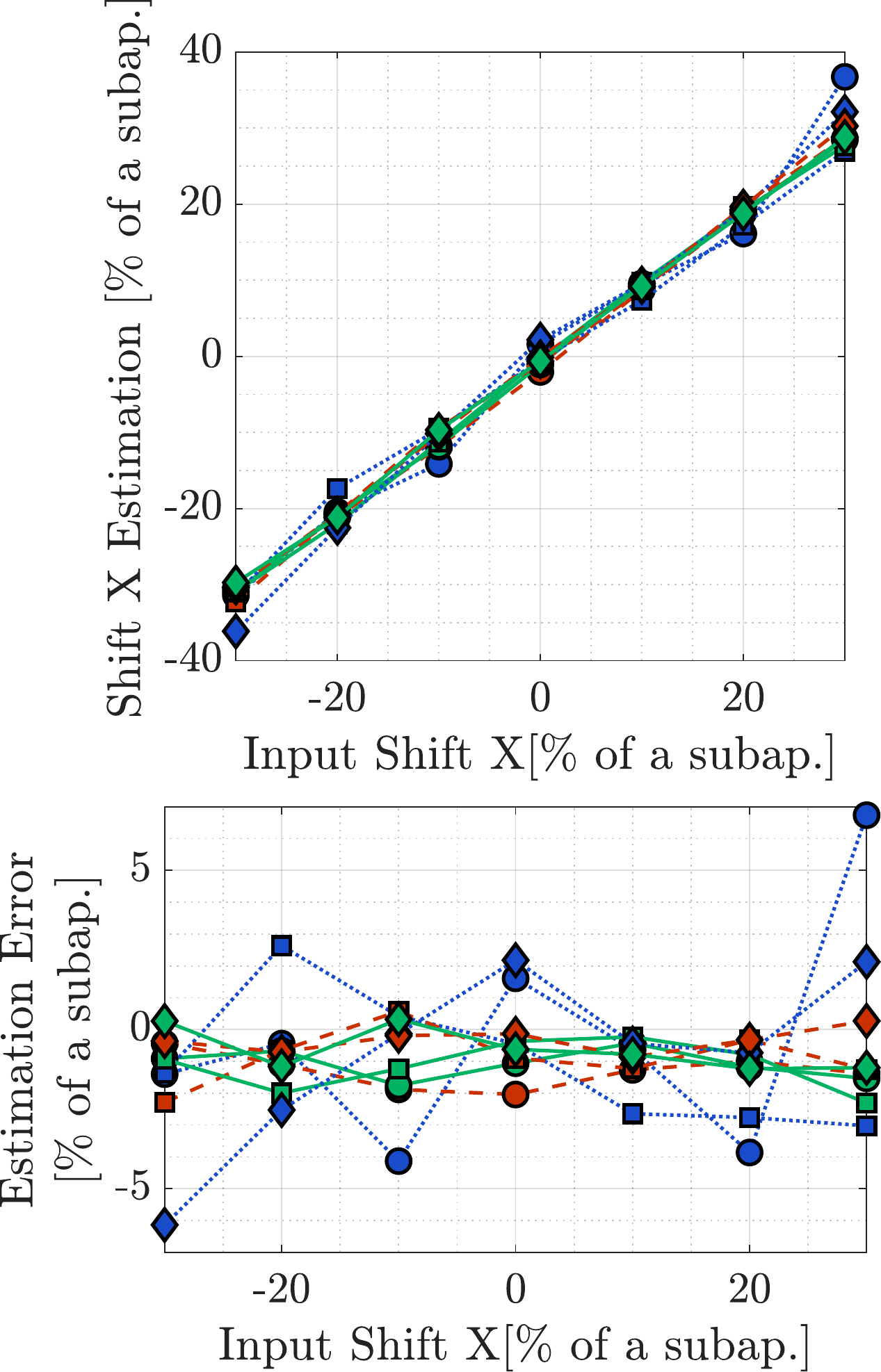}}}{ \includegraphics[width = 0.2 cm]{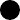} : Wind Speed 10 m/s }\hspace{0.1cm}
    \stackunder[10pt]{\subfloat[Push-Pull Amplitude 20 nm]{\includegraphics[width=0.31\textwidth]{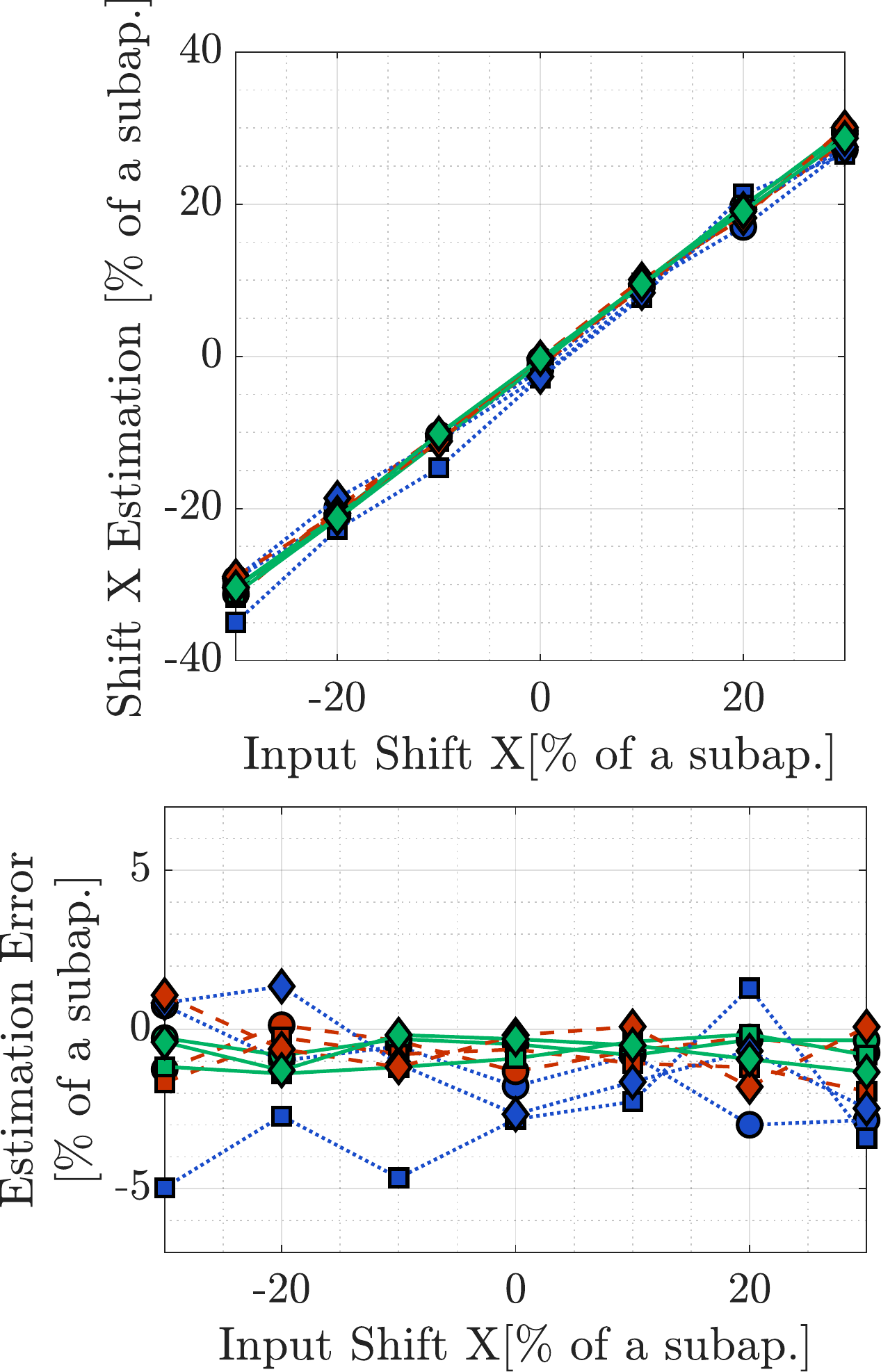}}}{\includegraphics[width = 0.2 cm]{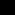} : Wind Speed 20 m/s}\hspace{0.1cm}
    \stackunder[10pt]{\subfloat[Push-Pull Amplitude 50 nm]{\includegraphics[width=0.31\textwidth]{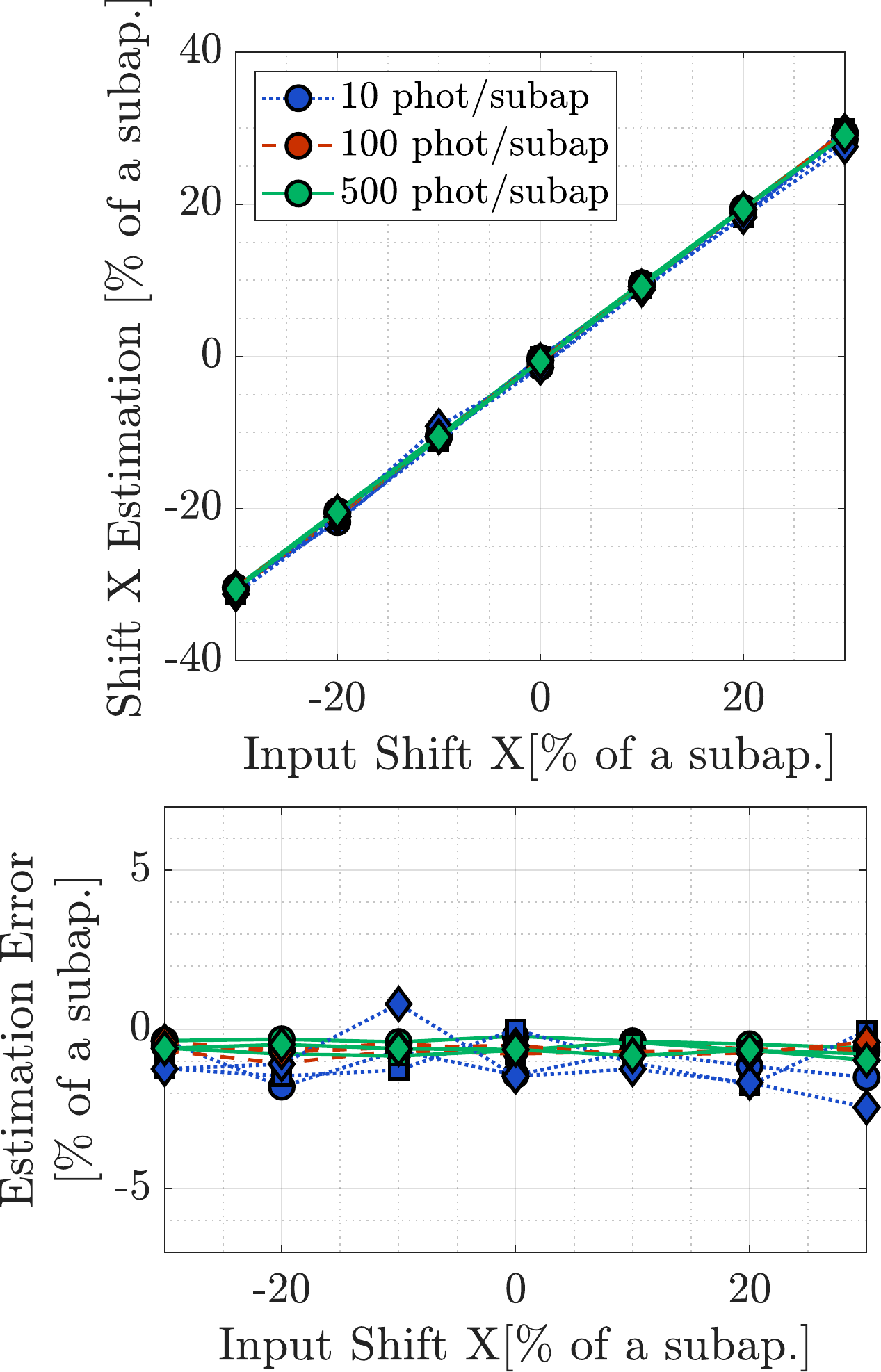}}}{\includegraphics[width = 0.2 cm]{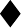} : Wind Speed 30 m/s }%
    \vspace{0.05 cm}
    \caption{Shift X estimation using a PWFS (top) and corresponding estimation error (bottom) as a function of the input shift X for a push-pull amplitude of 10 nm (a), 20 nm (b) and 50 nm (c) using a PWFS. The results are given for different noise regimes (dotted blue, dashed red and solid green lines). The markers correspond to different wind-speeds (bullet, squares and diamonds) }
    \label{fig:invasive_ramp_X}
    \end{center}{}
\end{figure*}
\begin{figure*}
    \begin{center}
    \stackunder[10pt]{\subfloat[Push-Pull Amplitude 10 nm]{\includegraphics[width=0.31\textwidth]{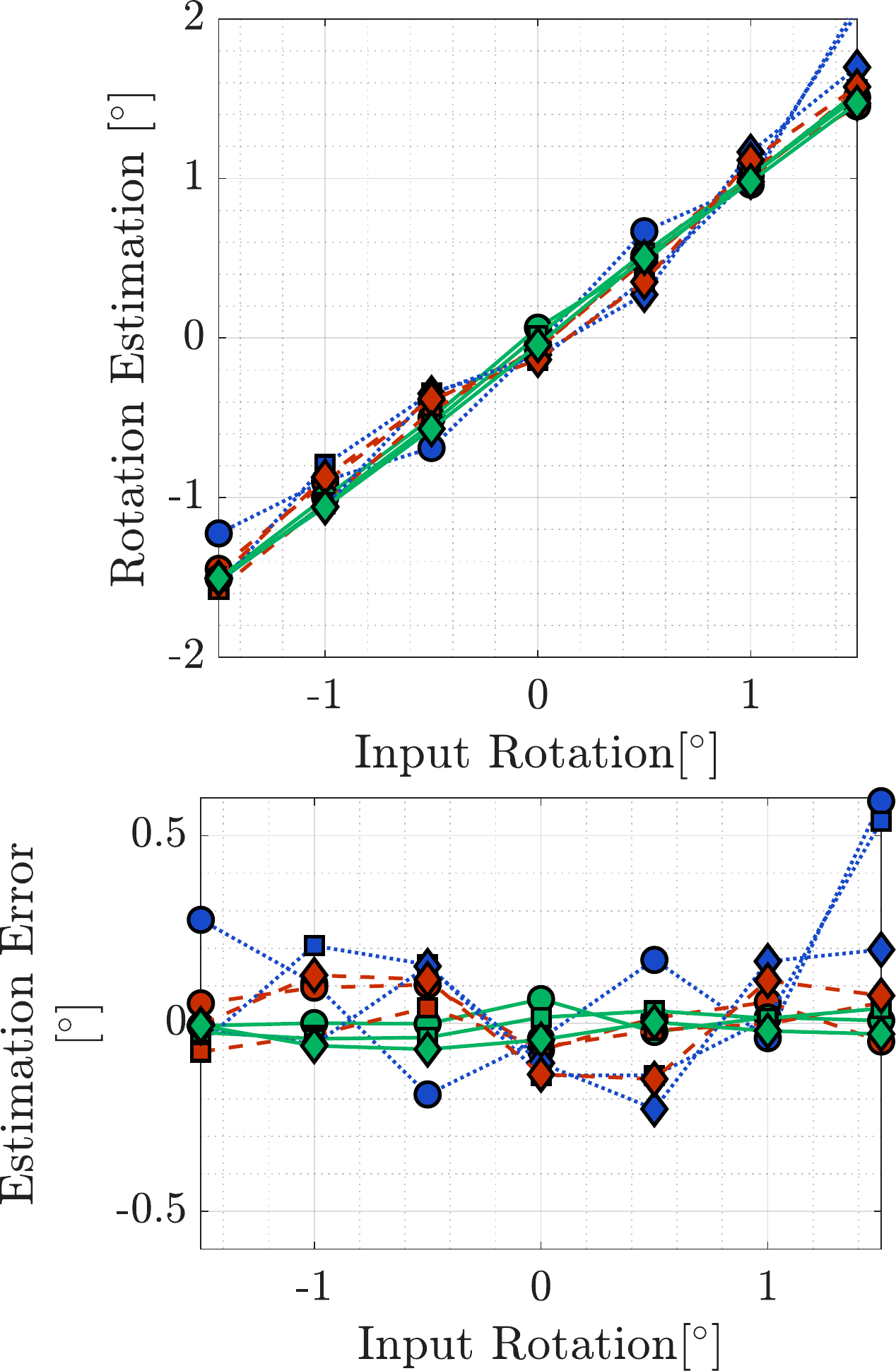}}}{ \includegraphics[width = 0.2 cm]{plots/circle.pdf} : Wind Speed 10 m/s }\hspace{0.1cm}
    \stackunder[10pt]{\subfloat[Push-Pull Amplitude 20 nm]{\includegraphics[width=0.31\textwidth]{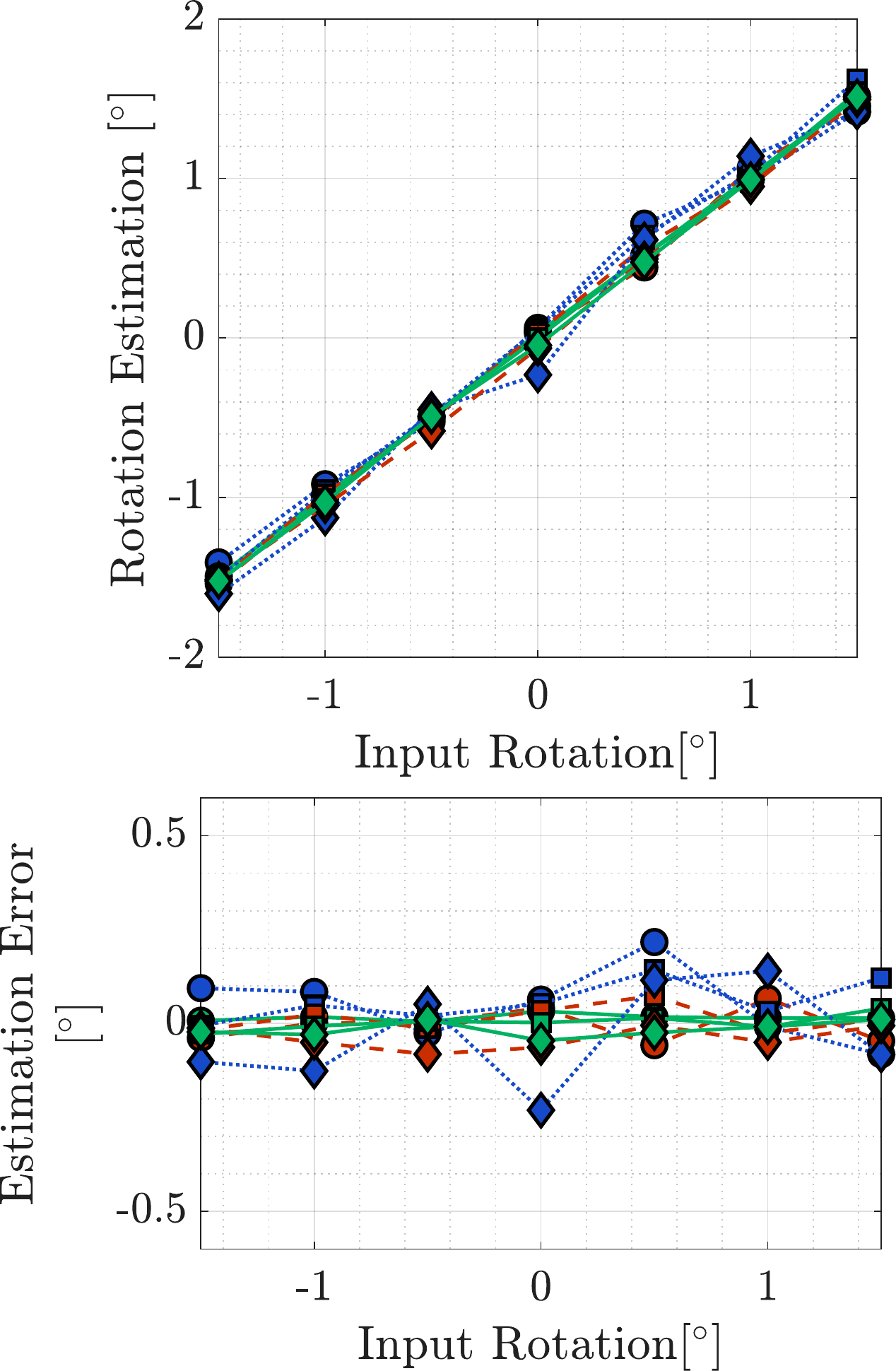}}}{\includegraphics[width = 0.2 cm]{plots/square.pdf} : Wind Speed 20 m/s}\hspace{0.1cm}
    \stackunder[10pt]{\subfloat[Push-Pull Amplitude 50 nm]{\includegraphics[width=0.31\textwidth]{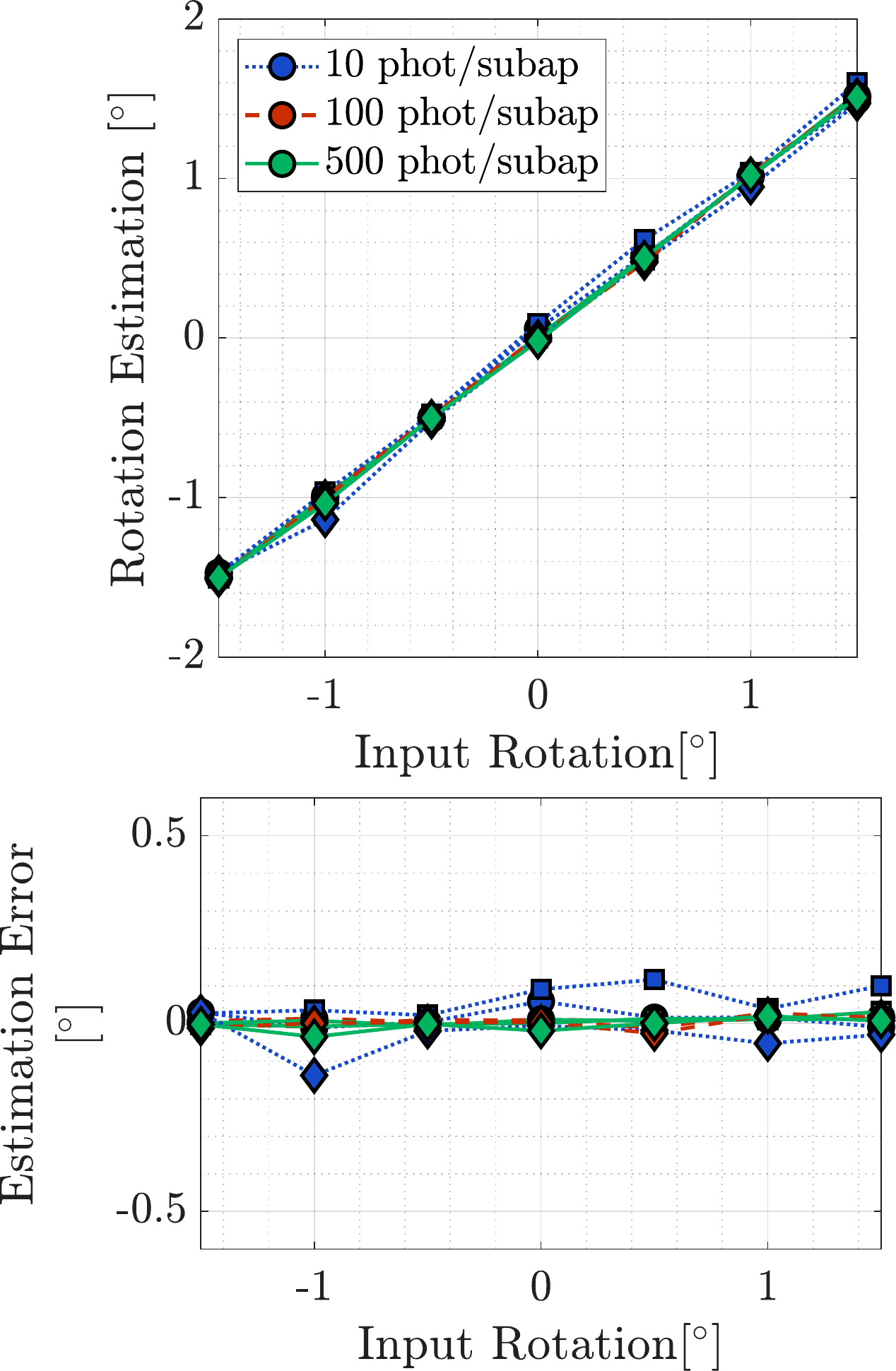}}}{\includegraphics[width = 0.2 cm]{plots/diamond.pdf} : Wind Speed 30 m/s }%
    \vspace{0.05 cm}
    \caption{Rotation estimation using a PWFS (top) and corresponding estimation error (bottom) as a function of the input shift X for a push-pull amplitude of 10 nm (a), 20 nm (b) and 50 nm (c). The results are given for different noise regimes (dotted blue, dashed red and solid green lines). The markers correspond to different wind-speeds (bullet, squares and diamonds) }
    \label{fig:invasive_ramp_rot}
    \end{center}{}
\end{figure*}
\begin{figure*}
    \begin{center}
    \includegraphics[width=1\textwidth]{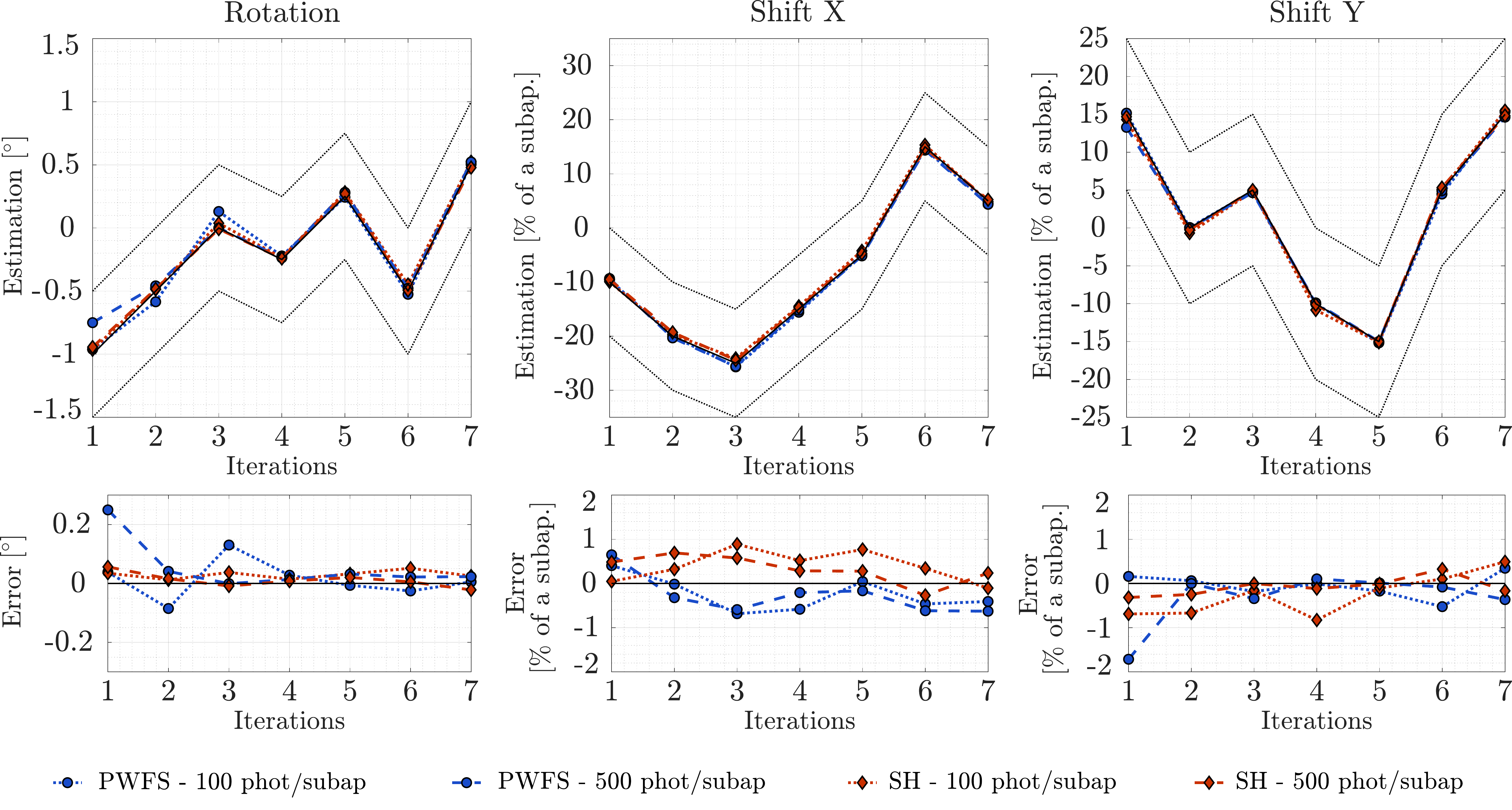}
    \caption{Mis-registration parameters estimation (top) and corresponding estimation error (bottom) as a function of the number of iterations. The results are given for the rotation (left), shift X (middle) and shift Y (right). The black solid lines represent the actual true evolution of the mis-registration parameters and the dashed black lines are the accuracy target (10\% of a sub-aperture shift). The blue dashed lines represent the identified values of the mis-registration parameters when using a PWFS, while the red ones correspond to a SH WFS.}
    \label{fig:invasive_multi}
    \end{center}{}
\end{figure*}
\subsection{Impact of the Signal to Noise Ratio}
\label{subsection_SNR}
The first analysis consists in studying the impact of the SNR of the on-sky signals on the estimation of the parameters. We consider here an AO system with a SH-WFS since the same results were obtained using a PWFS. As detailed in section \ref{subsection_push_pull}, the accuracy of the estimation depends on the amplitude, noise level and number of modes considered. We propose to consider a system with a static mis-registration of 20 \% of a sub-aperture shift X and investigate the accuracy reached by the algorithm for different noise regimes, amplitude of the signals and number of signals considered. We consider one, three and five PCA modes for each type of mis-registration making the total number of modes acquired on sky multiplied by a factor three (Rotation, Shift X and Shift Y). 

Figure \ref{fig:invasive_impact_SNR} gives the estimation error of the algorithm for different SNR conditions. It shows that reducing the whole interaction matrix to a few well selected modes allows to extract the mis-registration parameters. The accuracy of the estimation depends both on the SNR of the signals (amplitude and number of measurement) and the number of modes considered. As expected, for a given SNR, the higher the number of modes the better the accuracy.

This implies that considering a larger number of modes allows to reduce the number of measurements required to reach convergence and provides a better estimation of the parameters for a given SNR. But the impact on the scientific path and the time required for the calibration becomes larger. However, using the minimum number of modes (3 PCA modes), Figure \ref{fig:invasive_impact_SNR} shows that with no more than 20 push-pull measurements, the method provides an accuracy better than 5\% of a sub-aperture in all the cases considered. 

The results presented in the previous figures are summarized in Table \ref{tab:convergence_PCA_nAveraged} that gives the total number of push pull measurements required to reach convergence for the estimation of a shift X of 20\% of a subaperture. To optimize the identification of the mis-registrations, a trade-off is required, using either many modes with a small number of measurements or using few modes with a larger number of measurements. Overall, the PCA modes allow to reduce significantly the acquisition time necessary to measure the experimental signals required by the identification algorithm and estimate accurately the mis-registration parameters. In the following, we always consider the use of 3 PCA modes and 50 push/pull measurements to make sure that the estimation of the mis-registration parameters is close enough to the convergence value.

\subsection{Ramps of mis-registrations}
\label{subsection_ramps_misReg}
The second analysis consists in applying a ramp of a given mis-registration to a closed loop AO system and acquire 50 push-pull measurements of 3 PCA modes (see section \ref{subsection_SNR}). The mis-registrations applied here are chosen to maintain the system in a stable closed loop regime (Figure \ref{wfe_shift}).

To improve the readability of the paper, this section only shows the result corresponding to a PWFS. The same simulation results obtained with a SH-WFS are provided in Appendix \ref{appendix_SH} and they show the same behavior and performance. Figure \ref{fig:invasive_ramp_X} and Figure \ref{fig:invasive_ramp_rot} provide the mis-registration parameters estimation corresponding to a ramp of shift X and of rotation. For both cases and for all the closed-loop conditions explored (wind speed and noise regime), the accuracy reached with the method is very good with a maximal error of 7\% of a subaperture for the shift and 0.6 degrees for the rotation. For both cases, this maximal error is obtained for the highest level of noise and smallest amplitude of push-pull. For this system, the shift on the border of the pupil corresponding to a rotation of 0.6 degrees is 10.5\% of a sub-aperture. 

Excluding this specific case, the estimation reached by the method is well below the accuracy required: all the estimation errors are better than 5\% of a sub-aperture for the shift and better than 0.25 degree for the rotation. Considering the highest amplitude of push-pull and the lowest noise regime allows to improve the accuracy of the estimation to less than 1\% of a subaperture for the shift and less than 0.05 degrees for the rotation. 

These results are consistent with the considerations presented in the section \ref{subsection_push_pull} and confirm the trends expected: the estimation of the parameters is more accurate when the SNR of the on-sky signal is higher. In particular, even for the lowest flux considered (10 photons per subaperture per frame), the method presented provides still a very accurate estimation of the parameter (under 5\% of a subaperture), adjusting the amplitude of the signals to 20 nm for example. In addition, the method appears to be robust to the different conditions of observation and to be independent on the mis-registration value to identify.

\subsection{Multiple mis-registrations evolving dynamically}
\label{subsection_multiple_misReg}

A more realistic situation consists in applying multiple mis-registrations at the same time (rotation and shifts) evolving dynamically with time.  In the following results, one iteration corresponds to one acquisition sequence of 50 push pull measurements. We make the reasonable assumption that the mis-registrations remain static during one full acquisition sequence. The number of observing parameters considered is reduced to two different flux regimes for an amplitude of 20 nm RMS and a wind-speed of 15 m/s to limit the number of plots.

The results are provided for both PWFS and SH-WFS in Figure \ref{fig:invasive_multi} that shows that the mis-registration parameters estimation in this realistic case are once again very accurate with a maximum error of about 0.2 degrees for the rotation and 2 \% of a subaperture and most of the estimations better than 0.1 degree and 1 \% shift error. This shows that the algorithm does not suffer from strong coupling between the parameters considered. In addition, the methods performs equally for both WFS.

\begin{figure}
    \begin{center}
    \subfloat[Estimation and Estimation Error]{\includegraphics[width=0.35\textwidth]{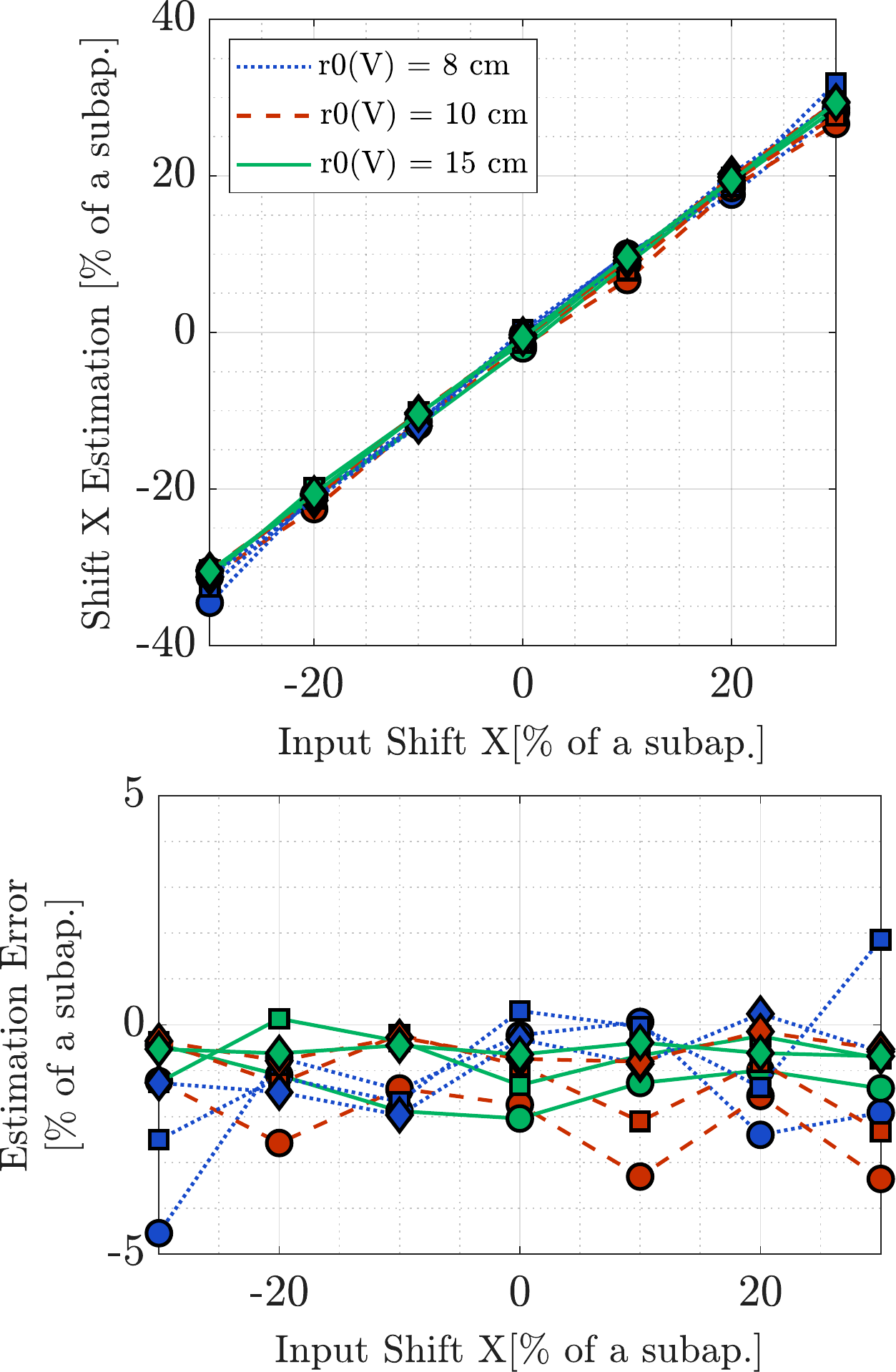}\label{sx_vs_r0_est}}\\
    \subfloat[Optical Gains]{\includegraphics[width=0.35\textwidth]{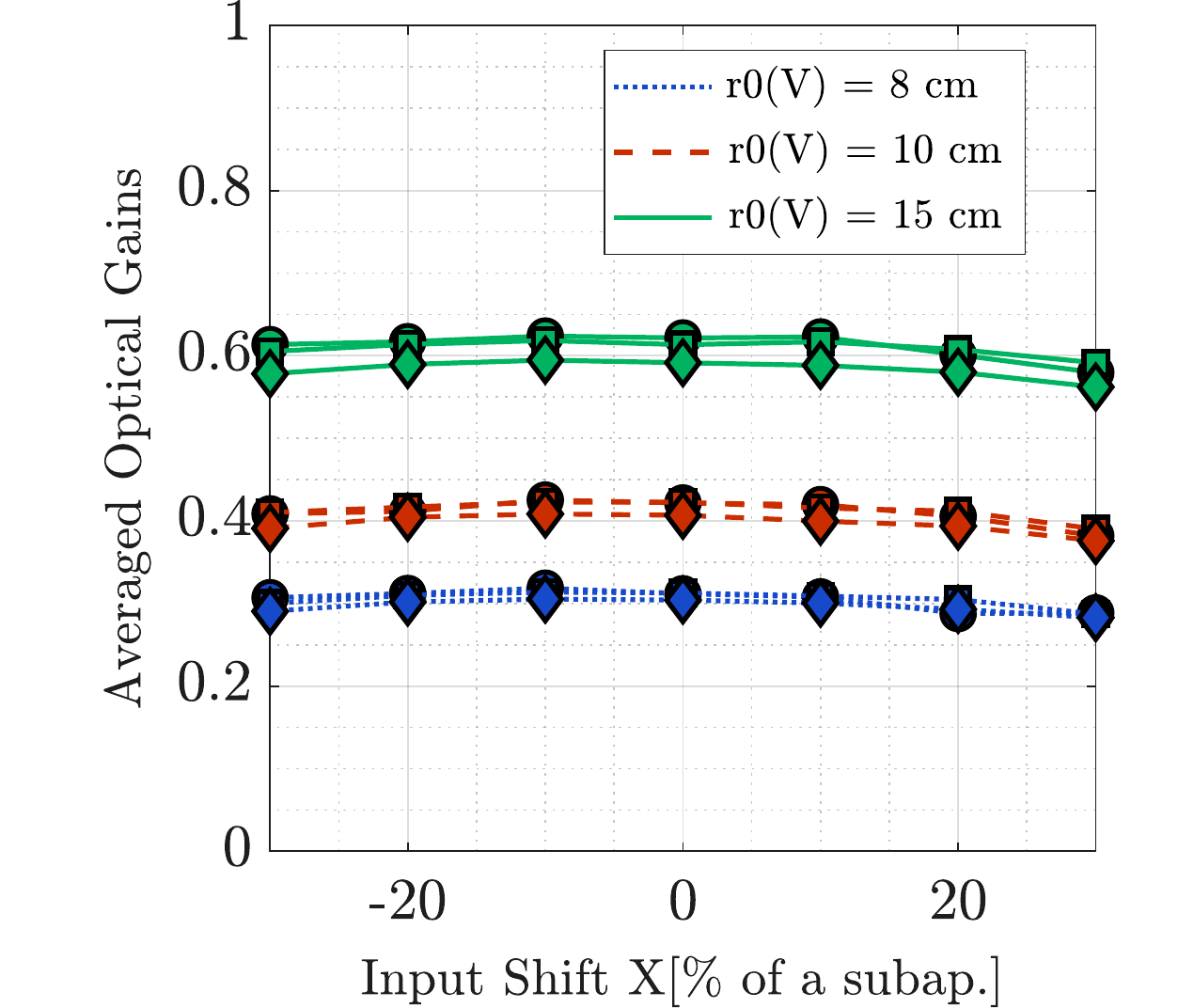}\label{sx_vs_r0_og}}
    \caption{Shift X estimation and estimation error (a) and corresponding Optical Gains (b) as a function of the input shift X for different seeing conditions.The markers correspond to a push-pull amplitude of 10 nm (circles), 20 nm (squares) and 50 nm (diamonds).}
    \label{fig:invasive_ramp_r0}
    \end{center}{}
\end{figure}
\subsection{Sensitivity to seeing conditions}
\label{subsection_sensitivity_seeing}
In addition to the previous results, the algorithm has been validated against different seeing conditions to challenge the PWFS loss of sensitivity. This study does not include seeing variations during the acquisition of the signals. To limit the number of plots, we consider only a ramp of shift X for a wind-speed of 10 m/s and a flux regime of 100 and 500 photons per subaperture. Once again we consider 50 push-pull measurements.

In that case, the estimation of the mis-registration parameters and of the corresponding optical gains \ggamma for different Fried Parameter $r_0$ are provided in Figure \ref{fig:invasive_ramp_r0}. In this Figure, the markers correspond to different push-pull amplitudes (10, 20 and 50 nm RMS). Figure \ref{sx_vs_r0_est} shows that the algorithm is very robust against seeing variations, with a maximum error better than 5 \% of a subaperture which corresponds to the worst case considered (lowest $r_0$ value and lowest push pull amplitude). As expected, the best estimation of the parameters is obtained for the largest $r_0$ value and largest push pull amplitude. In that case the maximum error is better than 1\% of a subaperture.

In addition, Figure \ref{sx_vs_r0_og} gives the averaged optical gains identified by the algorithm (we consider here the averaged optical gains for the 3 PCA modes, \textit{e.g.} the mean value of the diagonal of \ggamma). This plot shows that we retrieve the typical attenuation expected for mid-order modes of a modulated PWFS operating in I band as a function of $r_0$ with an attenuation of 40\%, 60\% and 70\% for a Fried parameter of respectively 15, 10 and 8 cm in the visible (\citealt{deo2018modal}). This plot also exhibits a slightly lower gain for larger values of mis-registrations. This is consistent with the fact that the AO residuals are slightly higher in the case of an imperfect DM/PWFS alignment, impacting thus the sensitivity of the PWFS.

\begin{figure*}
    \begin{center}
    \subfloat[No Disturbance]{\includegraphics[width=0.95\textwidth]{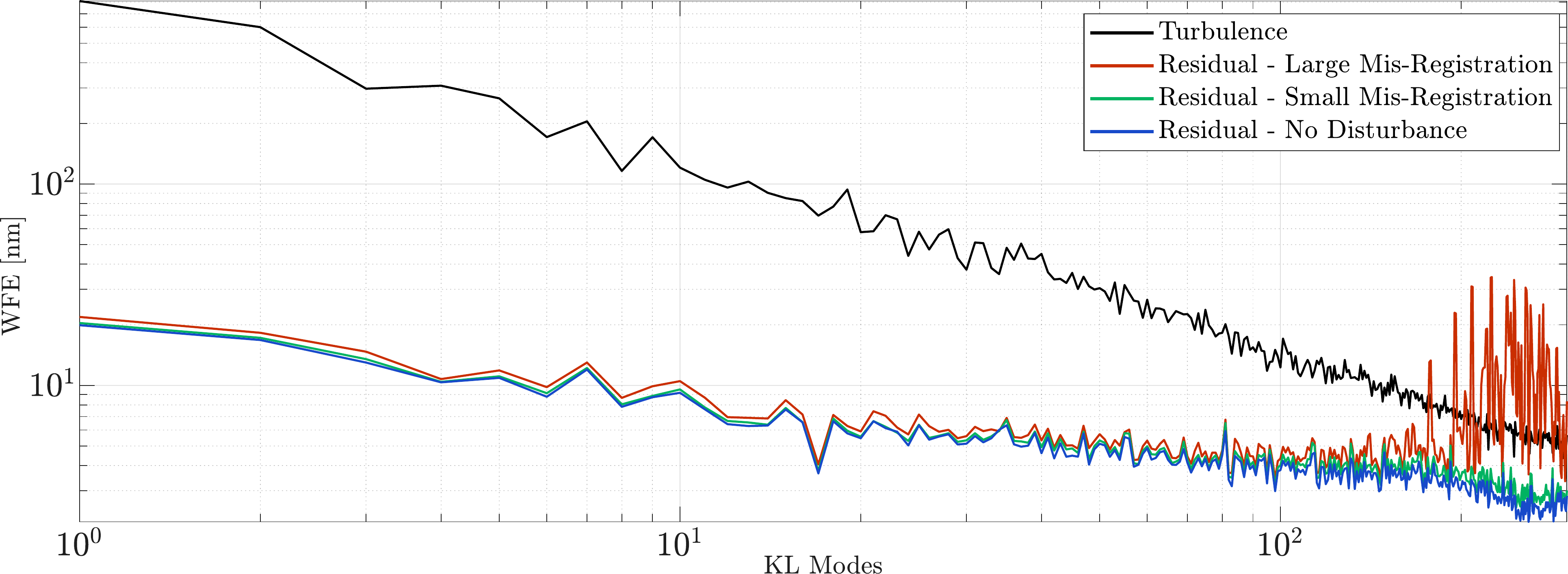}\label{impact_misReg}}\\
    \subfloat[Disturbance PCA$_{\text{rot}}$]{\includegraphics[width=0.95\textwidth]{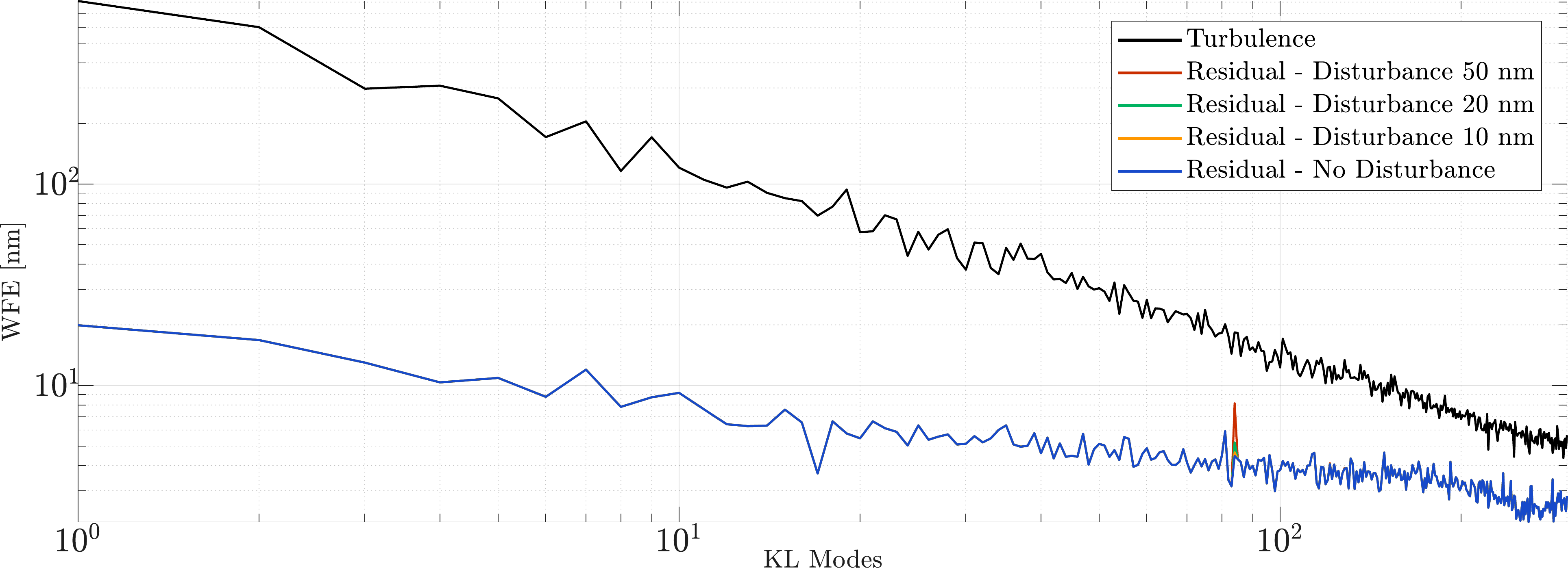}\label{impact_PCArot}}\\
    \subfloat[Disturbance PCA$_{\text{X}}$]{\includegraphics[width=0.95\textwidth]{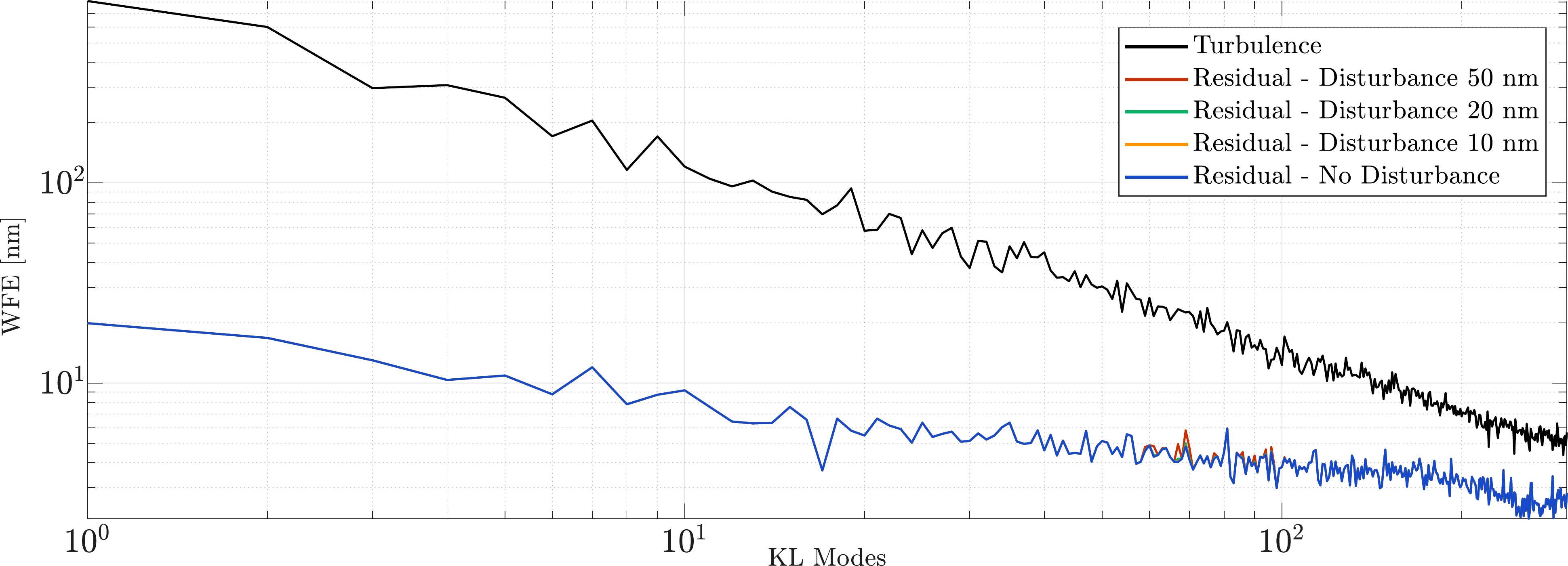}\label{impact_PCAx}}\\
    \caption{Modal PSD corresponding to the cases of Table \ref{tab:summary_impact}. The open loop turbulence (solid black line) and reference case (solid blue line) are given for each plot.}
    \label{fig:impact_science_psd}
    \end{center}
\end{figure*}
\begin{figure*}
    \begin{center}
    \subfloat[]{\includegraphics[width=0.3\textwidth]{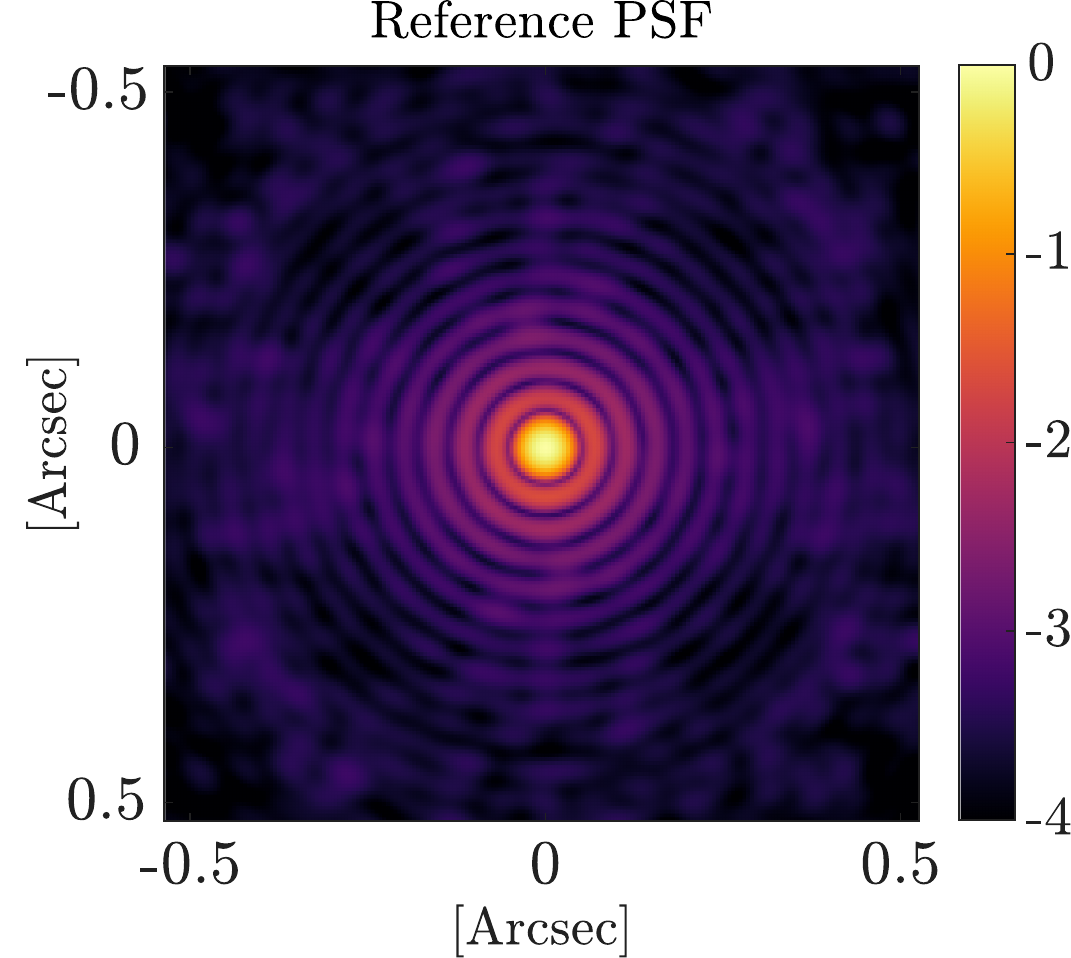}\label{psf_ref}}
    \subfloat[Small Mis-Registration]{\includegraphics[width=0.3\textwidth]{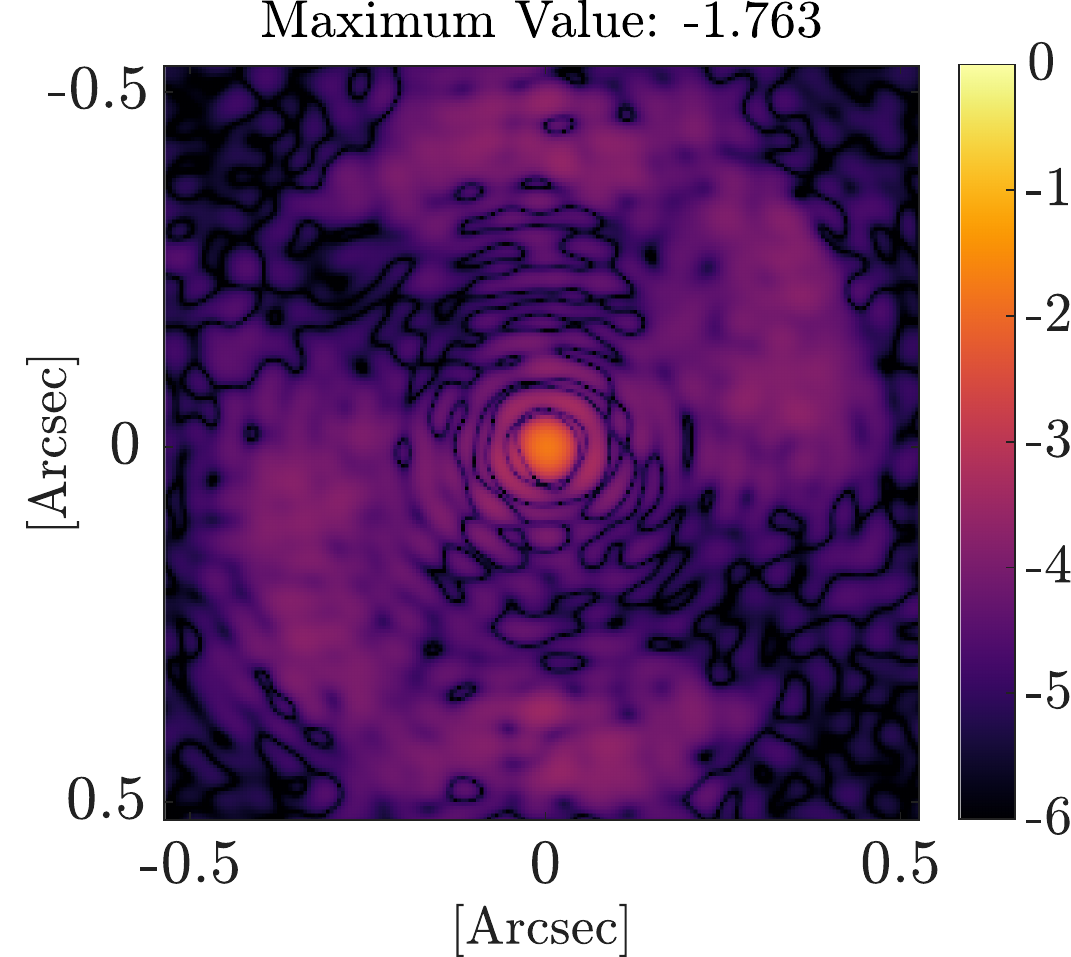}\label{psf_small_mis_reg}}
    \subfloat[Large Mis-Registration]{\includegraphics[width=0.3\textwidth]{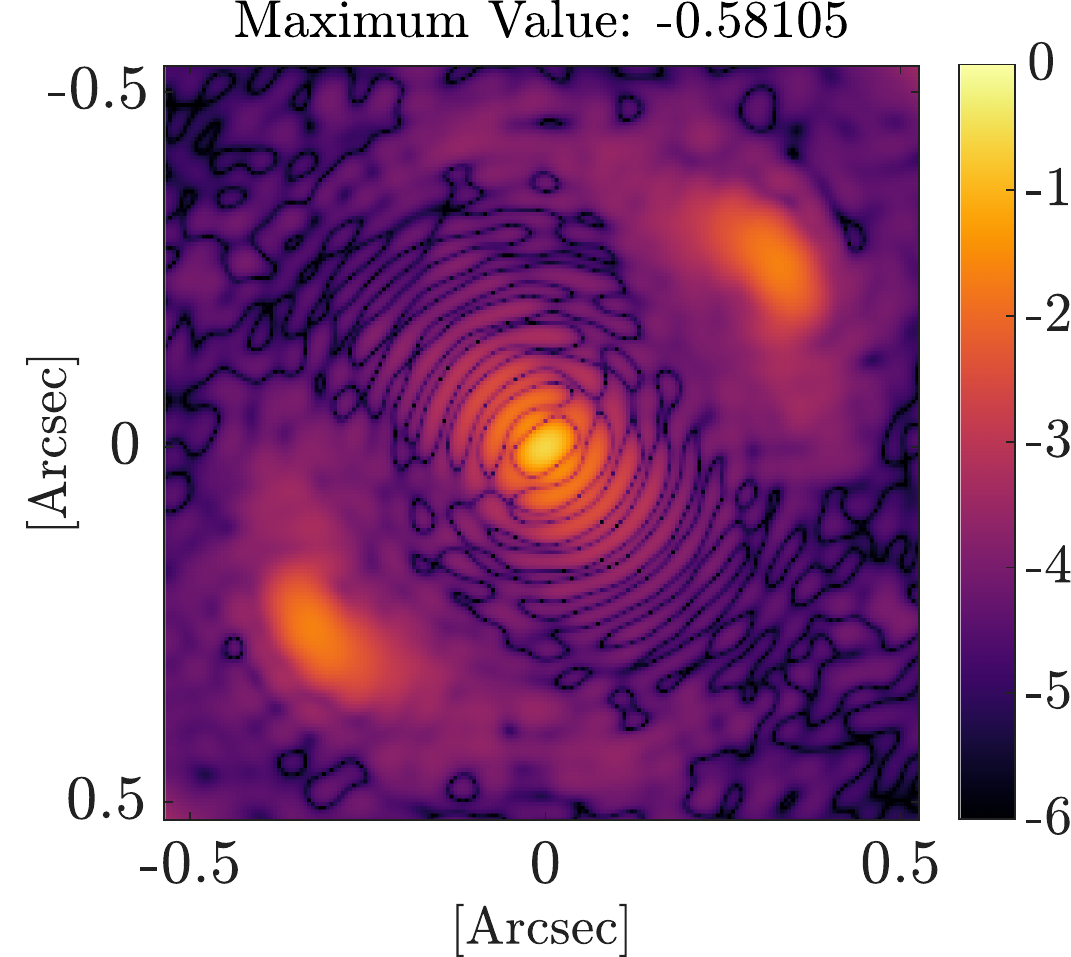}\label{psf_large_mis_reg}}\\
    \subfloat[PCA$_{\text{rot}}$ - 10 nm]{\includegraphics[width=0.3\textwidth]{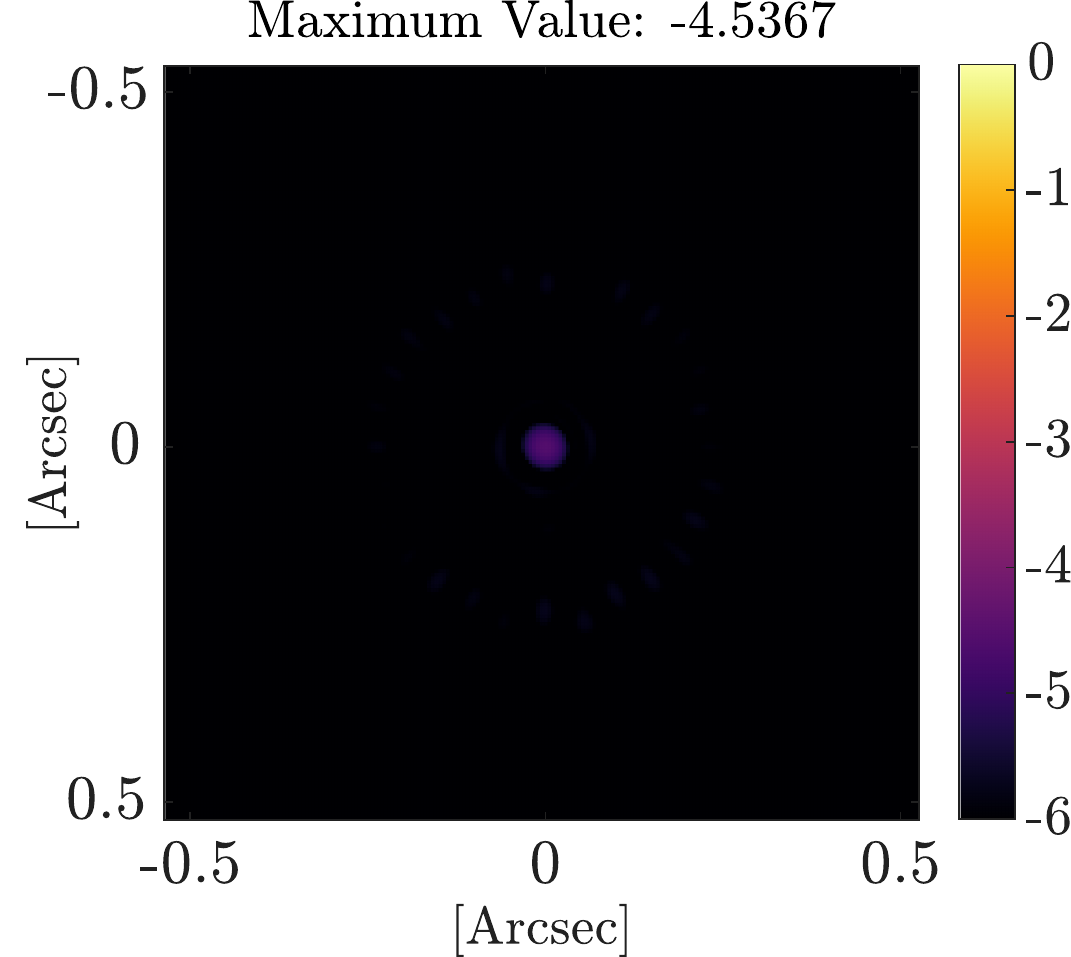}\label{psf_PCA_rot_10}}
    \subfloat[PCA$_{\text{rot}}$ - 20 nm]{\includegraphics[width=0.3\textwidth]{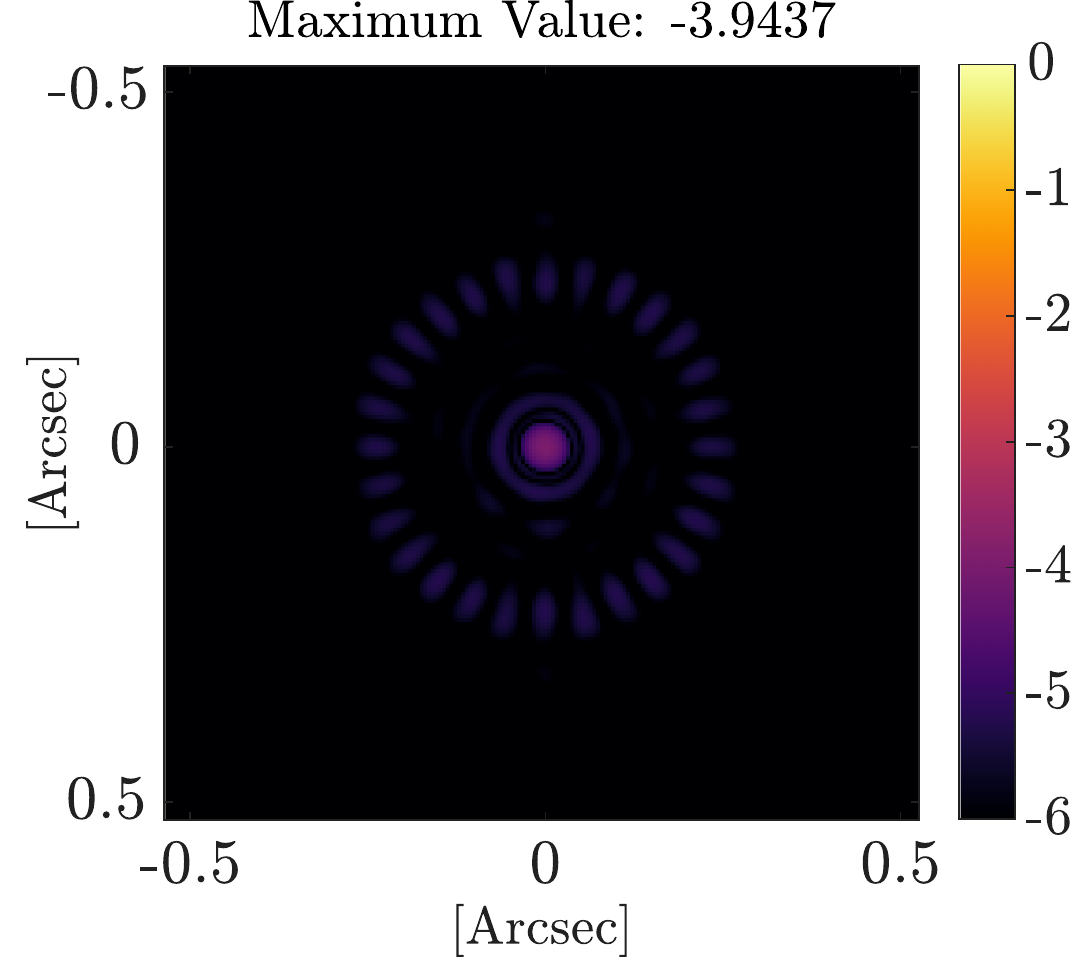}\label{psf_PCA_rot_20}}
    \subfloat[PCA$_{\text{rot}}$ - 50 nm]{\includegraphics[width=0.3\textwidth]{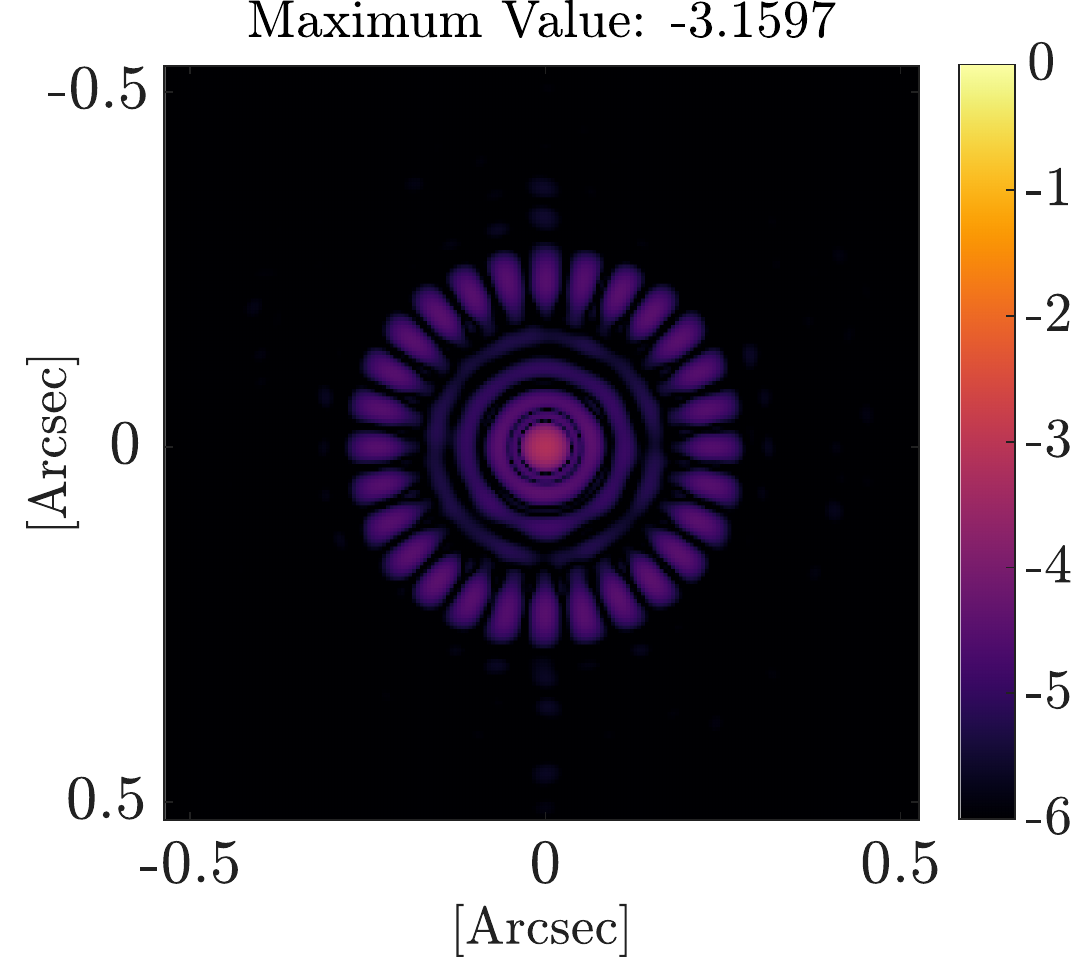}\label{psf_PCA_rot_50}}\\
    \subfloat[PCA$_{\text{X}}$ - 10 nm]{\includegraphics[width=0.3\textwidth]{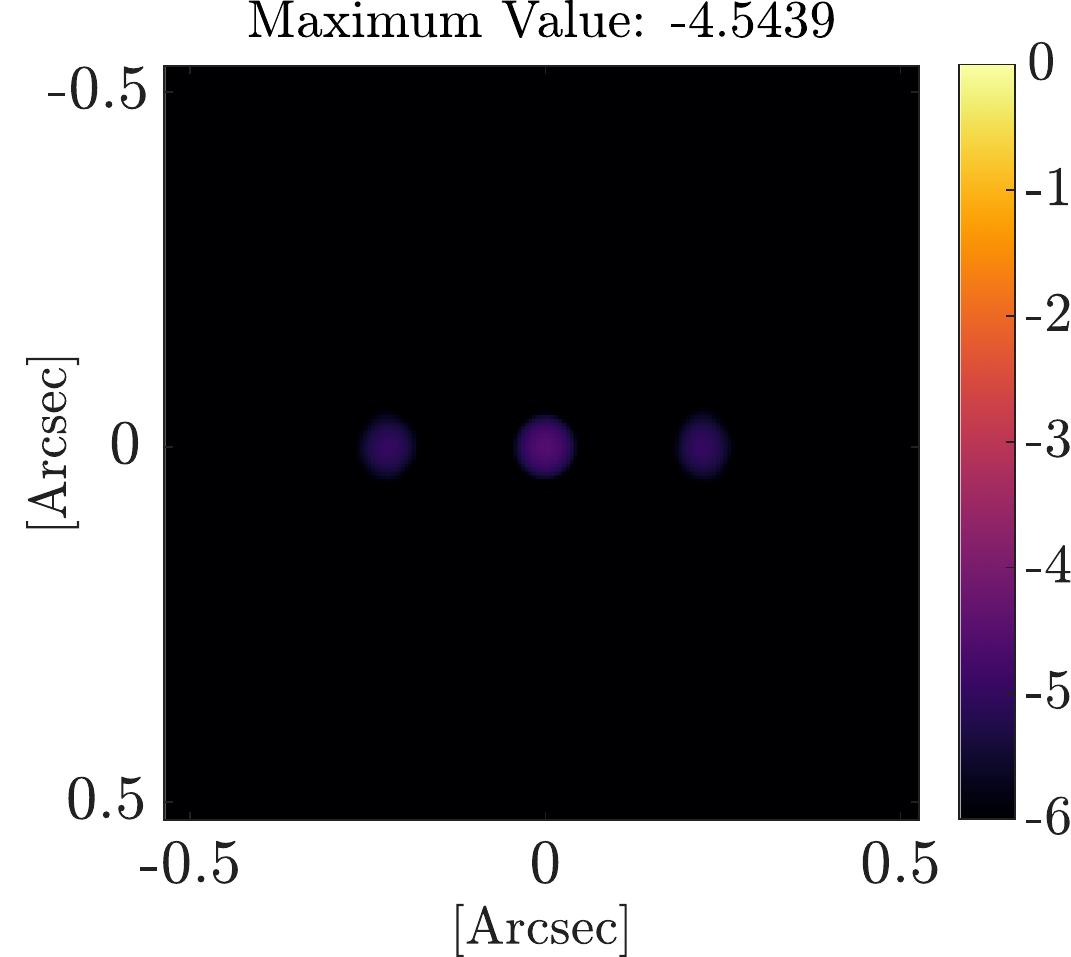}\label{psf_PCA_X_10}}
    \subfloat[PCA$_{\text{X}}$ - 20 nm]{\includegraphics[width=0.3\textwidth]{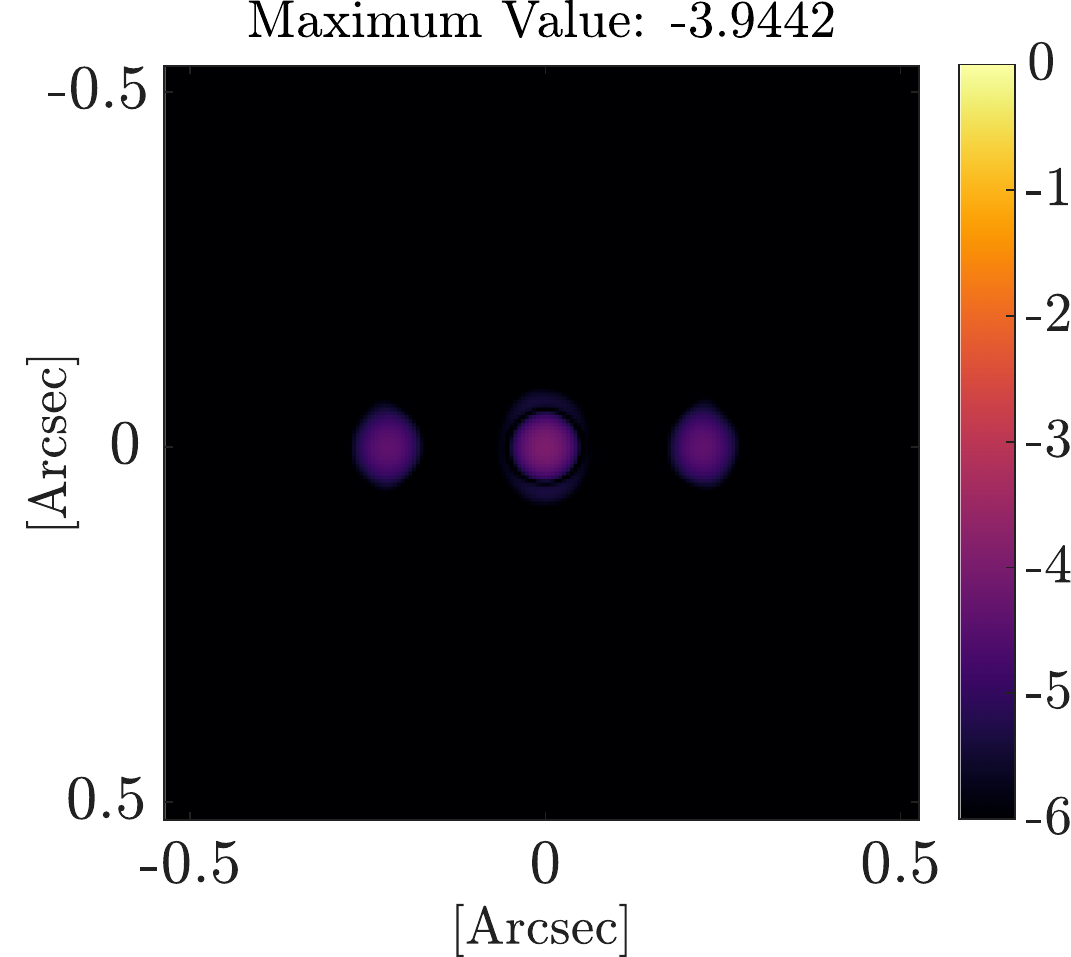}\label{psf_PCA_X_20}}
    \subfloat[PCA$_{\text{X}}$ - 50 nm]{\includegraphics[width=0.3\textwidth]{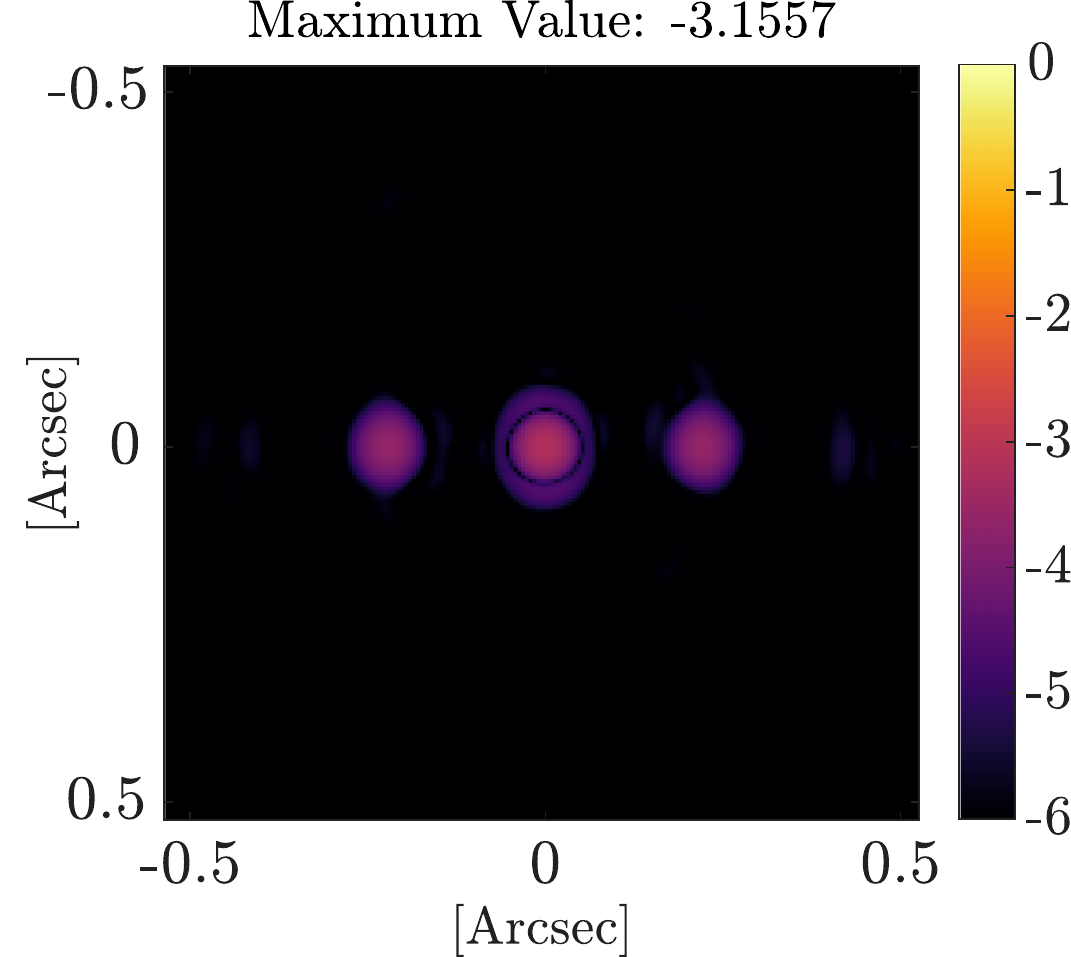}\label{psf_PCA_X_50}}\\
    
    \caption{Comparison of impact on the PSF. (a): Normalized H-Band PSF (in logarithm scale) corresponding to the reference case. From (b) to (i): Difference between the reference PSF and the cases listed in Table \ref{tab:summary_impact}.}
    \label{fig:impact_science_psf}
    \end{center}
\end{figure*}
\subsection{Impact on the scientific path }
\label{subsection_impact_science}
\subsubsection{Quantitative Analysis}
\label{quantitative_impact}
The purpose of this section is to quantify the cost of applying the push/pull measurements on the scientific path taking the cases presented in sections \ref{subsection_ramps_misReg}, \ref{subsection_multiple_misReg} and \ref{subsection_sensitivity_seeing} using 50 push/pull measurements every 50 frames at 1 kHz of one given PCA mode. The summary of the cases considered with the corresponding long exposure Strehl Ratio is given in Table \ref{tab:summary_impact}. 
\begin{table}
\small{
\begin{centering}
    \begin{tabular}{|c|c|c|c|}	
	\cline{1-4}
    \multirow{2}{*}{\textbf{Case}}&\multirow{2}{*}{\textbf{Disturbance}}&{\textbf{Mis-Registrations} }& \multirow{2}{*}{SR(H)}\\
	    &&[$\alpha_{\text{rot}}$, $\alpha_\text{X}$, $\alpha_\text{Y}$] & \\
	\cline{1-4}
	0&None&[$0^\circ$, 0 \%, 0 \%] & 79.71\\
	\cline{1-4}
	1&None&[0.3$^\circ$, -15 \%, 20 \%] & 78.25\\
	\cline{1-4}
	2&None&[0.5$^\circ$, -20 \%, 20 \%] & 55.77\\
	\cline{1-4}
	3&PCA$_{\text{rot}}$ - 10 nm&[$0^\circ$, 0 \%, 0 \%] & 79.71\\
	\cline{1-4}	
	4&PCA$_{\text{rot}}$ - 20 nm&[$0^\circ$, 0 \%, 0\%]  & 79.70\\
	\cline{1-4}
	5&PCA$_{\text{rot}}$ - 50 nm&[$0^\circ$, 0 \%, 0 \%]  & 79.65\\
	\cline{1-4}	
	6&PCA$_\text{X}$ - 10 nm&[$0^\circ$, 0 \%, 0 \%]  &79.71\\
	\cline{1-4}	
	7&PCA$_\text{X}$ - 20 nm&[$0^\circ$, 0 \%, 0 \%]  &79.70 \\
	\cline{1-4}
	8&PCA$_\text{X}$ - 50 nm&[$0^\circ$, 0 \%, 0 \%]  &79.65\\
	\cline{1-4}
	\end{tabular}
    \end{centering}}
    \caption{Summary of the cases considered to study the impact of the method in the scientific path. The units of the shifts applied ($\alpha_\text{X}$ and $\alpha_\text{Y}$) are in percentage of a subaperture. }
    \label{tab:summary_impact}
\end{table}
We limit here the study to one PCA mode at a time corresponding to the rotation (cases 3 to 5 in Table \ref{tab:summary_impact}) and to the shift X (cases 6 to 8 in Table \ref{tab:summary_impact}) with different push-pull amplitudes. Based on the results presented in the previous section, we assume that the mis-registration monitoring strategy allows to perfectly compensate for the mis-registrations so that the system with disturbance is operating around its nominal working point and only suffers from the perturbation introduced by the PCA modes. The reference case for the nominal performance consists of a closed loop system with no disturbance applied and no mis-registration (case 0 in Table \ref{tab:summary_impact}). In addition, we propose to compare the impact of our invasive strategy with respect to a small and a large mis-registration (cases 1 and 2 in Table \ref{tab:summary_impact}) that remains uncorrected. 

Table \ref{tab:summary_impact} shows that the disturbances have a negligible impact on the performance of the AO system with a maximum loss of 0.05\% of Strehl Ratio in H Band in the worst case (larger amplitude of the signals). As a comparison, the mis-registrations considered cause a loss of respectively 0.75 \% and 15\% of Strehl-Ratio. 

To provide a more detailed analysis, the modal PSD corresponding to the different cases considered in Table \ref{tab:summary_impact} are provided in Figure \ref{fig:impact_science_psd}. This Figure confirms the performance presented in Table \ref{tab:summary_impact}, exhibiting a strong impact of the mis-registrations and a negligible impact of the disturbance applied using PCA modes. In particular, we clearly identify that the disturbance due to the actuation of the PCA modes is very localised on one or a few modes. For low push-pull amplitudes, the effect is even hardly visible on the modal PSD as the curves overlap (\ref{impact_PCArot} and \ref{impact_PCAx}). On the contrary, the effect of uncompensated mis-registrations affects all the modes and is clearly visible on the modal PSD (\ref{impact_misReg}). In the case of large mis-registrations, we retrieve the behaviour introduced in Figure \ref{impact_misReg} with instabilities for the higher spatial frequencies. 

In addition, the corresponding effects on the scientific PSF are provided in Figure \ref{fig:impact_science_psf}. We consider long exposure closed loop PSF in H band (2.5 seconds integration at 1 kHZ with no source of noise) as a reference. This reference PSF is normalized to 1 and all the other PSF are normalized using the same normalization factor to provide a relative comparison with respect to this nominal case. The PSF displayed in Figure \ref{psf_small_mis_reg} to \ref{psf_PCA_rot_50} correspond to the difference between the reference PSF \ref{psf_ref} and the PSF obtained in the different cases listed in Table \ref{tab:summary_impact}. For each case, the maximum value of the residual PSF is indicated.

This Figure shows that the effects of the mis-registrations considered are well visible inside the correction zone of the PSF. The shape of the PSF results to be quite altered, especially in the case of a large mis-registration where the maximum value of the delta PSF is around $10^{-0.6}$. 
On the contrary, the impact of the disturbance applied using the PCA modes appears to be of a much smaller amplitude with a maximum value of about $10^{-3}$ for the highest amplitude. This represents a factor 1000 with respect to the small mis-registered case and this factor goes to 10 000 when considering the smallest push/pull amplitude. Figure \ref{fig:impact_science_psf} allows also to quantify clearly how the PSF is spatially impacted by the modes selected as a function of the amplitude of the signal.

The results provided in this section give confidence into using the proposed measurement strategy even during the observation as the impact on the observation appears negligible. This conclusion is however system dependent and the impact should be carefully evaluated beforehand. The next section proposes to define a methodology to select the modes used to track the mis-registrations. 

\subsubsection{Qualitative Analysis}
Section \ref{quantitative_impact} has shown that the impact of the perturbation introduced in the scientific path depends on the amplitude and spatial properties of the signals. In terms of operation, the choice of the signals properties has to be tailored to the observing conditions (level of noise?, turbulence?, level of AO correction?), on the accuracy requirements and on the type of scientific observation (impact in the focal plane? on the performance?). The measurement strategy results then of a trade-off between all these different considerations and will be system dependent. We propose here to define a methodology to identify the modes that are the most relevant to estimate the mis-registrations. We recall the expression of the measurement noise $\boldsymbol{\xi_k}$:
\begin{equation}
\boldsymbol{\xi_k}=\frac{-\wfs.\gDelta \phiRes{k}+\boldsymbol{\eta_{k}}-\boldsymbol{\eta_{k+1}}}{2a}
\label{push_pull_noise_2}
\end{equation}\\
The procedure should include:
\begin{itemize}
    \item Definition of the requirements in terms of accuracy for the mis-registration parameters with a sensitivity analysis of the system (see Figure \ref{fig:misRegImpactScience}). 
    \item Identification of the observing conditions: level of noise $\boldsymbol{\eta}$, level of turbulence (\gDelta\phiRes{}) and level of AO correction.
    \item Investigation of the impact on the science path (impact of the modes on the PSF? AO performance? Acquisition time $T$ allocated to identify the parameters?) to identify the constraints for the amplitude $a$ and the spatial properties of the modes considered.
    \item Determine the trade-off between number of modes $N_{modes}$, the acquisition time allocated $T$, amplitude $a$ required to reach the accuracy targeted.
\end{itemize}
If the amplitudes of the modes are small (typically 20 nm RMS), and the measurement time required as well (hundreds of frames for a few PCA modes), the impact will be negligible on long-exposures PSF as shown in \ref{quantitative_impact}. By contrast if higher amplitudes are required, the dithering could be applied during the read-out time of the detectors, when the scientific shutter is closed. This would provide a way to regularly acquire high SNR signals during the operations without impacting the scientific path. 


\section{Conclusion}

To address the question of a regular tracking of the DM/WFS mis-registrations during scientific operations, we introduced a strategy based on an invasive approach. This strategy is inspired from the state of the art in terms of on-sky calibration and consists in applying these calibration techniques (modulation/demodulation) to few well selected modes and estimate from them the mis-registration parameters. We demonstrated that the method is applicable for both SH-WHS and PWFS as the algorithm includes a compensation for the PWFS optical gains. 

Our research is oriented to minimize the number of modes required by the algorithm to estimate accurately the mis-registration parameters. This is by identifying the most sensitive modes to the mis-registrations using a Principal Component Analysis of the sensitivity interaction matrices. 

Using on-sky push-pull measurements, we investigated the accuracy achieved with only 3 PCA modes exploring different observing conditions. Even in very low flux conditions, it is always possible to tune the push/pull amplitude and the measurement time to reach a very good estimation accuracy. In addition, it has been shown that the impact of these on-sky disturbance on the quality of the science PSF is fully negligible. 

At last, we demonstrated that this procedure is performing extremely well for various mis-registrations evolving dynamically at the same time By using only 3 PCA modes with an amplitude of 20 nm RMS, we could provide a tracking of the mis-registration parameters with an accuracy better than 1\% of a subaperture. 

The future step will require an experimental validation of the method, implementing it on an existing facility equipped with a secondary adaptive mirror. In addition, it will be relevant to investigate if this calibration strategy allows to retrieve other parameters such as optical gains for the PWFS, for instance to compensate properly Non Common Path Aberrations (\citealt{esposito2020sky}). 

In addition, it will be required to study how this novel method performs taking into account more complex closed-loop effects specific to large adaptive telescopes equipped with a large adaptive secondary mirror. The couplings with pupil fragmentation effects due to the presence of thick spiders, deformable mirror saturation and the segmentation of the primary mirror will have to be investigated.
In particular, in this paper, we put light on the coupling between the optical gains of PWFS and the presence of mis-registrations. In the context of the ELT, PWFS optical gains are expected to exhibit large variations (\citealt{deo2018modal}) which will require aggressive compensation strategies (\citealt{deo2019close}, \citealt{deo2021correlation}). An accurate tracking and compensation of the mis-registration will be required to prevent any bias in the optical gains estimation that would lead to loop instabilities or over/under compensation of Non-Common Path Aberrations (\citealt{esposito2020sky}). 

\section{acknowledgements}
We thank the referee for constructive comments that helped strengthen the paper.
This document has been prepared as part of the activities of OPTICON H2020 (2017-2020) Work Package 1 (Calibration and test tools for AO assisted E-ELT instruments). OPTICON is supported by the Horizon 2020 Framework Programme of  the  European  Commission’s  (Grant  number  730890). This work was also supported by the Action Spécifique Haute Résolution Angulaire (ASHRA) of CNRS/INSU co-funded by CNES and benefited from the support of the WOLF project ANR-18-CE31-0018 of the French National Research Agency (ANR).

\section*{Data Availability}

The data and simulation code underlying this article will be shared on reasonable request to the corresponding author.



\bibliographystyle{mnras}
\bibliography{main} 

\begin{thebibliography}{}
\makeatletter
\relax
\def\mn@urlcharsother{\let\do\@makeother \do\$\do\&\do\#\do\^\do\_\do\%\do\~}
\def\mn@doi{\begingroup\mn@urlcharsother \@ifnextchar [ {\mn@doi@}
  {\mn@doi@[]}}
\def\mn@doi@[#1]#2{\def\@tempa{#1}\ifx\@tempa\@empty \href
  {http://dx.doi.org/#2} {doi:#2}\else \href {http://dx.doi.org/#2} {#1}\fi
  \endgroup}
\def\mn@eprint#1#2{\mn@eprint@#1:#2::\@nil}
\def\mn@eprint@arXiv#1{\href {http://arxiv.org/abs/#1} {{\tt arXiv:#1}}}
\def\mn@eprint@dblp#1{\href {http://dblp.uni-trier.de/rec/bibtex/#1.xml}
  {dblp:#1}}
\def\mn@eprint@#1:#2:#3:#4\@nil{\def\@tempa {#1}\def\@tempb {#2}\def\@tempc
  {#3}\ifx \@tempc \@empty \let \@tempc \@tempb \let \@tempb \@tempa \fi \ifx
  \@tempb \@empty \def\@tempb {arXiv}\fi \@ifundefined
  {mn@eprint@\@tempb}{\@tempb:\@tempc}{\expandafter \expandafter \csname
  mn@eprint@\@tempb\endcsname \expandafter{\@tempc}}}

\bibitem[\protect\citeauthoryear{Arsenault et~al.,}{Arsenault
  et~al.}{2008}]{arsenault2006AOF}
Arsenault R.,  et~al., 2008. pp 7015 -- 7015

\bibitem[\protect\citeauthoryear{B{\'e}chet, Kolb, Madec, Tallon  \&
  Thi{\'e}baut}{B{\'e}chet et~al.}{2011}]{bechet1a2011identification}
B{\'e}chet C.,  Kolb J.,  Madec P.-Y.,  Tallon M.,   Thi{\'e}baut E.,  2011,
  AO4ELT II Conference

\bibitem[\protect\citeauthoryear{B{\'e}chet, Tallon  \&
  Thi{\'e}baut}{B{\'e}chet et~al.}{2012}]{bechet2012optimization}
B{\'e}chet C.,  Tallon M.,   Thi{\'e}baut E.,  2012, in SPIE Astronomical
  Telescopes+ Instrumentation. pp 84472C--84472C

\bibitem[\protect\citeauthoryear{Bertrou-Cantou et~al.,}{Bertrou-Cantou
  et~al.}{2020}]{bertrou2020petalometry}
Bertrou-Cantou A.,  et~al., 2020, in Adaptive Optics Systems VII. p. 1144812

\bibitem[\protect\citeauthoryear{Biasi, Gallieni, Salinari, Riccardi  \&
  Mantegazza}{Biasi et~al.}{2010}]{biasi2010contactless}
Biasi R.,  Gallieni D.,  Salinari P.,  Riccardi A.,   Mantegazza P.,  2010, in
  Adaptive Optics Systems II. p. 77362B

\bibitem[\protect\citeauthoryear{Bonnefond, Tallon, Le~Louarn  \&
  Madec}{Bonnefond et~al.}{2016}]{bonnefond2016wavefront}
Bonnefond S.,  Tallon M.,  Le~Louarn M.,   Madec P.-Y.,  2016, in Adaptive
  Optics Systems V. p. 990972

\bibitem[\protect\citeauthoryear{Bonnet et~al.,}{Bonnet
  et~al.}{2018}]{bonnet2018adaptive}
Bonnet H.,  et~al., 2018, in Adaptive Optics Systems VI. p. 1070310

\bibitem[\protect\citeauthoryear{Boyer, Michau  \& Rousset}{Boyer
  et~al.}{1990}]{boyer1990adaptive}
Boyer C.,  Michau V.,   Rousset G.,  1990, in Adaptive Optics and Optical
  Structures. pp 63--82

\bibitem[\protect\citeauthoryear{Briguglio et~al.,}{Briguglio
  et~al.}{2018a}]{briguglio2018optical}
Briguglio R.,  et~al., 2018a, Scientific Reports, 8, 1

\bibitem[\protect\citeauthoryear{Briguglio, Pariani, Xompero, Riccardi,
  Tintori, Gallieni  \& Biasi}{Briguglio et~al.}{2018b}]{briguglio2018possible}
Briguglio R.,  Pariani G.,  Xompero M.,  Riccardi A.,  Tintori M.,  Gallieni
  D.,   Biasi R.,  2018b, in Adaptive Optics Systems VI. p. 1070379

\bibitem[\protect\citeauthoryear{Chambouleyron, Fauvarque, Janin-Potiron,
  Correia, Sauvage, Schwartz, Neichel  \& Fusco}{Chambouleyron
  et~al.}{2020}]{chambouleyron2020pyramid}
Chambouleyron V.,  Fauvarque O.,  Janin-Potiron P.,  Correia C.,  Sauvage
  J.-F.,  Schwartz N.,  Neichel B.,   Fusco T.,  2020, arXiv preprint
  arXiv:2006.08294

\bibitem[\protect\citeauthoryear{Cheffot, Vigan, L{\'e}v{\^e}que  \&
  Hugot}{Cheffot et~al.}{2020}]{cheffot2020measuring}
Cheffot A.-L.,  Vigan A.,  L{\'e}v{\^e}que S.,   Hugot E.,  2020, Optics
  Express, 28, 12566

\bibitem[\protect\citeauthoryear{Chiuso, Muradore  \& Marchetti}{Chiuso
  et~al.}{2010}]{chiuso2010dynamic}
Chiuso A.,  Muradore R.,   Marchetti E.,  2010, IEEE Transactions on Control
  Systems Technology, 18, 705

\bibitem[\protect\citeauthoryear{Cirasuolo et~al.,}{Cirasuolo
  et~al.}{2018}]{cirasuolo2018elt}
Cirasuolo M.,  et~al., 2018, The Messenger, 171, 20

\bibitem[\protect\citeauthoryear{Conan, Correia  et~al.}{Conan
  et~al.}{2014}]{conan2014object}
Conan R.,  Correia C.,   et~al., 2014.

\bibitem[\protect\citeauthoryear{Deo, Gendron, Rousset, Vidal  \& Buey}{Deo
  et~al.}{2018}]{deo2018modal}
Deo V.,  Gendron {\'E}.,  Rousset G.,  Vidal F.,   Buey T.,  2018, in Adaptive
  Optics Systems VI. p. 1070320

\bibitem[\protect\citeauthoryear{Deo et~al.,}{Deo et~al.}{2019a}]{deo2019close}
Deo V.,  et~al., 2019a, in 6th AO4ELT conference-Adaptive Optics for Extremely
  Large Telescopes.

\bibitem[\protect\citeauthoryear{Deo, Gendron, Rousset, Vidal, Sevin, Ferreira,
  Gratadour  \& Buey}{Deo et~al.}{2019b}]{deo2019telescope}
Deo V.,  Gendron {\'E}.,  Rousset G.,  Vidal F.,  Sevin A.,  Ferreira F.,
  Gratadour D.,   Buey T.,  2019b, Astronomy \& Astrophysics, 629, A107

\bibitem[\protect\citeauthoryear{Deo et~al.,}{Deo
  et~al.}{2021}]{deo2021correlation}
Deo V.,  et~al., 2021, arXiv preprint arXiv:2103.09921

\bibitem[\protect\citeauthoryear{Esposito, Pinna, Puglisi, Agapito, Veran  \&
  Herriot}{Esposito et~al.}{2015}]{esposito2015NCPA}
Esposito S.,  Pinna E.,  Puglisi A.,  Agapito G.,  Veran J.,   Herriot G.,
  2015, in Adaptive Optics for Extremely Large Telescopes 4--Conference
  Proceedings.

\bibitem[\protect\citeauthoryear{Esposito, Puglisi, Pinna, Agapito,
  Quir{\'o}s-Pacheco, V{\'e}ran  \& Herriot}{Esposito
  et~al.}{2020}]{esposito2020sky}
Esposito S.,  Puglisi A.,  Pinna E.,  Agapito G.,  Quir{\'o}s-Pacheco F.,
  V{\'e}ran J.,   Herriot G.,  2020, Astronomy \& Astrophysics, 636, A88

\bibitem[\protect\citeauthoryear{Fauvarque, Janin-Potiron, Correia,
  Br{\^u}l{\'e}, Neichel, Chambouleyron, Sauvage  \& Fusco}{Fauvarque
  et~al.}{2019}]{fauvarque2019kernel}
Fauvarque O.,  Janin-Potiron P.,  Correia C.,  Br{\^u}l{\'e} Y.,  Neichel B.,
  Chambouleyron V.,  Sauvage J.-F.,   Fusco T.,  2019, JOSA A, 36, 1241

\bibitem[\protect\citeauthoryear{Fusco, Conan, Rousset, Mugnier  \&
  Michau}{Fusco et~al.}{2001}]{fusco2001optimal}
Fusco T.,  Conan J.,  Rousset G.,  Mugnier L.,   Michau V.,  2001, JOSA A, 18,
  2527

\bibitem[\protect\citeauthoryear{Gendron}{Gendron}{1995}]{gendron1995optimisation}
Gendron E.,  1995, PhD thesis

\bibitem[\protect\citeauthoryear{Gilmozzi \& Spyromilio}{Gilmozzi \&
  Spyromilio}{2007}]{gilmozzi2007european}
Gilmozzi R.,  Spyromilio J.,  2007, The Messenger

\bibitem[\protect\citeauthoryear{Hartmann}{Hartmann}{1900}]{hartmann1900bermerkungen}
Hartmann J.,  1900, Zeitschrift {f\"ur} Instrumentenkunde, 20, 47

\bibitem[\protect\citeauthoryear{Heritier et~al.,}{Heritier
  et~al.}{2017}]{heritier2017overview}
Heritier C.~T.,  et~al., 2017.

\bibitem[\protect\citeauthoryear{Heritier et~al.,}{Heritier
  et~al.}{2018}]{heritier2018new}
Heritier C.~T.,  et~al., 2018, Monthly Notices of the Royal Astronomical
  Society

\bibitem[\protect\citeauthoryear{Kolb, Madec, Louarn, Muller  \&
  B{\'e}chet}{Kolb et~al.}{2012}]{kolb2012calibration}
Kolb J.,  Madec P.-Y.,  Louarn M.~L.,  Muller N.,   B{\'e}chet C.,  2012, in
  Proc. of SPIE Vol. pp 84472D--1

\bibitem[\protect\citeauthoryear{Korkiakoski, V{\'e}rinaud  \&
  Le~Louarn}{Korkiakoski et~al.}{2008}]{korkiakoski2008improving}
Korkiakoski V.,  V{\'e}rinaud C.,   Le~Louarn M.,  2008, Applied optics, 47, 79

\bibitem[\protect\citeauthoryear{Lai, Chun, Dungee, Lu  \& Carbillet}{Lai
  et~al.}{2020}]{lai2020crime}
Lai O.,  Chun M.,  Dungee R.,  Lu J.,   Carbillet M.,  2020, Monthly Notices of
  the Royal Astronomical Society

\bibitem[\protect\citeauthoryear{Le~Louarn, Madec, Marchetti, Bonnet  \&
  Esselborn}{Le~Louarn et~al.}{2016}]{le2016simulations}
Le~Louarn M.,  Madec P.-Y.,  Marchetti E.,  Bonnet H.,   Esselborn M.,  2016,
  in Adaptive Optics Systems V. p. 990975

\bibitem[\protect\citeauthoryear{Madec}{Madec}{2012}]{madec2012overview}
Madec P.-Y.,  2012, in Adaptive Optics Systems III. p. 844705

\bibitem[\protect\citeauthoryear{Marquardt}{Marquardt}{1963}]{marquardt1963algorithm}
Marquardt D.~W.,  1963, Journal of the society for Industrial and Applied
  Mathematics, 11, 431

\bibitem[\protect\citeauthoryear{Meimon, Delavaquerie, Cassaing, Fusco, Mugnier
   \& Michau}{Meimon et~al.}{2008}]{meimon2008phasing}
Meimon S.,  Delavaquerie E.,  Cassaing F.,  Fusco T.,  Mugnier L.~M.,   Michau
  V.,  2008, in Ground-based and Airborne Telescopes II. p. 701214

\bibitem[\protect\citeauthoryear{Neichel, Parisot, Petit, Fusco  \&
  Rigaut}{Neichel et~al.}{2012}]{neichel2012identification}
Neichel B.,  Parisot A.,  Petit C.,  Fusco T.,   Rigaut F.,  2012, in Proc.
  SPIE. p. 84475N

\bibitem[\protect\citeauthoryear{Oberti, H~Bonnet, Fedrigo, Ivanescu, Kasper
  \& Paufique}{Oberti et~al.}{2004}]{oberti2004calibration}
Oberti S.,  H~Bonnet H.,  Fedrigo E.,  Ivanescu L.,  Kasper M.,   Paufique J.,
  2004, in Advancements in Adaptive Optics. pp 139--151

\bibitem[\protect\citeauthoryear{Oberti et~al.,}{Oberti
  et~al.}{2006}]{oberti2006large}
Oberti S.,  et~al., 2006, in Advances in Adaptive Optics II. p. 627220

\bibitem[\protect\citeauthoryear{Oberti et~al.,}{Oberti
  et~al.}{2018}]{oberti2018AOinTheAOF}
Oberti S.,  et~al., 2018. p.~20

\bibitem[\protect\citeauthoryear{Pearson}{Pearson}{1901}]{pearson1901principal}
Pearson K.,  1901, The London, Edinburgh, and Dublin Philosophical Magazine and
  Journal of Science, 6, 559

\bibitem[\protect\citeauthoryear{Pieralli, Puglisi, Quir{\'o}s-Pacheco  \&
  Esposito}{Pieralli et~al.}{2008}]{pieralli2008sinusoidal}
Pieralli F.,  Puglisi A.,  Quir{\'o}s-Pacheco F.,   Esposito S.,  2008,
  Adaptive Optics Systems N. Norbert Hubin and EM Claire and and PL Wizinowich
  and eds, 7015, 70153A

\bibitem[\protect\citeauthoryear{Pinna et~al.,}{Pinna
  et~al.}{2012}]{pinna2012first}
Pinna E.,  et~al., 2012, in Proc. SPIE. p. 84472B

\bibitem[\protect\citeauthoryear{Ragazzoni}{Ragazzoni}{1996}]{ragazzoni1996pupil}
Ragazzoni R.,  1996, Journal of modern optics, 43, 289

\bibitem[\protect\citeauthoryear{Riccardi et~al.,}{Riccardi
  et~al.}{2010}]{riccardi2010adaptive}
Riccardi A.,  et~al., 2010, in Proc. SPIE. p. 77362C

\bibitem[\protect\citeauthoryear{Schwartz et~al.,}{Schwartz
  et~al.}{2018}]{schwartz2018analysis}
Schwartz N.,  et~al., 2018, in Adaptive Optics Systems VI. p. 1070322

\bibitem[\protect\citeauthoryear{Shack}{Shack}{1971}]{shack1971production}
Shack R.~V.,  1971, J. Opt. Soc. Am., 61, 656

\bibitem[\protect\citeauthoryear{Southwell}{Southwell}{1980}]{southwell1980wave}
Southwell W.~H.,  1980, JOSA, 70, 998

\bibitem[\protect\citeauthoryear{Vernet, Cayrel, Hubin, Mueller, Biasi,
  Gallieni  \& Tintori}{Vernet et~al.}{2012}]{vernet2012specifications}
Vernet E.,  Cayrel M.,  Hubin N.,  Mueller M.,  Biasi R.,  Gallieni D.,
  Tintori M.,  2012, in Adaptive Optics Systems III.

\bibitem[\protect\citeauthoryear{Vernet, Cayrel, Hubin, Biasi, Gallieni  \&
  Tintori}{Vernet et~al.}{2014}]{vernet2014way}
Vernet E.,  Cayrel M.,  Hubin N.,  Biasi R.,  Gallieni D.,   Tintori M.,  2014,
  in Adaptive Optics Systems IV. p. 914824

\bibitem[\protect\citeauthoryear{Wallner}{Wallner}{1983}]{wallner1983optimal}
Wallner E.,  1983, JOSA, 73, 1771

\makeatother
\end{thebibliography}



\newpage
\section*{Appendix}
\appendix
\section{Complementary Results With Shack-Hartmann Wave-Front Sensor}
\label{appendix_SH}
\begin{figure*}
    \begin{center}
\stackunder[10pt]{\subfloat[Push-Pull Amplitude 10 nm]{\includegraphics[width=0.31\textwidth]{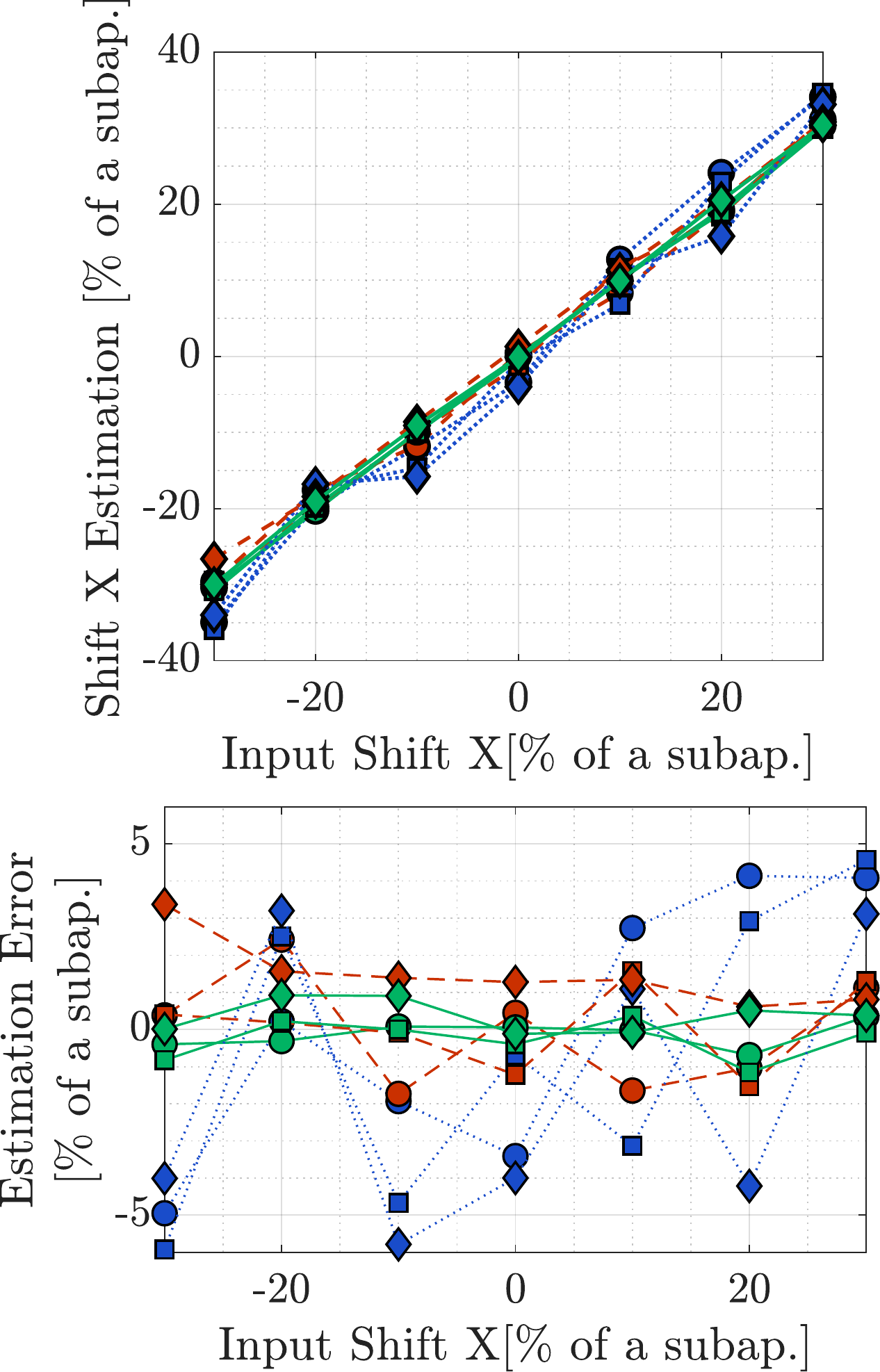}}}{ \includegraphics[width = 0.2 cm]{plots/circle.pdf} : Wind Speed 10 m/s }\hspace{0.1cm}
    \stackunder[10pt]{\subfloat[Push-Pull Amplitude 20 nm]{\includegraphics[width=0.31\textwidth]{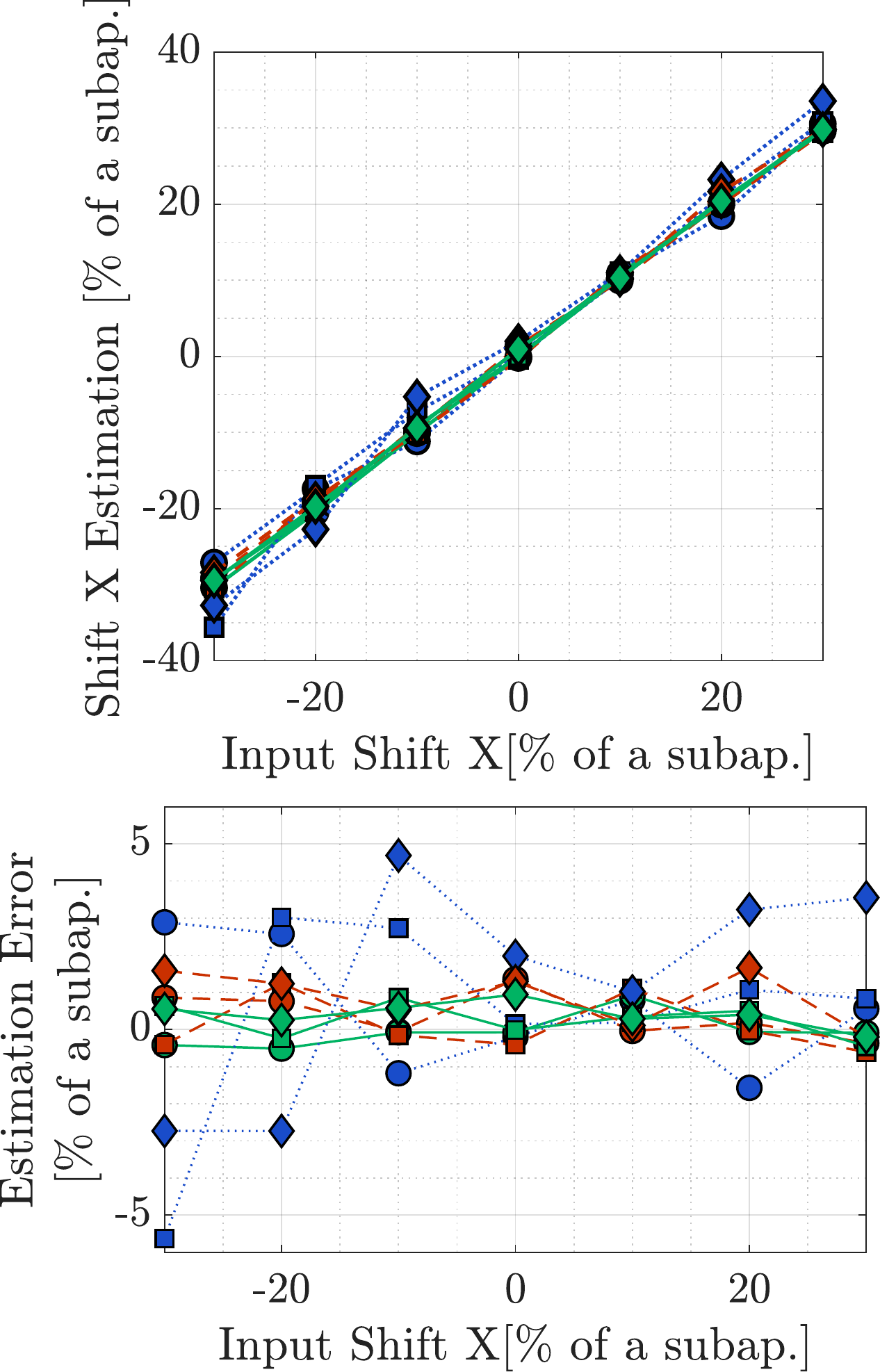}}}{\includegraphics[width = 0.2 cm]{plots/square.pdf} : Wind Speed 20 m/s}\hspace{0.1cm}
    \stackunder[10pt]{\subfloat[Push-Pull Amplitude 50 nm]{\includegraphics[width=0.31\textwidth]{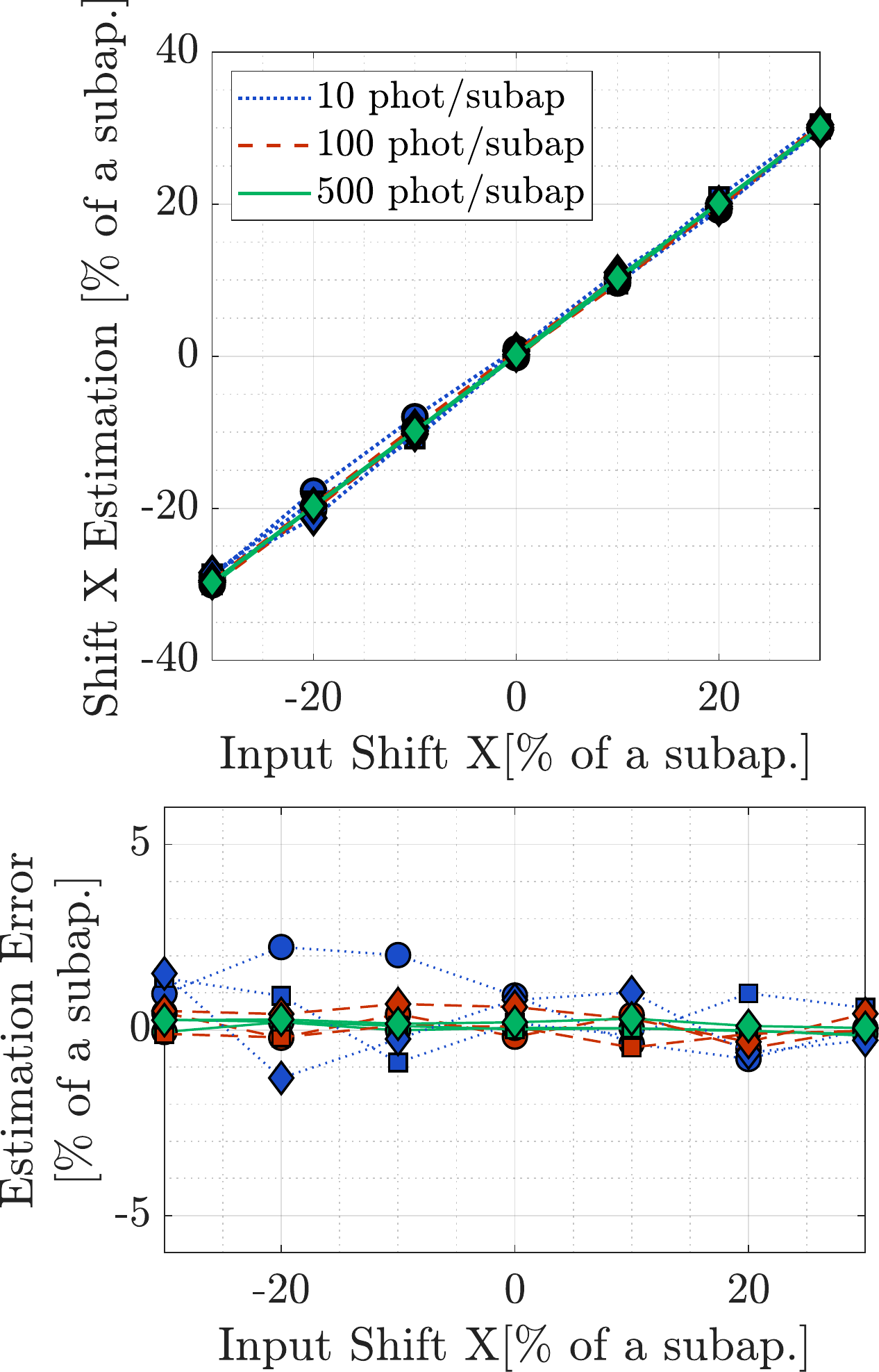}}}{\includegraphics[width = 0.2 cm]{plots/diamond.pdf} : Wind Speed 30 m/s }%
    \vspace{0.05 cm}
    \caption{Shift X estimation (top) and corresponding estimation error (bottom) as a function of the input shift X for a push-pull amplitude of 10 nm (a), 20 nm (b) and 50 nm (c) using a SH-WFS. The results are given for different noise regimes (dotted blue, dashed red and solid green lines). The markers correspond to different wind-speeds (bullet, squares and diamonds) }
    \label{fig:invasive_ramp_X_SH}
    \end{center}{}
\end{figure*}

\begin{figure*}
    \begin{center}
    \stackunder[10pt]{\subfloat[Push-Pull Amplitude 10 nm]{\includegraphics[width=0.31\textwidth]{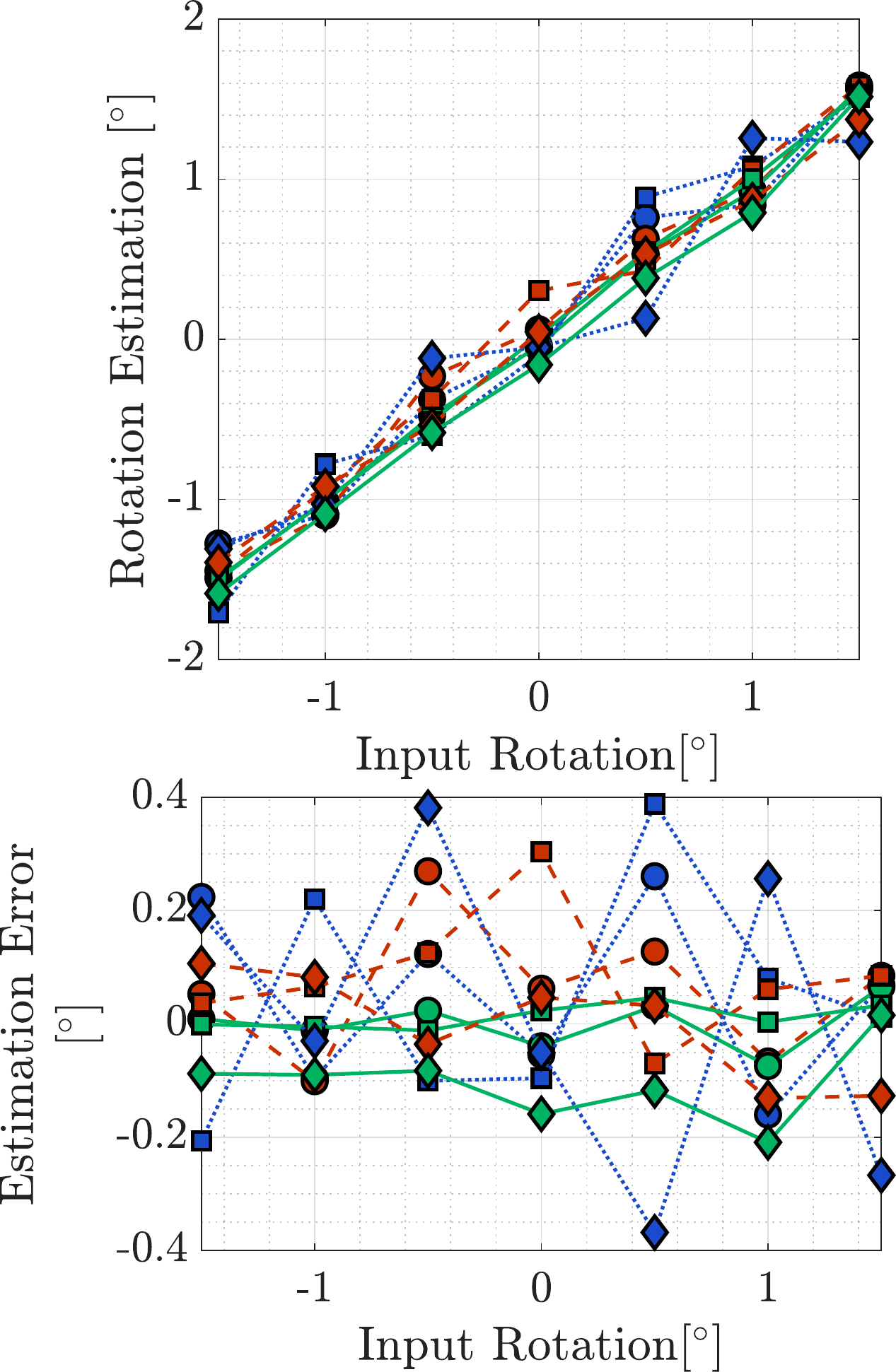}}}{ \includegraphics[width = 0.2 cm]{plots/circle.pdf} : Wind Speed 10 m/s }\hspace{0.1cm}
    \stackunder[10pt]{\subfloat[Push-Pull Amplitude 20 nm]{\includegraphics[width=0.31\textwidth]{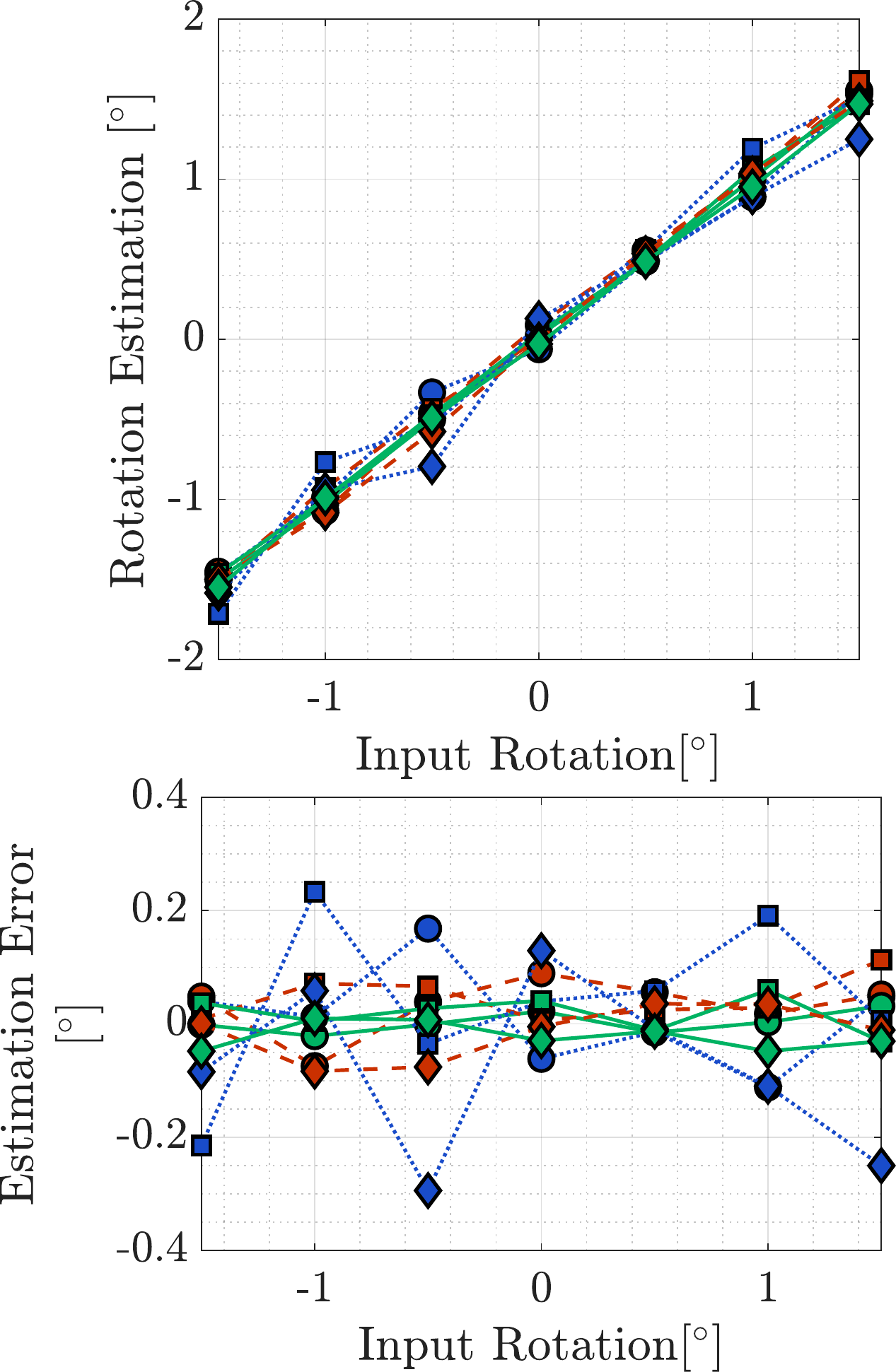}}}{\includegraphics[width = 0.2 cm]{plots/square.pdf} : Wind Speed 20 m/s}\hspace{0.1cm}
    \stackunder[10pt]{\subfloat[Push-Pull Amplitude 50 nm]{\includegraphics[width=0.31\textwidth]{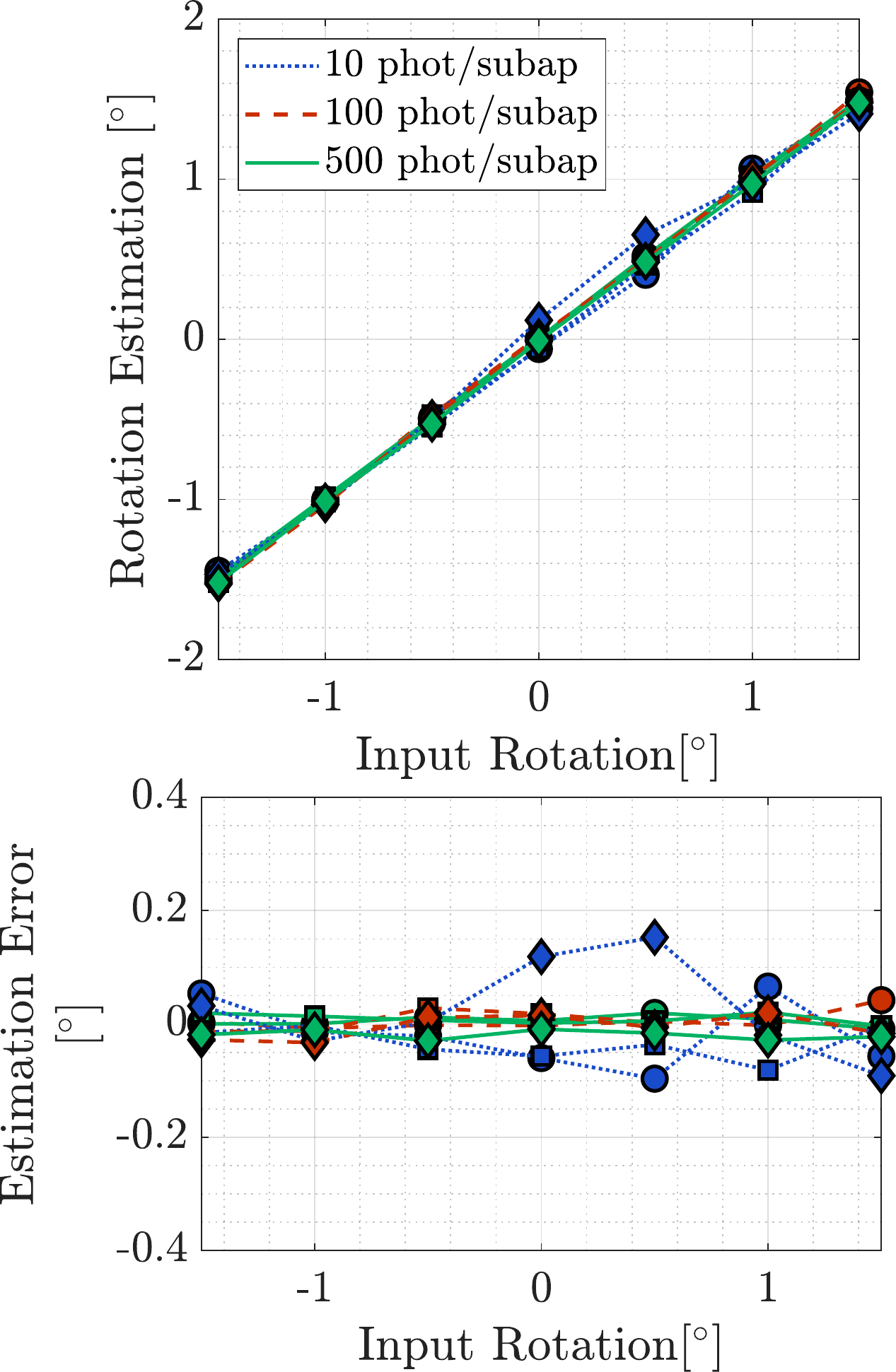}}}{\includegraphics[width = 0.2 cm]{plots/diamond.pdf} : Wind Speed 30 m/s }%
    \vspace{0.05 cm}
    \caption{Rotation estimation (top) and corresponding estimation error (bottom) as a function of the input shift X for a push-pull amplitude of 10 nm (a), 20 nm (b) and 50 nm (c) using a SH-WFS. The results are given for different noise regimes (dotted blue, dashed red and solid green lines). The markers correspond to different wind-speeds (bullet, squares and diamonds) }
    \label{fig:invasive_ramp_rot_SH}
    \end{center}{}
\end{figure*}


\bsp	
\label{lastpage}
\end{document}